\documentclass[journal ]{new-aiaa}
\usepackage[utf8]{inputenc}
\usepackage{textcomp}

\usepackage{multirow}
\usepackage{subfigure}

\usepackage{graphicx}
\usepackage{amsmath}
\usepackage[version=4]{mhchem}
\usepackage{siunitx}
\usepackage{longtable,tabularx}
\title{Deep Learning for Multi-Fidelity Aerodynamic Distribution Modeling from Experimental and Simulation Data}

\author{Kai Li \footnote{Graduate Student, School of Aeronautics, International Joint Institute of Artificial Intelligence on Fluid Mechanics; likay@mail.nwpu.edu.cn.}, Jiaqing Kou.\footnote{Graduate Student, School of Aeronautics, International Joint Institute of Artificial Intelligence on Fluid Mechanics; koujiaqing93@163.com.}, and Weiwei Zhang \footnote{Professor, School of Aeronautics, International Joint Institute of Artificial Intelligence on Fluid Mechanics; aeroelastic@nwpu.edu.cn (Corresponding Author).}}
\affil{Northwestern Polytechnical University, 710072 Xi’an, People’s Republic of China}

\begin{document}

\maketitle

\begin{abstract}
The wind-tunnel experiment plays a critical role in the design and development phases of modern aircraft, which is limited by prohibitive cost. In contrast, numerical simulation, as an important alternative paradigm, mimics complex flow behaviors but is less accurate compared to experiment. This leads to the recent development and emerging interest in applying data fusion for aerodynamic prediction. In particular, the accurate prediction of aerodynamic with lower computational cost can be achieved by fusing experimental (high-fidelity) and computational (low-fidelity) aerodynamic data. Currently, existing works on aerodynamic model using data fusion mainly concern integral data (lift, drag, etc.). In this paper, a multi-fidelity aerodynamic model based on Deep Neural Network (DNN), where both numerical and experimental data are introduced in the loss function with proper weighting factor to balance the overall accuracy, is developed for aerodynamic distribution over the wing surface. Specially, the proposed approach is illustrated by modeling the surface pressure distribution of ONERA M6 wing in transonic flow, including different secnarios where the flow condition varies or there is less high-fidelity data. The results demonstrate that the proposed approach with decent number of low-fidelity data and a few high-fidelity data can accurately predict the surface pressure distribution on transonic wing. The outperformance of the proposed model over other DNNs from only high-fidelity or low-fidelity, has been reported.
\end{abstract}

\section*{Nomenclature}

{\renewcommand\arraystretch{1.0}
\noindent\begin{longtable*}{@{}l @{\quad=\quad} l@{}}
$Ma$  & Mach number \\
$\alpha$ &    angle of attack, deg \\
$\bf{\theta}$  & weight of neural network \\
$J$ &    loss function \\
$\rho$ &    coefficient of low-fidelity data \\
$C_p$& pressure coefficient \\
$b$   & semi-span, m \\
$c$   & chord, m     \\
$X,Y$   & coordinate for the wing surface, m     \\
\multicolumn{2}{@{}l}{Subscripts}\\
$exp$ & experiment\\
$com$ & computation
\end{longtable*}}

\section{Introduction}
\lettrine{T}{he} research in fluid dynamics mainly consists of experimental fluids dynamics, theoretical fluids dynamics, and computational fluids dynamics (CFD) \cite{Anderson1985computational}. Among these disciplines, wind-tunnel experiment and CFD are two main sources to obtain aerodynamic data,which can be divided into two types: integral quantities (lift, drag, and moment coefficients) and distributed data (surface pressure and skin friction distributions). Wind-tunnel experiment plays a pivotal role in the design and development phases of modern aircraft. However, it is extremely difficult to obtain abundant and high-fidelity aerodynamic distribution due to the limitation of pressure measuring holes and experimental cost \cite{zhao_research_2021}. In contrast, there has been tremendous progress in the accuracy and efficiency of CFD, but the results from CFD still need to be validated and verified by wind-tunnel experiment \cite{OBERKAMPF2002209}.

Due to the vast and increasing volume of data over the past few years, many researchers are capable of constructing data-driven aerodynamic model using data from experiments, field measurements, and numerical simulations. So far, the data-driven model has attracted great attention owing to a good balance between the efficiency and accuracy. The data-driven model can be divided into three groups: semi-empirical models, reduced-order models (ROMs), and surrogate models. Semi-empirical models utilizes linear and nonlinear difference Equations to describe the physical characterises of flow, hereby obtains the lift, drag, and moment coefficients of airfoils at dynamic condition \cite{leishman1989semi}. ROMs \cite{lucia2004reduced, KOU2021100725} are classified into system-identification-based model including auto-regressive with exogenous input model (ARX)  \cite{cowan2001accelerating, zhang2015mechanism}, Volterra series model \cite{raveh2001reduced, marzocca2004nonlinear}, Kriging \cite{glaz2010reduced, da2011generation}, neural network model (NN) \cite{zhang2012efficient, mannarino2014nonlinear} and feature-extraction-based model including Proper Orthogonal Decomposition (POD) \cite{berkooz1993proper, taira2020modal} and Dynamic Mode Decomposition (DMD) \cite{rowley2009spectral, schmid2010dynamic}. Surrogate models are widely applied on steady aerodynamic \cite{jeong2005efficient}, the input of which usually consists of flow conditions, geometric parameters, etc. However, the input of ROMs is primarily composed of motions at different time step for unsteady aerodynamic. Currently, these data-driven models have been widely used in flow reconstruction \cite{pawar2019deep, Han2019A}, flight dynamics \cite{ghoreyshi2014reduced}, aeroelasticity \cite{gao2020transonic}, flow control \cite{brunton2015closed, gao2017active}, and optimization design \cite{yondo2018review}.

However, existing data-driven approaches are constructed from a single source of data (experiment or numerical simulations), whose predictive performance depends on the amount and accuracy of available data. To model transonic aerodynamic loads with strongly nonlinear characterises, large-scale and reliable data is needed to represent flow physics with a high level of accuracy. Nevertheless, It is time-consuming and difficult to obtain higher fidelity aerodynamic data from experiments or numerical simulations that contain detailed flow structures, by making data-driven modeling from one source of data far from engineer applications. Therefore, data-driven modeling from abundant low-fidelity data from CFD and a few high-fidelity data from experiments has become a promising solution \cite{KOU2021100725}.

To handal multiple types of dataset, data fusion has received considerable attentions \cite{peherstorfer2018survey}. data fusion techniques combine data from multiple sources and related information from associated databases to achieve improved accuracy and more specific inferences than could be achieved by the use of a single sources alone \cite{castanedo_review_2013}. Currently, most studies for data fusion technique in aerodynamics have mainly been carried out in the fast simulations and optimization of steady aerodynamics to maintain the consistency between data from CFD and experiments and reduce the computational burden. Multi-fidelity aerodynamic modeling using steady aerodynamic data from tests and CFD has been developed by Ghoreyshi et al \cite{ghoreyshi2009accelerating}. CoKriging and gradient enhanced Kriging based on Kriging developed by Keane \cite{keane2012cokriging} and Han \cite{han2017weighted} respectively, might potentially be used with significant efficacy in data fusion. The reliability of aerodynamic data is enhanced by fusing data from different sources in these studies. In addition, a multi-kernel neural network is utilized by Kou and Zhang \cite{kou2019multi} for a extension of unsteady aerodynamic ROMs to allow multi-fidelity data, where Euler and Unsteady Reynolds-Averaged Navier-Stokes (RANS) solvers are used to provide low-fidelity and high-fidelity data, respectively. Furthermore, multi-fidelity or variable-fidelity models also have been proposed by some researchers for aerodynamic optimization design \cite{leifsson2010multi, han2012alternative}. The main interest in these studied is about the prediction of integral aerodynamic quantities.

Besides diverse applications on aerodynamic models for integral quantities, several attempts have been made on aerodynamic modeling with data fusion technique concerning distributed quantities. For example, Rokita et al. \cite{rokita2018multifidelity} developed a multi-fidelity model combining POD with co-Kriging for hypersonic aerodynamic load prediction, where POD bases are obtained by both low-fidelity and high-fidelity snapshots. Gappy POD with constrained least-squares method is utilized to fuse CFD simulation and sparse measurement of wind tunnel data by Mifsud et al. \cite{mifsud2019fusing}. Then, the surface load distribution is recovered with improved accuracy. Renganathan et al. \cite{renganathan2020aerodynamic} employed a Bayesian framework and an extension of the proper orthogonal decomposition with constraints to infer the true fields conditioned on measured quantities of interest. A multi-fidelity ROM based on manifold alignment, fusing inconsistent fields from high-fidelity and low-fidelity simulations, was constructed by Perron et al. \cite{perron2020development}. Wang et al. \cite{wang2020multi} proposed a multi-fidelity ROM for the reconstruction of steady flow field at different conditions, which outperforms some traditional methods in both interpolated and extrapolated conditions.  

Machine Learning (ML) has demonstrated remarkable power in recent years for numerous applications\cite{brunton2021data}, like image processing, video and speech recognition, genetics and disease diagnosis. Moreover, ML algorithms also provide a critical tool in the fluid mechanics research \cite{brunton2020machine}. NNs, as an vital component of ML, are able to approximate any continuous functions with high accuracy, and are flexible with fault tolerant ability. Therefore, NNs also have been blossoming in the fluid mechanics research \cite{zhu2021turbulence, sekar2019fast}. To date, there are several studies that have investigated multi-fidelity modeling based on NNs \cite{minisci2013robust}. Meng et al. \cite{meng2020composite} proposed a composite neural network constructed by physics-informed neural networks (PINNs) with multi-fidelity data sets to learn the correlation between the low-and high-fidelity data and produce the multi-fidelity approximation. Subsequently, a bayesian neural network to link together a data-driven deep neural network (DNN) and a PINN for uncertainty quantification in predictions, was presented by Meng et al. \cite{meng2021multi}. A multi-fidelity model based on DNN, mapping the integral quantities $C_L$ and $C_M$ calculated from Euler and Navier-Stokes equations respectively, was constructed by He et al. \cite{he2020multi}. Different from the previous works mapping the correlation to link together multi-fidelity data \cite{fernandez2016review}, a multi-fidelity aerodynamic based on modified DNN using CFD data as soft penalty constraints is proposed in this study. In particular, the CFD data ought to reflect the underlying physical principles that dictate their generation, and, in principle, can be used as a weak mechanism for embedding these principles into this multi-fidelity model during its training phase \cite{karniadakis2021physics}.

This work is structured as follows. In section \ref{section:sec1}, the deep neural network, modified loss function, and multi-fidelity aerodynamic distributions modeling are introduced. In section \ref{section:sec2}, we briefly illustrate the flow solver. In section \ref{section:sec3}, the proposed approach is discussed in detail. Data sets are introduced in section \ref{section:sec3.1}. From section \ref{section:sec3.2} to section \ref{section:sec3.5}, we discuss some secnarios where the flow condition varies or there is less high-fidelity data. Finally, conclusions are presented in section \ref{section:sec4}.

\section{Methods} \label{section:sec1}

DNNs are the quintessential ML models. The goal of a DNN is to approximate some function $f$. For instance, for a regression problem, $f$ maps an input data to an output data. This model defines a mapping $\mathbf{y} = f(\mathbf{x};\boldsymbol{\theta})$, and learns the value of the parameters $\boldsymbol{\theta}$ that result in the best function approximation \cite{goodfellow2016deep}. 

\subsection{Deep Neural Networks}
\begin{figure}[htbp]
\centering
\includegraphics[width=.8\textwidth]{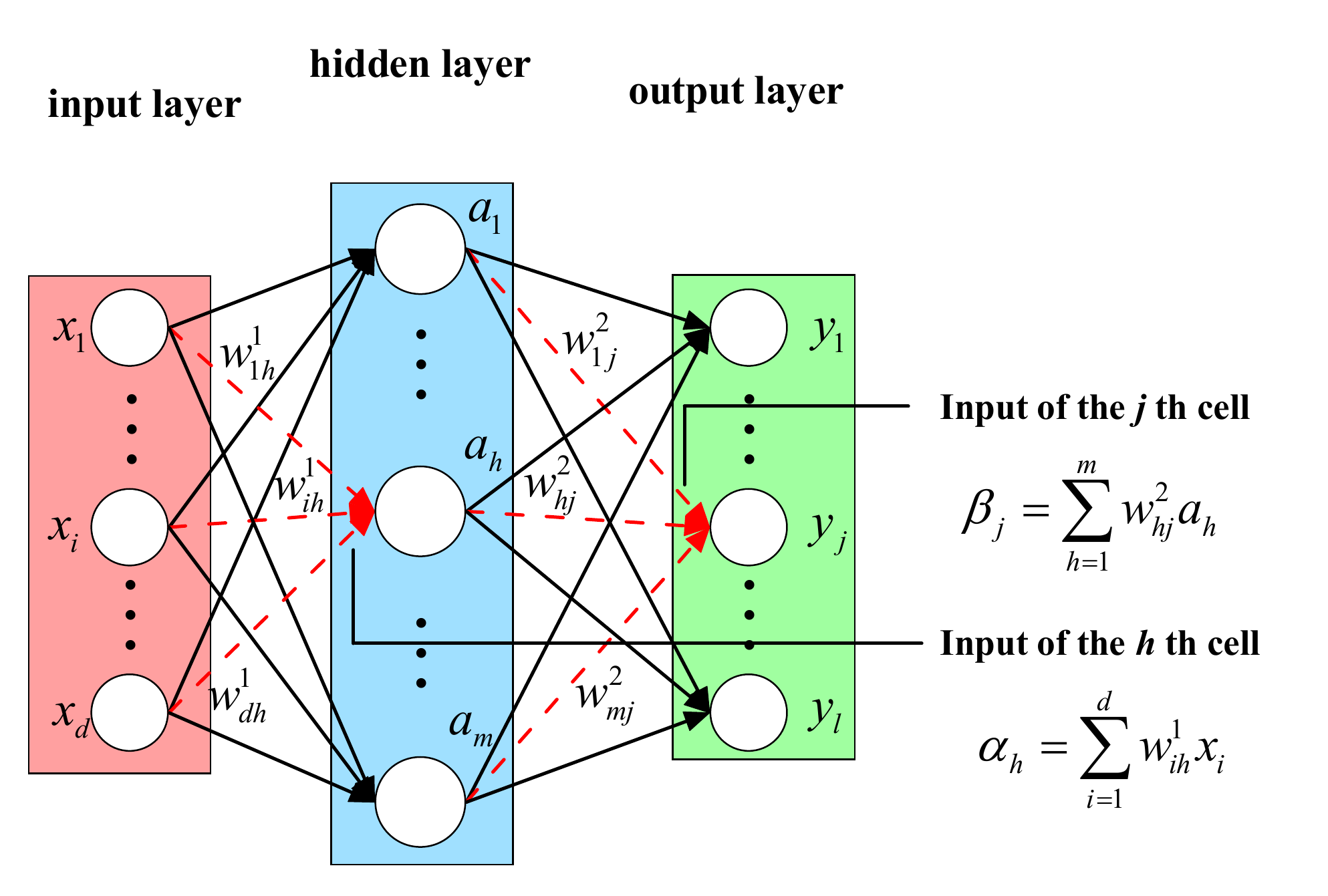}
\caption{The architecture of DNN.}
\label{fig:Fig1}
\end{figure}

The architecture of DNNs with strong generalization and universal approximation properties \cite{hornik1989multilayer} is simple. As shown in Fig.\ref{fig:Fig1}, the first and last layers of DNN are called input layer and output layer, respectively. The middle layer is called hidden layer. The main hyper-parameters include the depth (the number of hidden layer) and the width of the model (The dimensionality of each hidden layer). Moreover, hidden unit is a nonlinear or linear processing unit with local regulation, which only provides responds for a certain range of input values. The relationship between hidden layer and output layer is nonlinear or linear. For convenience, a architecture of DNN with single hidden layer and without biases is illustrated.  

\begin{figure}[htbp]
\centering
\includegraphics[width=.8\textwidth]{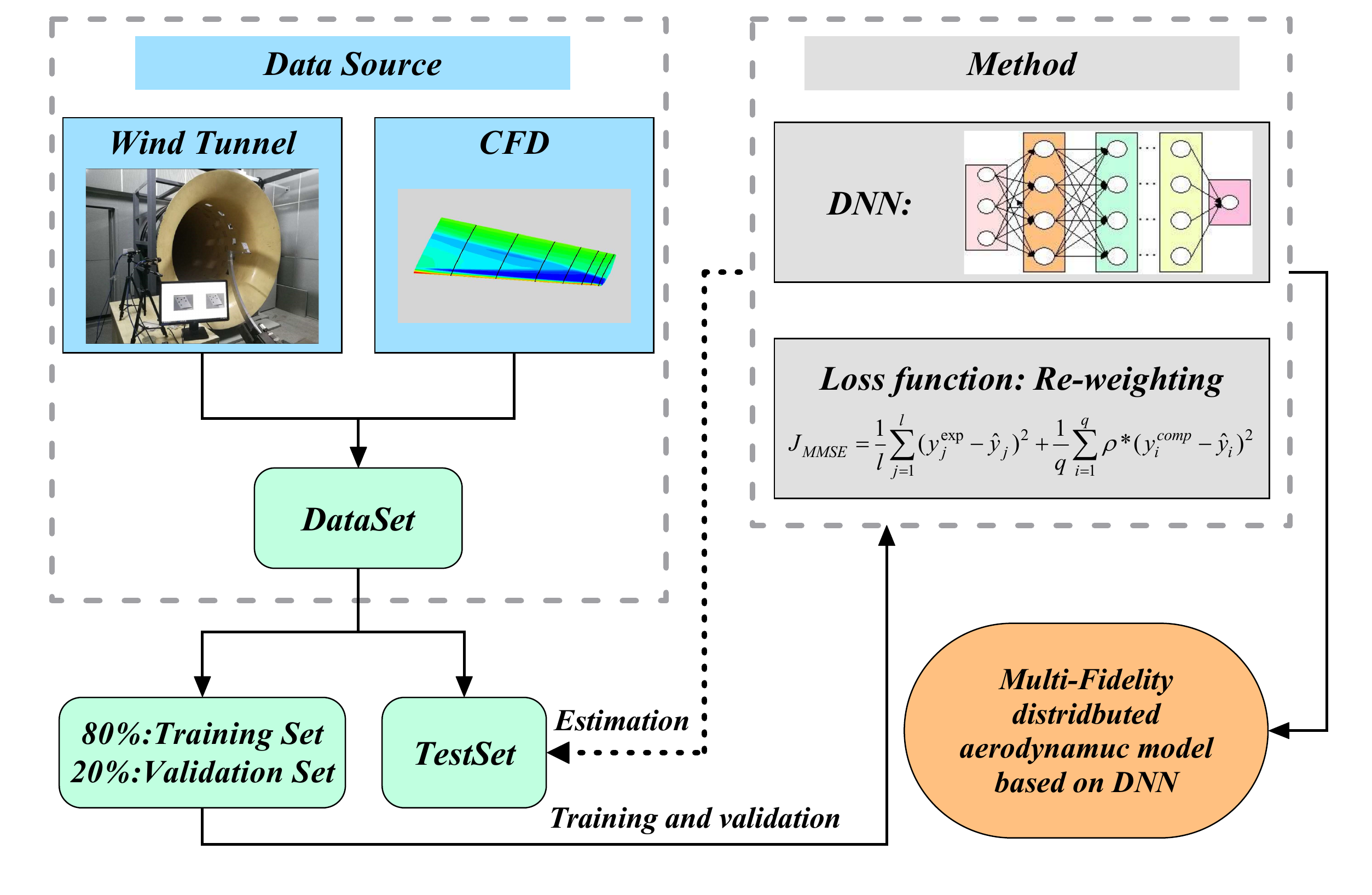}
\caption{Flowchart of multi-fidelity modeling based on DNN.}
\label{fig:Fig2}
\end{figure}

$d$ and $l$ are the dimension of input vector and output vector respectively; $m$ denotes unit number of the hidden layer; $w$ are the weights. The forward computation for this model is expressed following:

\begin{equation}
\label{equ:Equ1}
{y_j} = {f^2}\left\{ {\sum\limits_{h = 1}^m {w_{hj}^2\left[ {{f^1}(\sum\limits_{i = 1}^d {w_{ih}^1{x_i})} } \right]} } \right\}
\end{equation}
where $f^1$ and $f^2$ denotes the hidden unit activation functions. Generally, the output layer ($f^2$) has no activation function. The conventional activation functions include sigmoid, tanh functions. In this work, tanh function is used as following:

\begin{equation}
\label{equ:Equ2}
\begin{array}{l}
f(x) = \frac{{{e^x} - {e^{ - x}}}}{{{e^x} + {e^{ - x}}}}\\
f'(x) = 1 - {(f(x))^2}
\end{array}
\end{equation}

As illustrated in Eq.~\eqref{equ:Equ2}, the tanh activation function saturates to -1 when $x$ becomes very negative and saturates to 1 when $x$ becomes very positive, which is also a smooth and nonlinear function. 

The training for DNN can be summarized as the following steps:
\begin{enumerate}
\item Obtain the input data ${\bf{X = }}[{{\bf{x}}_1},{{\bf{x}}_2},...,{{\bf{x}}_p}]$ and output data ${\bf{Y = }}[{{\bf{y}}_1},{{\bf{y}}_2},...,{{\bf{y}}_p}]$;
\item Compute forward from the input layer to the hidden layer to the output layer, as shown in Eq.~\eqref{equ:Equ1};
\item Update the parameters of DNN using adaptive moment estimation (Adam) with adaptive learning rate, a stochastic gradient descent optimizer to reduce the training error and avoid local minimal point \cite{kingma2015adam}.
\end{enumerate}

\subsection{Modified Loss Function}
The difference between the predictions of model and the true values is reflected by the loss function, which determines the objective of model training. Given the input and output $({\bf{x}},{\bf{y}})$, the prediction can be calculated as ${\bf{\hat y}} = ({\hat y_1},{\hat y_2}, \ldots ,{\hat y_l})$ according to Eq.~\eqref{equ:Equ1}. For a regression problem, a mean squared error (MSE) loss function is selected as following:

\begin{equation}
\label{equ:Equ3}
J{(\boldsymbol{\theta})_{MSE}} = \frac{1}{l}\sum\limits_{j = 1}^l {{{({{\hat y}_j} - {y_j})}^2}}
\end{equation}

Based on the Re-weighting technique \cite{kang2019few} in ML, the MSE loss function is modified to allow dataset with multiple fidelities in this work, as illustrated in Eq.~\eqref{equ:Equ4}. Current efforts aim to impose such constraints using CFD data as prior information in a soft manner by appropriately penalizing the MSE loss function of conventional DNN approximation composed of low and high fidelity data, which is also capable of avoiding over-fitting. Thus, a modified mean squared error loss function is defined as:

\begin{equation}
\label{equ:Equ4}
J{(\boldsymbol{\theta})_{MMSE}} = \frac{1}{l}\sum\limits_{j = 1}^l {{{(y_j^{exp} - {{\hat y}_j})}^2}}  + \frac{1}{q}\sum\limits_{i = 1}^q {\rho *{{(y_i^{com} - {{\hat y}_i})}^2}}
\end{equation}
where $l$ and $q$ are the number of high-fidelity and low-fidelity samples, respectively; $\rho$ defines coefficient of the low-fidelity data, whose selection will be described in detail below. This is the main contribution and the key strategy used in the current work. In addition, it is noteworthy that the proposed approach is aso able to be used for problems with multiple fidelities, i.e. more than two fidelities.

\begin{figure}[htbp]
\centering
\subfigure[]{
\includegraphics[width=0.45\textwidth]{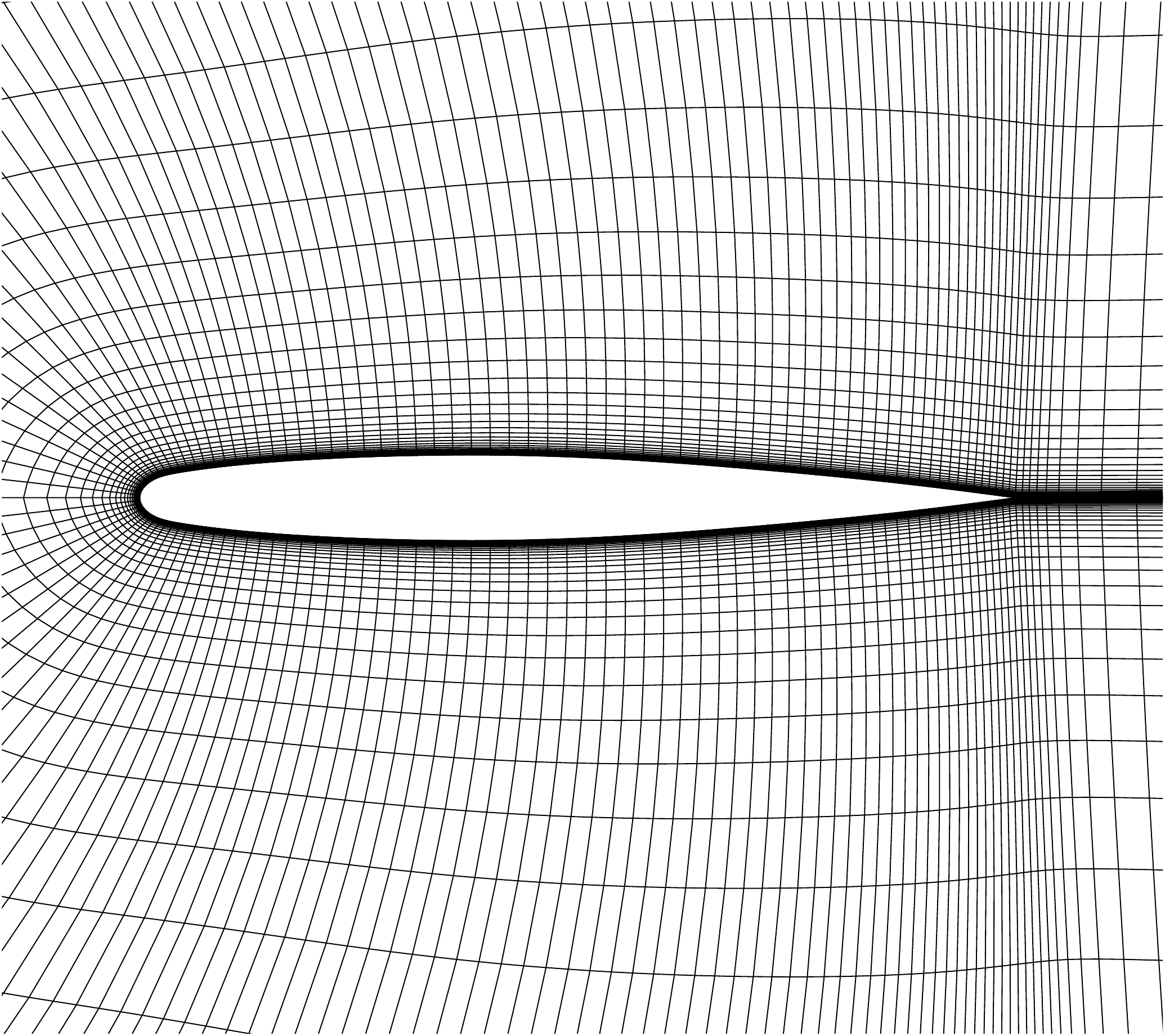}
}
\subfigure[]{
\includegraphics[width=0.45\textwidth]{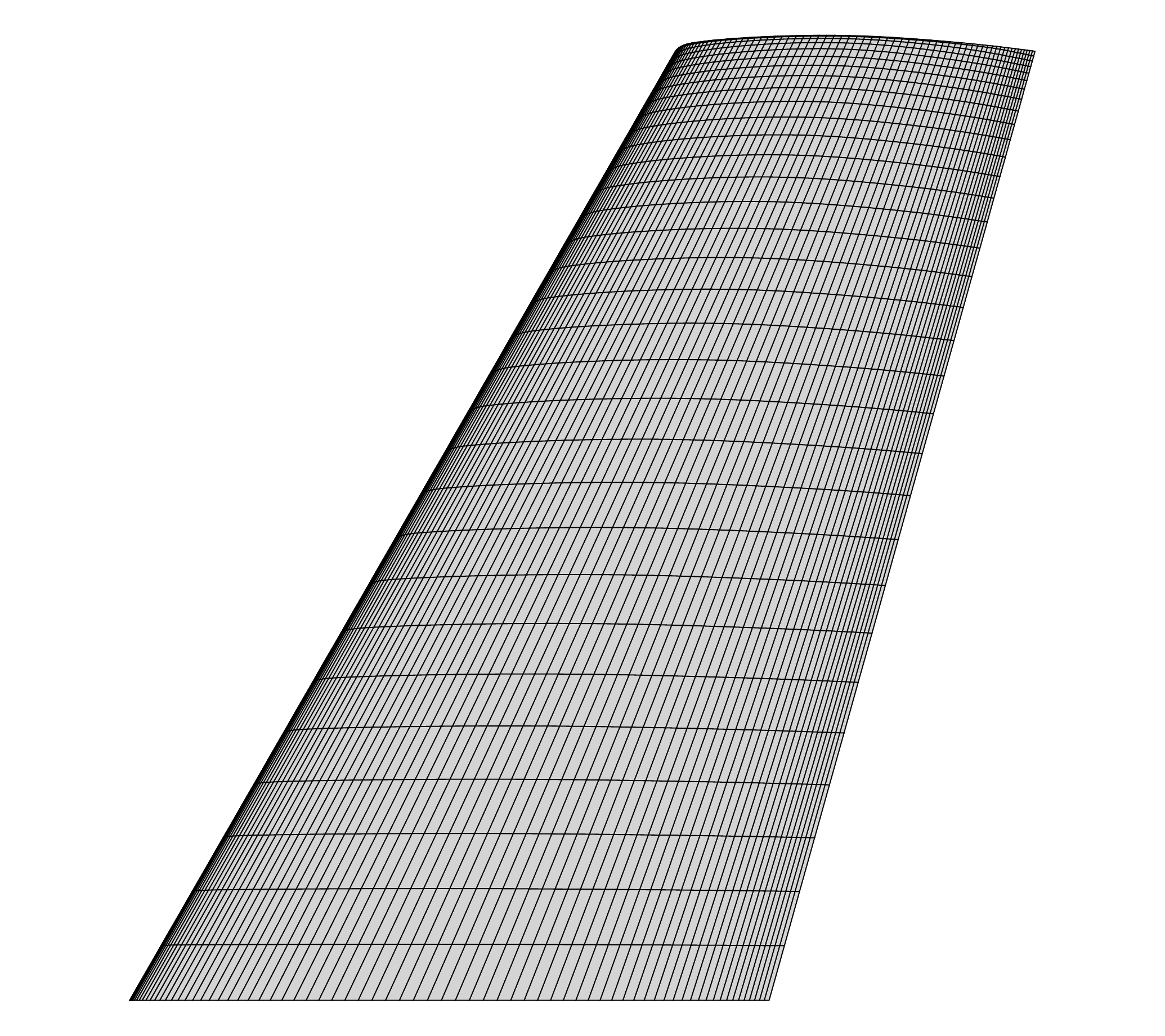}
}
\quad
\subfigure[]{
\includegraphics[width=0.45\textwidth]{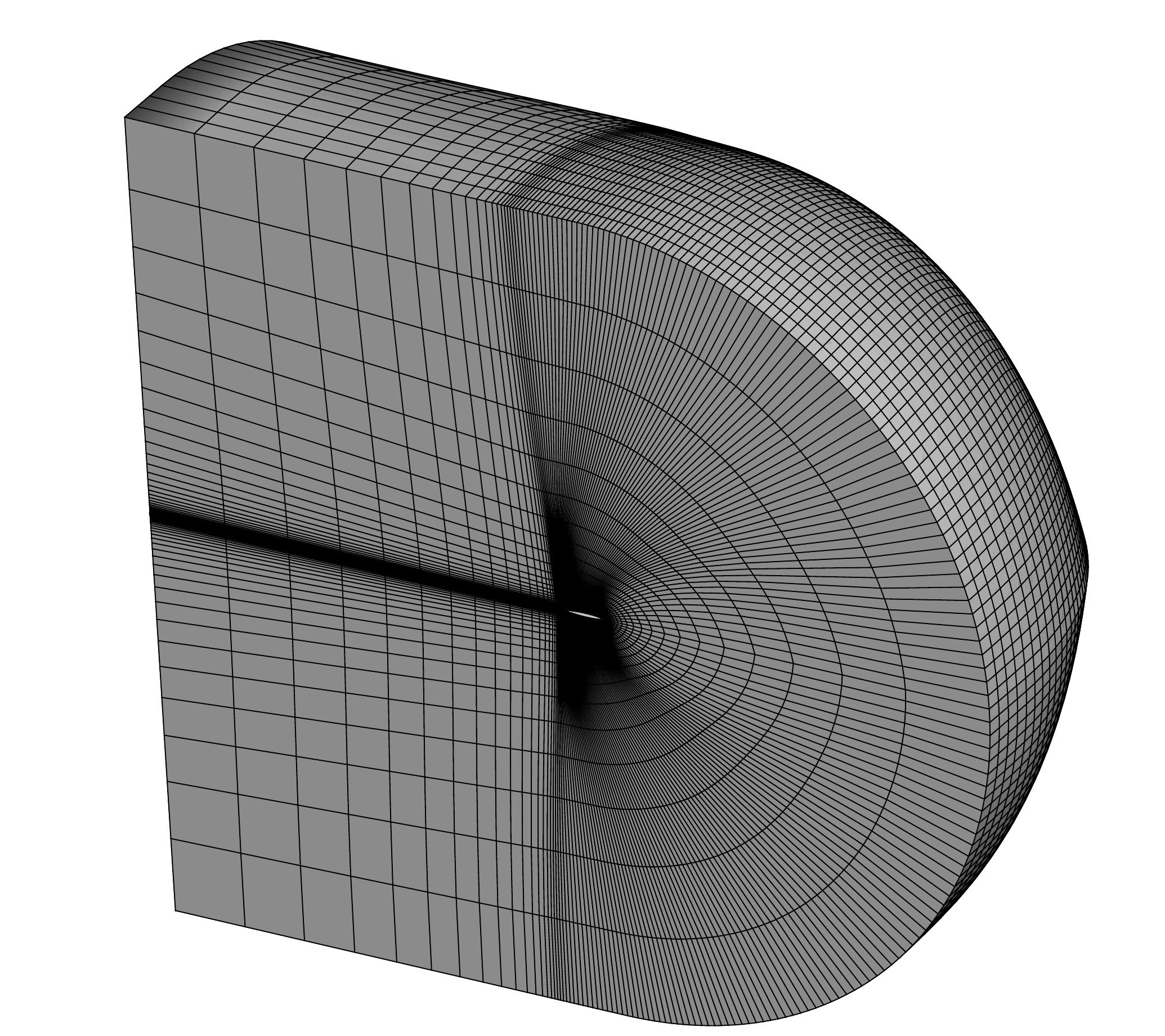}
}
\subfigure[]{
\includegraphics[width=0.45\textwidth]{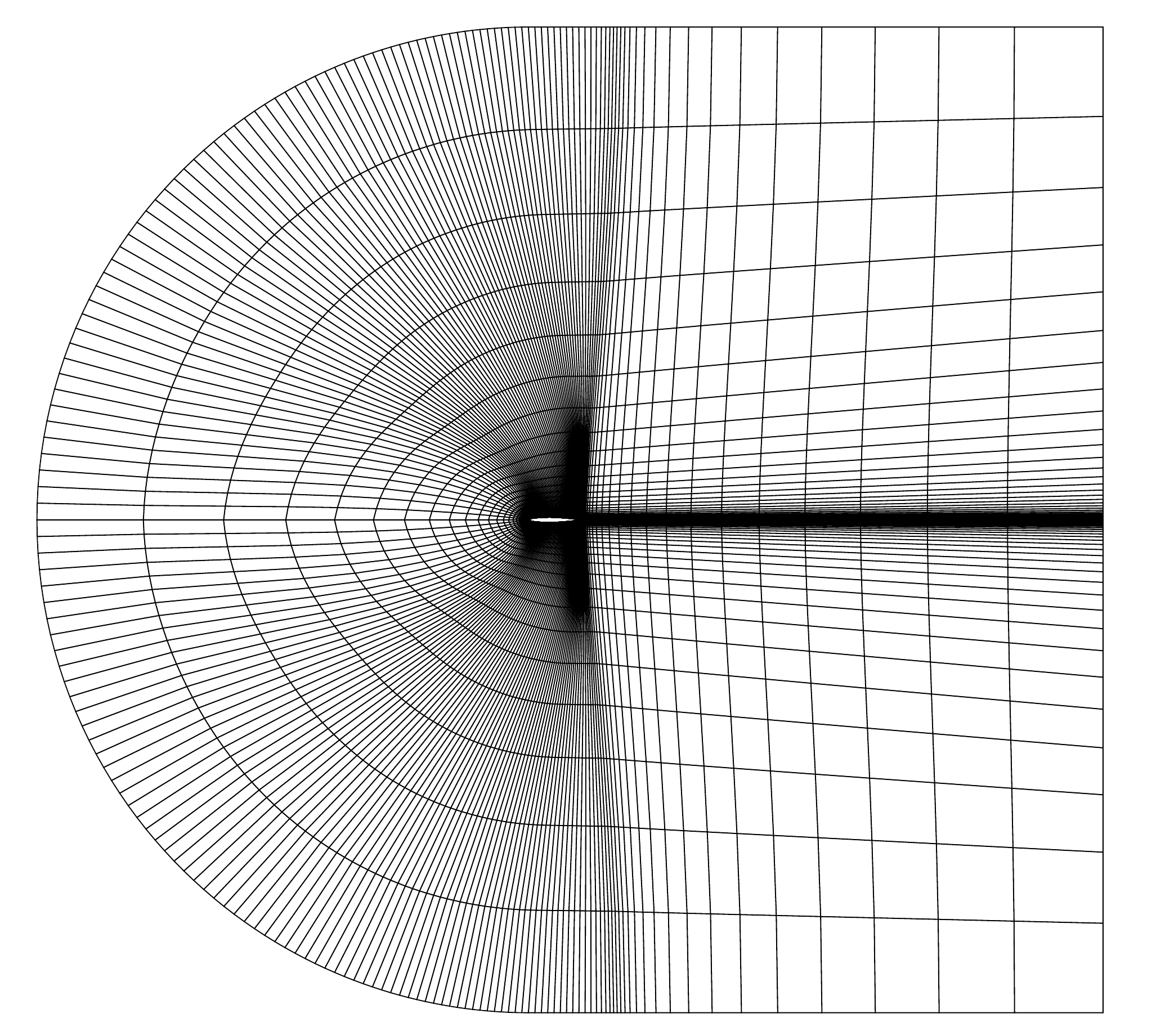}
}
\caption{The grid of ONERA M6 wing.}
\label{fig:Fig3}
\end{figure}

\subsection{Multi-Fidelity Aerodynamic Distributions Modeling}
In this paper, two flow conditions, including Mach number $Ma$ and angle of attack $\alpha$, are considered. Due to the significant difference between the upper and lower wing surface pressure distribution, the model is built separately for the upper and lower surface \cite{sekar2019inverse}. The four input variables of this model are Mach number $Ma$, angle of attack $\alpha$, the chord coordinate of wing surface $X/c$, the span coordinate of wing surface $Y/b$, and the output variable is surface pressure coefficient $C_p$ as:

\begin{equation}
\label{equ:Equ5}
\begin{array}{l}
{\bf{x}} = [Ma \quad \alpha \quad X/c \quad Y/b]\\
{\bf{y}} = [{C_p}]
\end{array}
\end{equation}

In the multi-fidelity modeling, the number of hidden neurons is given as 20 and the number of hidden layers is chosen to be 3. As a key hyper-parameter, $\rho$ is selected by the validation samples \cite{kou2016approach}.

A schematic of unsteady aerodynamic modeling is depicted in Fig.\ref{fig:Fig2}, and the modeling procedure is summarized as follows:
\begin{enumerate}
\item Sampling in the parameter space ($Ma$, $\alpha$ et al.) for training and validation data (Note that in this work this step is omitted since the experimental data is obtained from existing literatures \cite{schmitt1979pressure}.);
\item Calculate the low-fidelity aerodynamic data from CFD and obtain the high-fidelity aerodynamic data from wind-tunnel experiments;
\item Split the high-fidelity data into 80\% training sample to update the parameters of this model and 20\% validation samples to select $\rho$ in the modified loss function;
\item Estimate this model using test samples after training;
\item Use the proposed model for practical applications using different test dataset.
\end{enumerate}

\section{Flow Solver} \label{section:sec2}

\begin{figure}[htbp]
    \centering
    \begin{minipage}{0.66\textwidth}
        \includegraphics[width=0.5\linewidth]{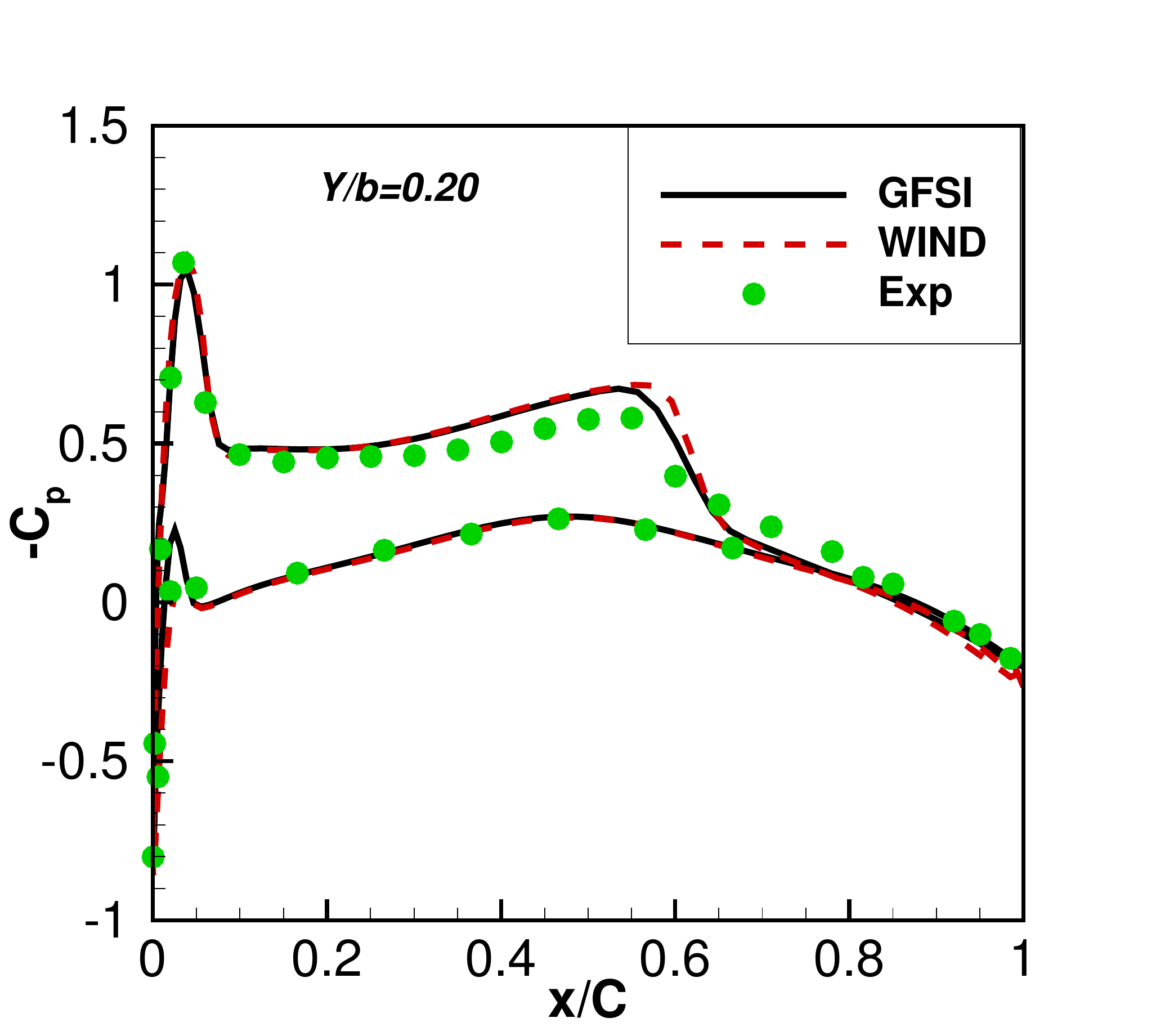}
        \includegraphics[width=0.5\linewidth]{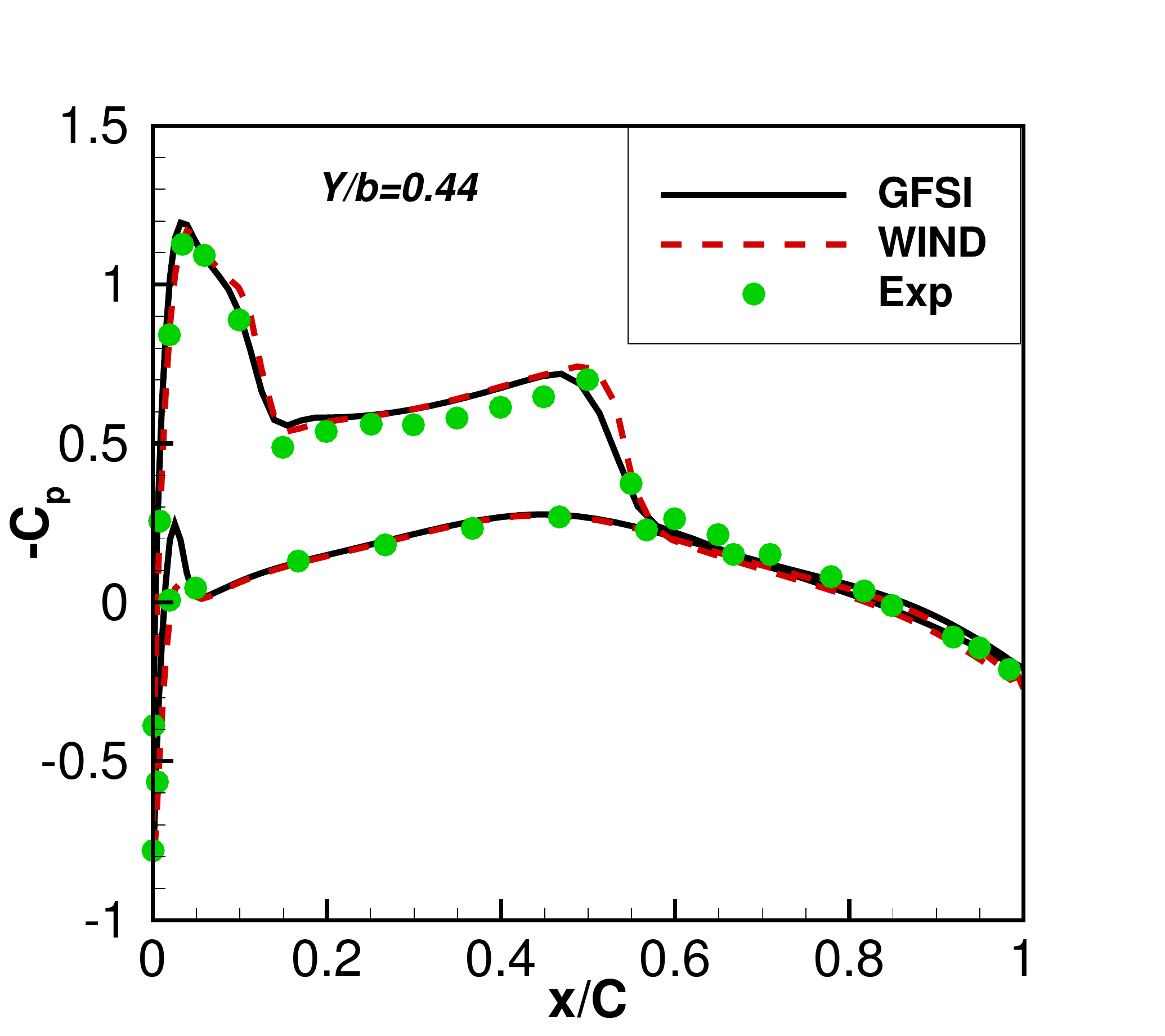}
        \quad
        \includegraphics[width=0.5\linewidth]{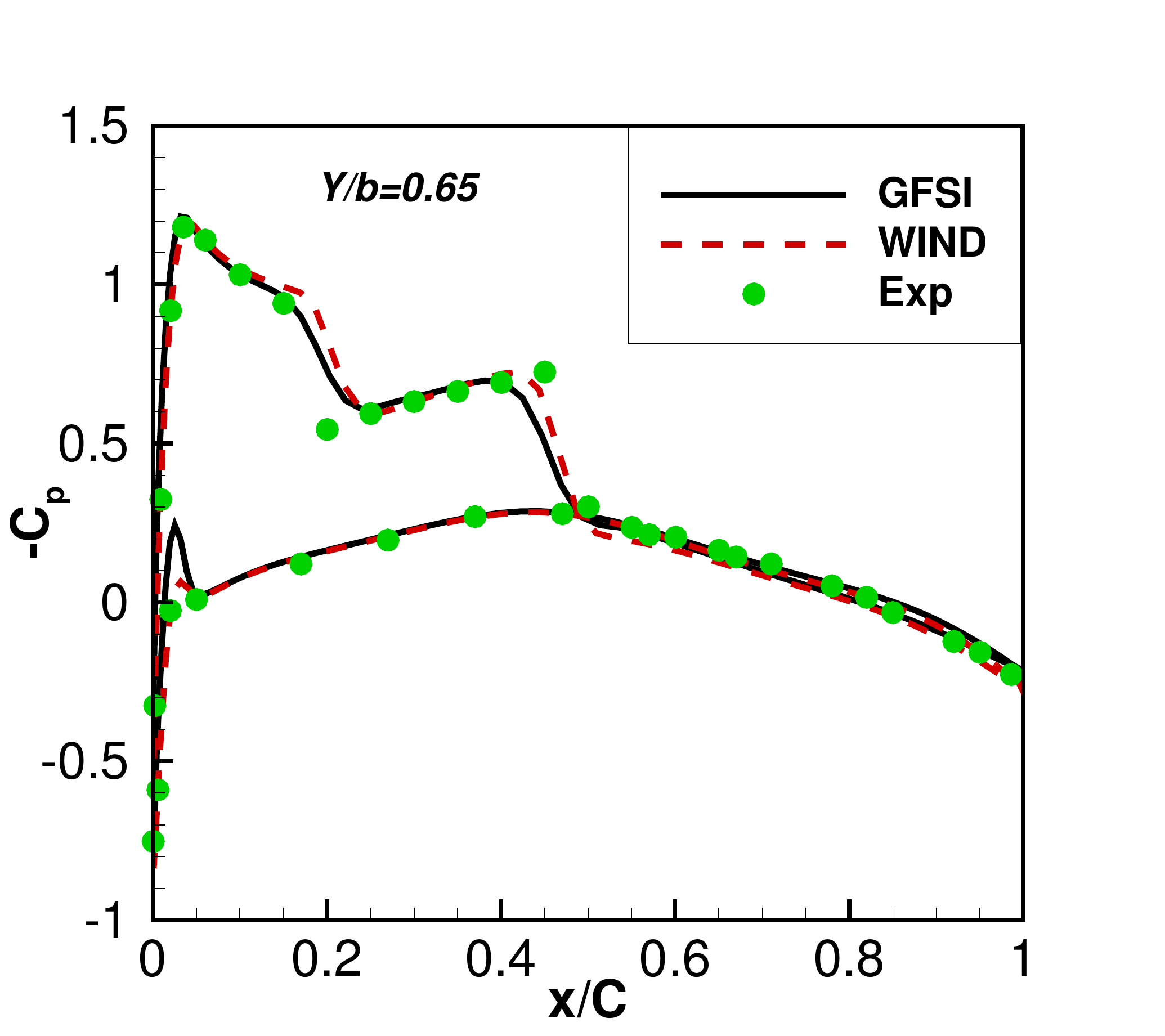}
        \includegraphics[width=0.5\linewidth]{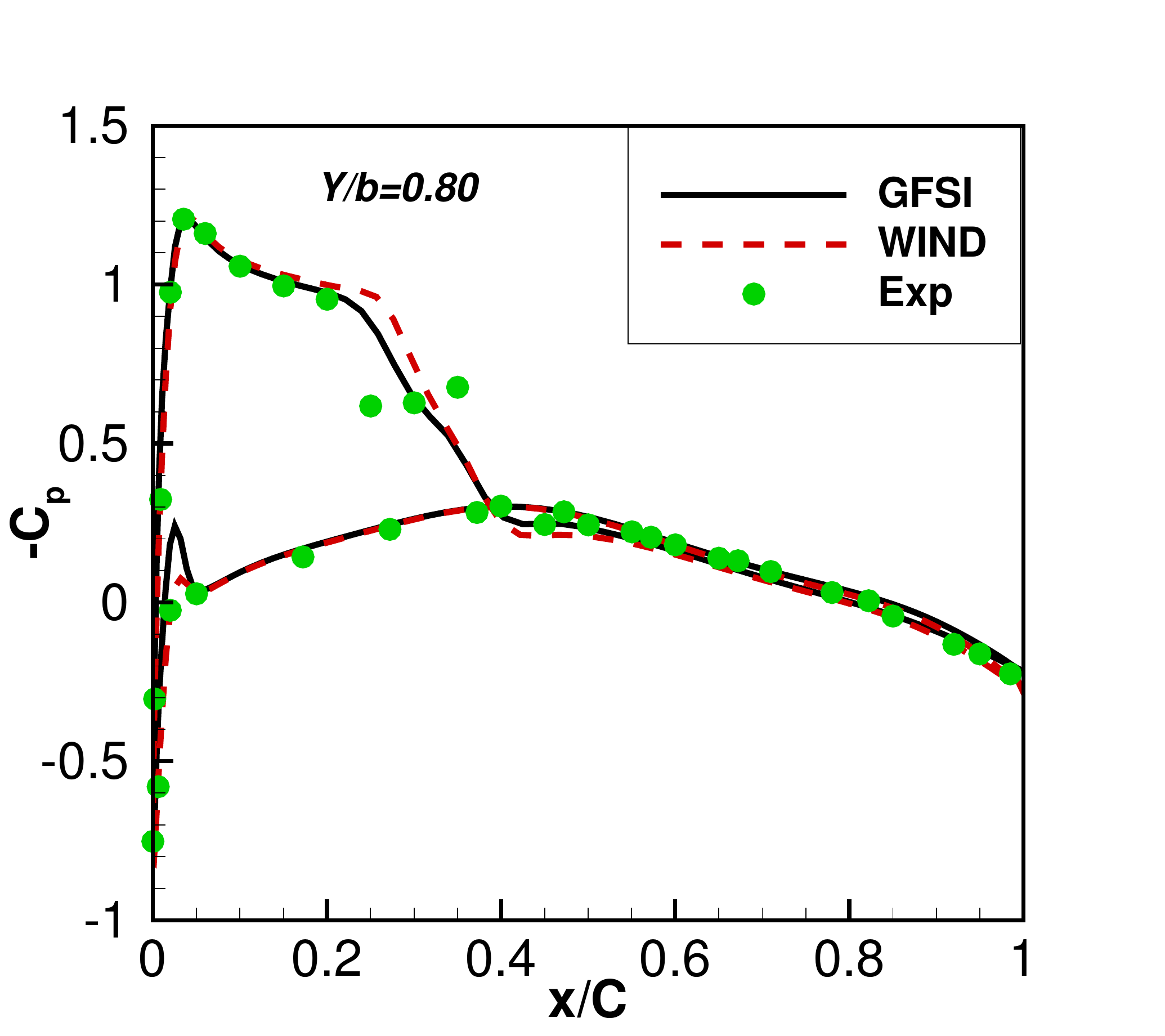}
    \end{minipage}
    \begin{minipage}{0.33\textwidth}
        \includegraphics[width=1\linewidth]{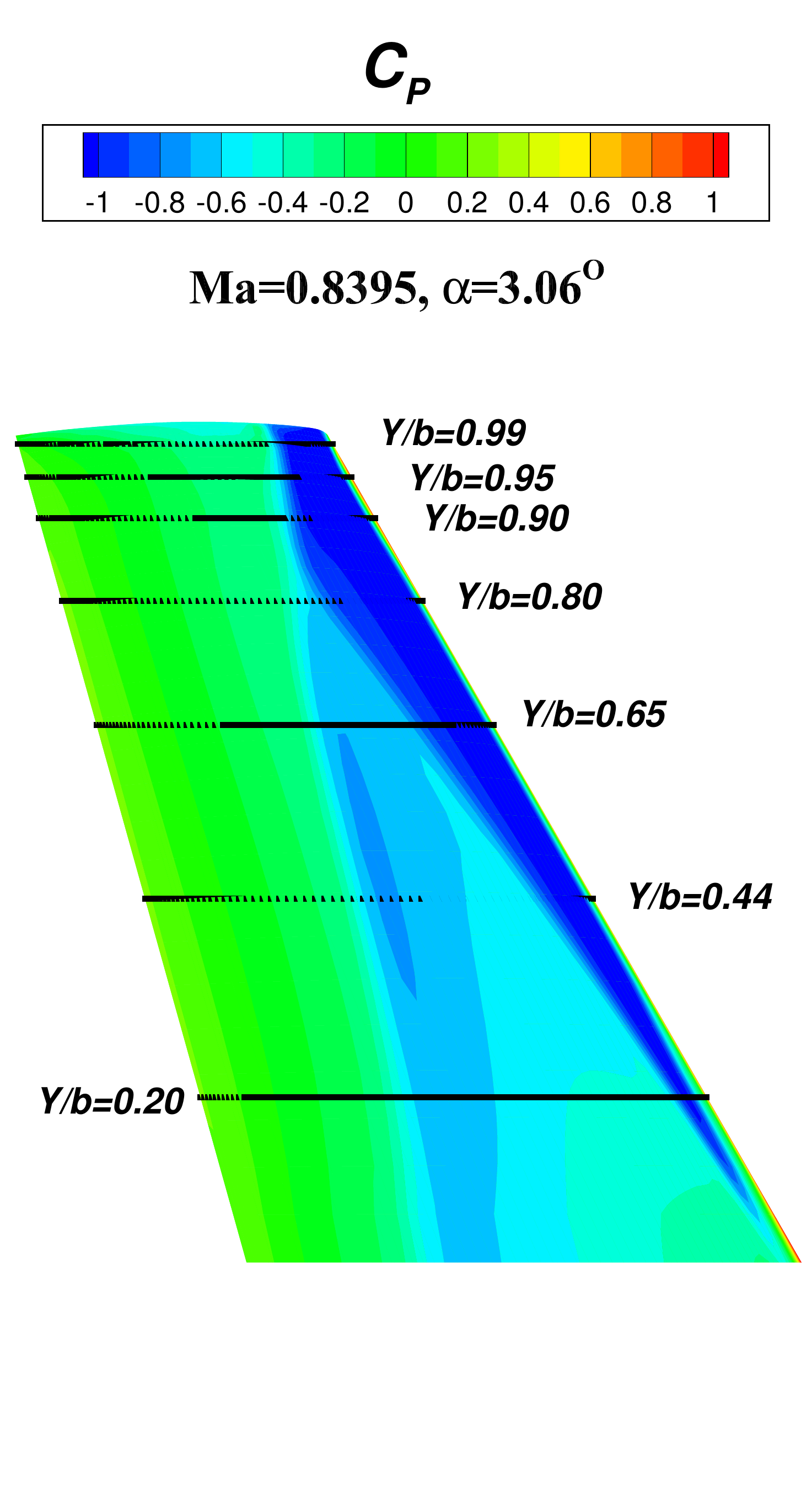}
    \end{minipage}
    \quad
    \begin{minipage}{1\textwidth}
        \includegraphics[width=0.33\linewidth]{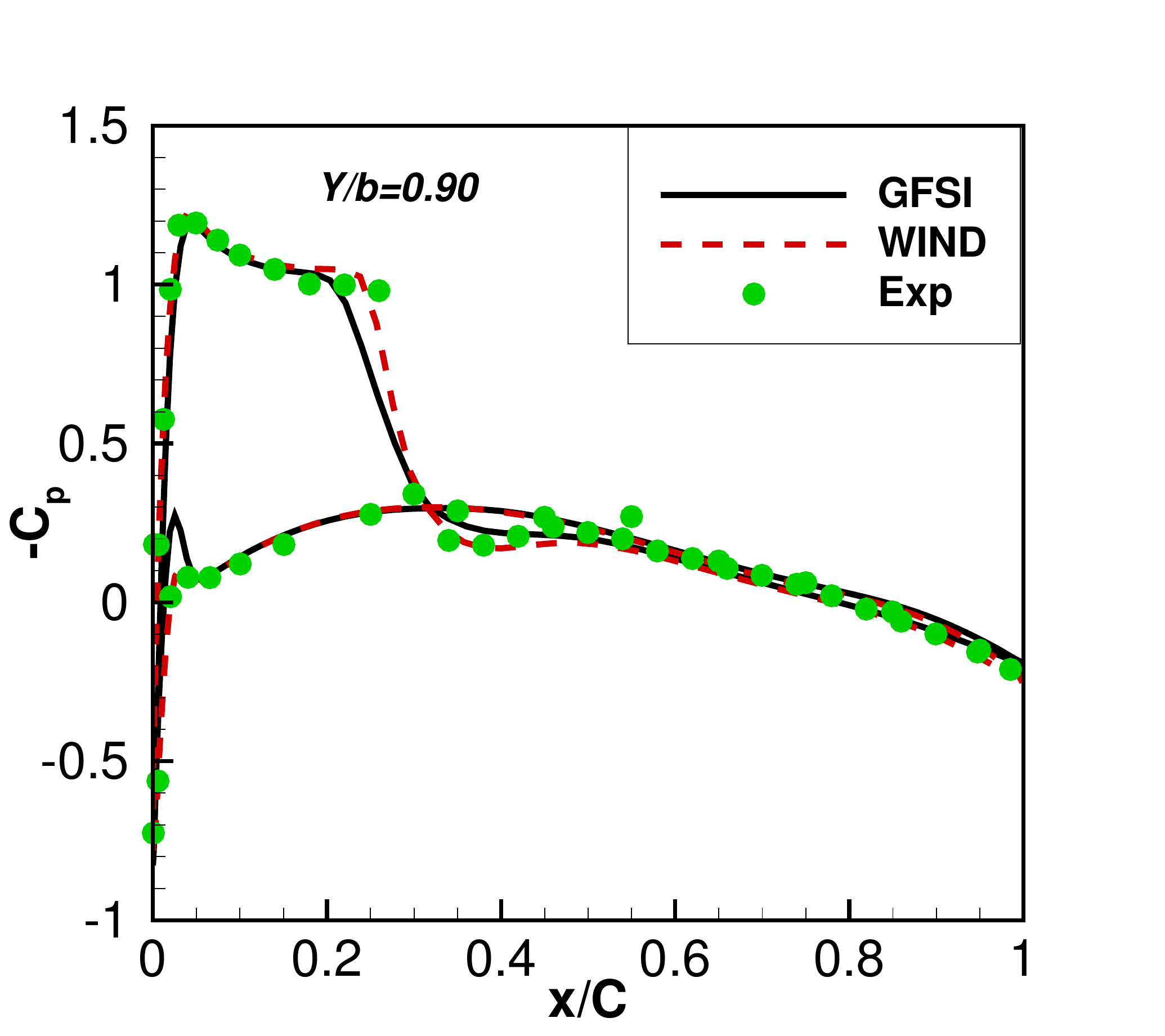}
        \includegraphics[width=0.33\linewidth]{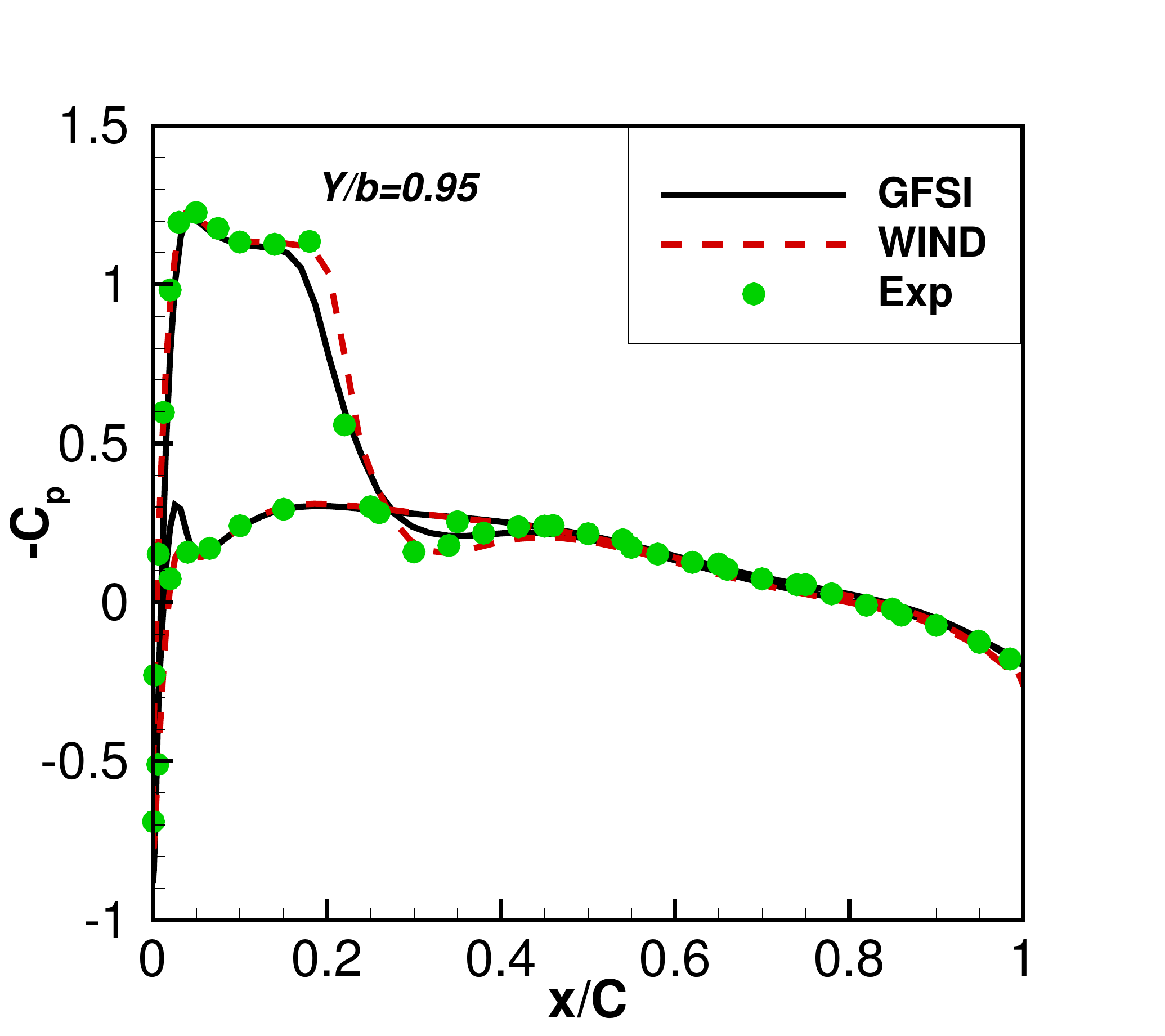}
        \includegraphics[width=0.33\linewidth]{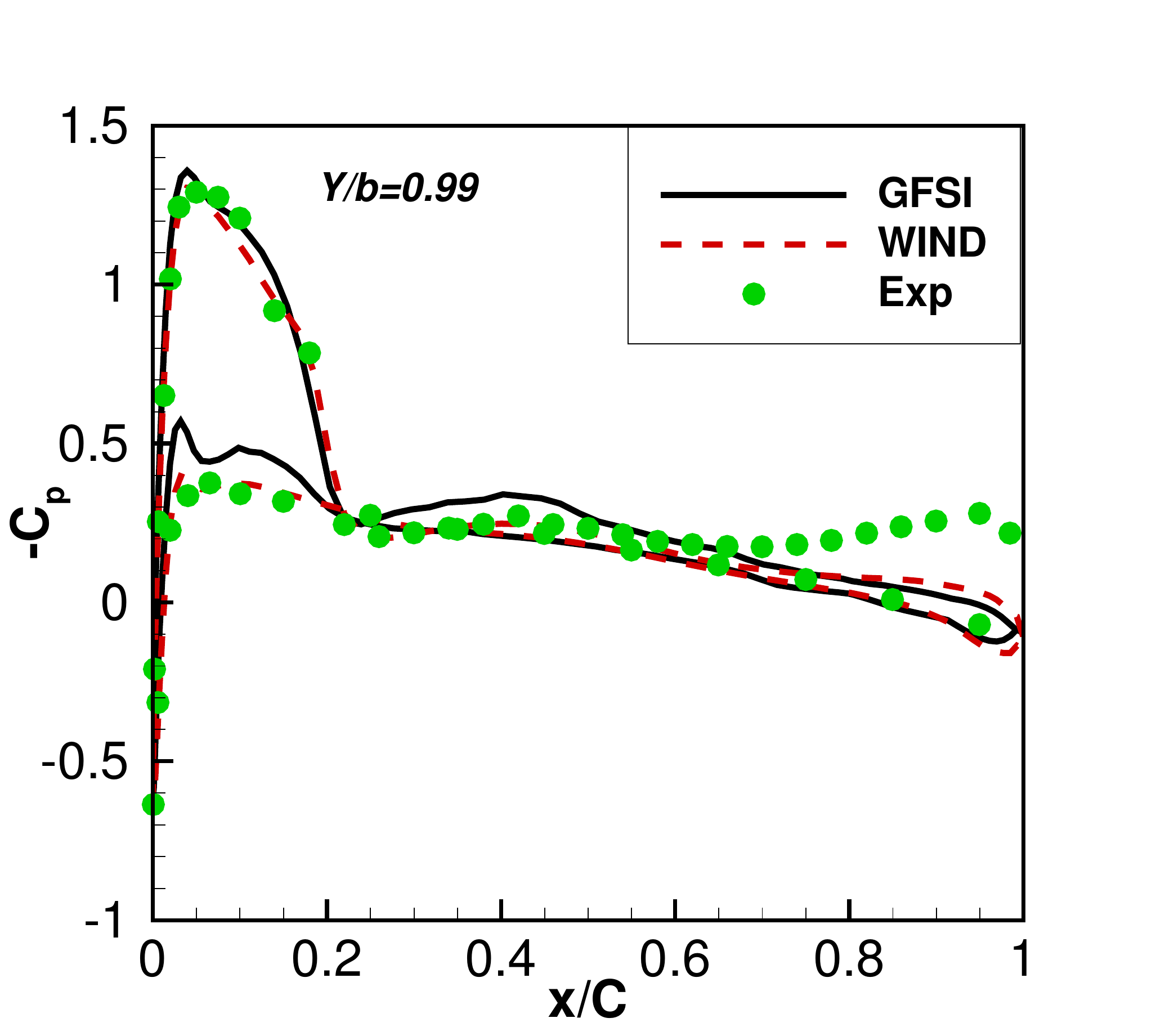}
    \end{minipage}
    \caption{Comparison of the computed $C_p$ with experimental data.}
    \label{fig:Fig4}
\end{figure}

In this study, we perform numerical simulations using an in-house CFD code GFSI which solves the three-dimensional compressible steady Reynolds-averaged Navier–Stokes (RANS) equations with the S–A turbulence model using a cell-centred finite volume approach. An adiabatic no-slip boundary condition is used on the wing surface. The second-order Roe scheme is used for spatial discretization. The computational results of WIND and the grid as shown in Fig.\ref{fig:Fig3} are obtained from the website\footnote{\url{https://www.grc.nasa.gov/WWW/wind/valid/m6wing/m6wing01/m6wing01.html}}. To verify the accuracy of the CFD solver, A typical case of ONERA M6 wing with $Ma=0.8395$ and $\alpha=3.06^o$ is computed and compared with WIND and the experimental results as depicted in Fig.\ref{fig:Fig4}. It is noticed that there is a dramatic difference between the computational and experimental results especially for $Y/b=0.80$ section, due to a ”lambda“ shock and flow separation on the upper surface of the wing.

\section{Results} \label{section:sec3}

\begin{figure}
    \centering
    \subfigure[$\alpha=0.04^o$]{
        \begin{minipage}[b]{0.3\textwidth}
        \includegraphics[width=1\textwidth]{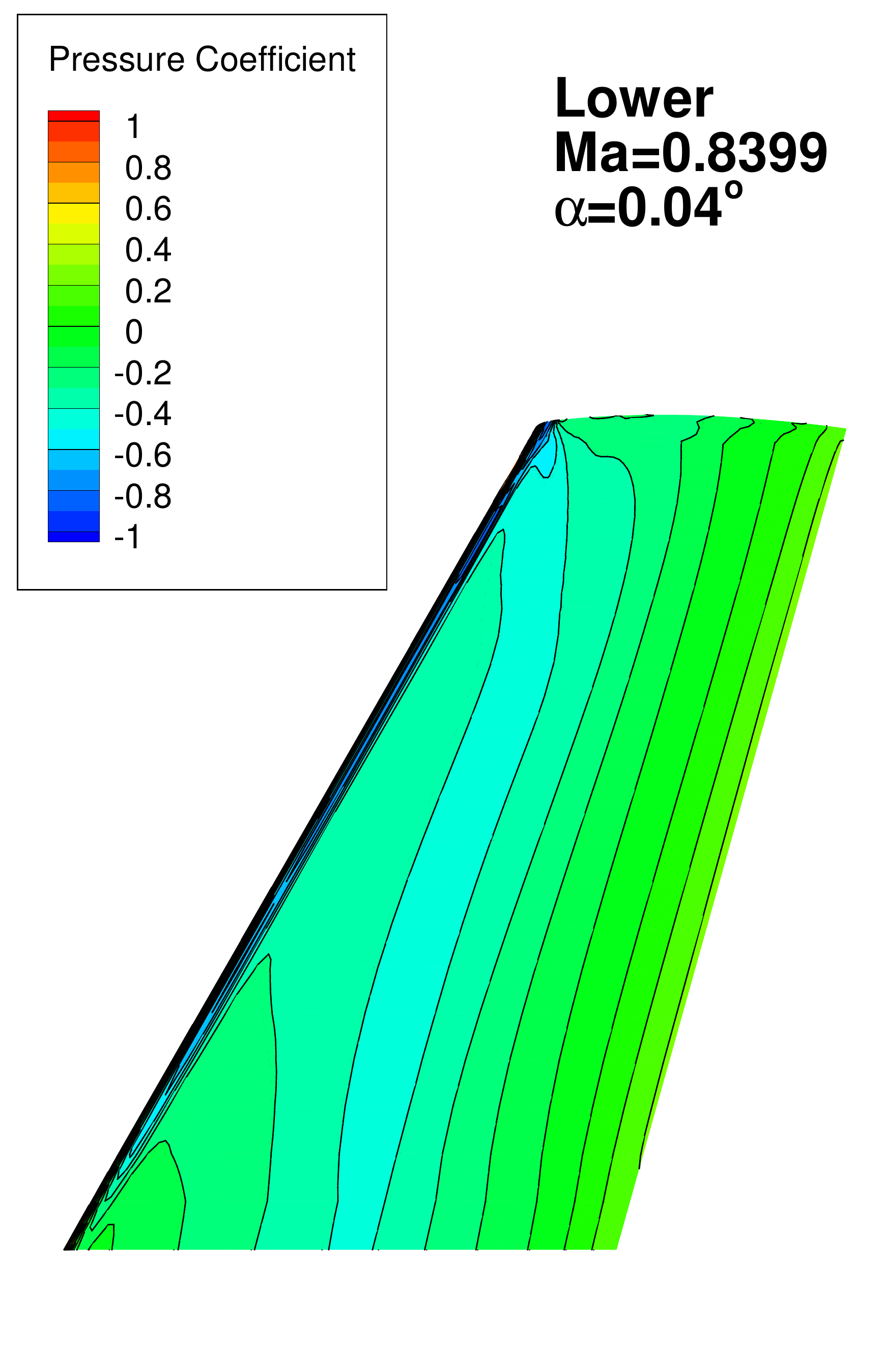} \\
        \includegraphics[width=1\textwidth]{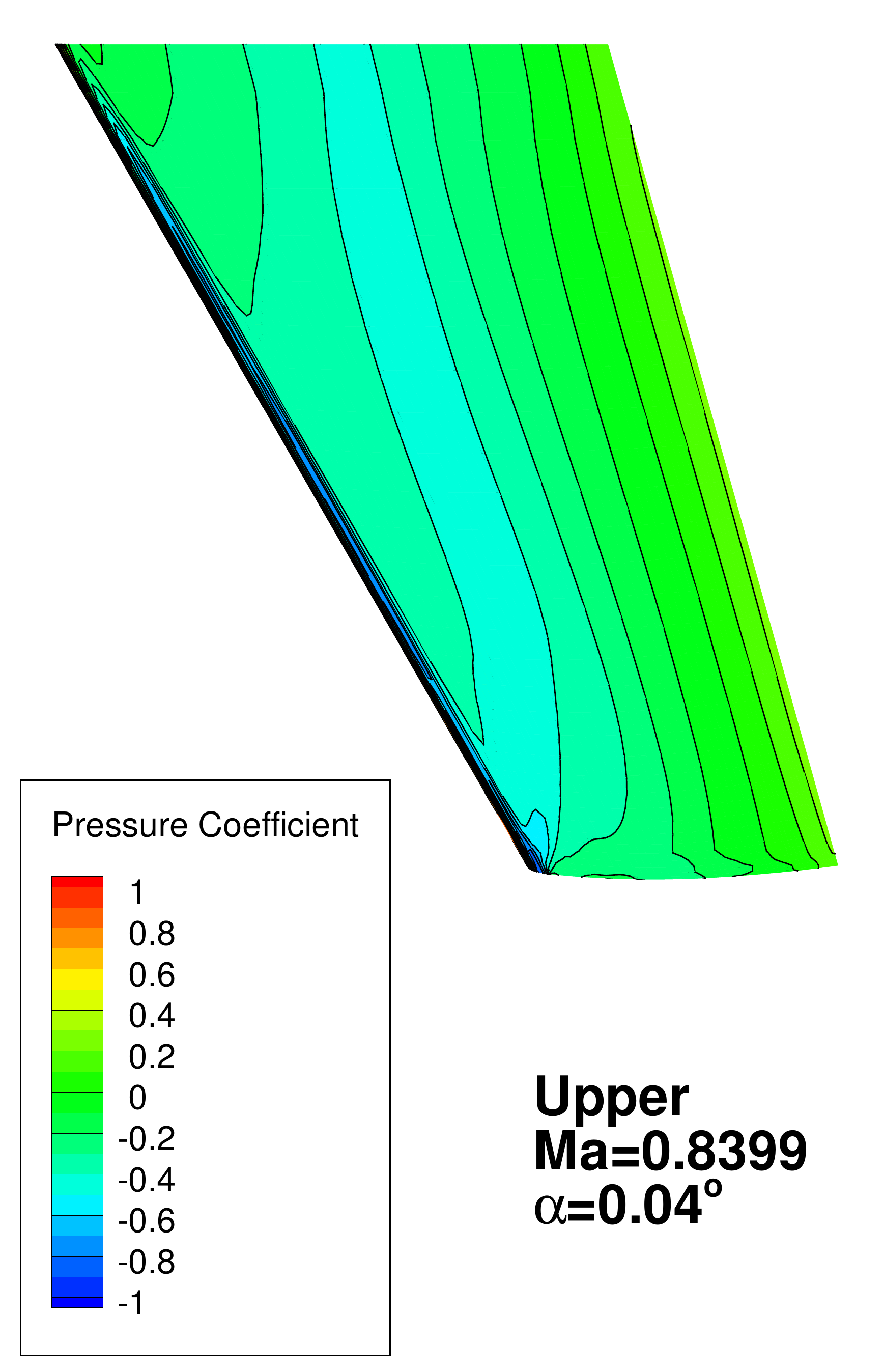}
        \end{minipage}
    }
    \subfigure[$\alpha=2.06^o$]{
        \begin{minipage}[b]{0.3\textwidth}
        \includegraphics[width=1\textwidth]{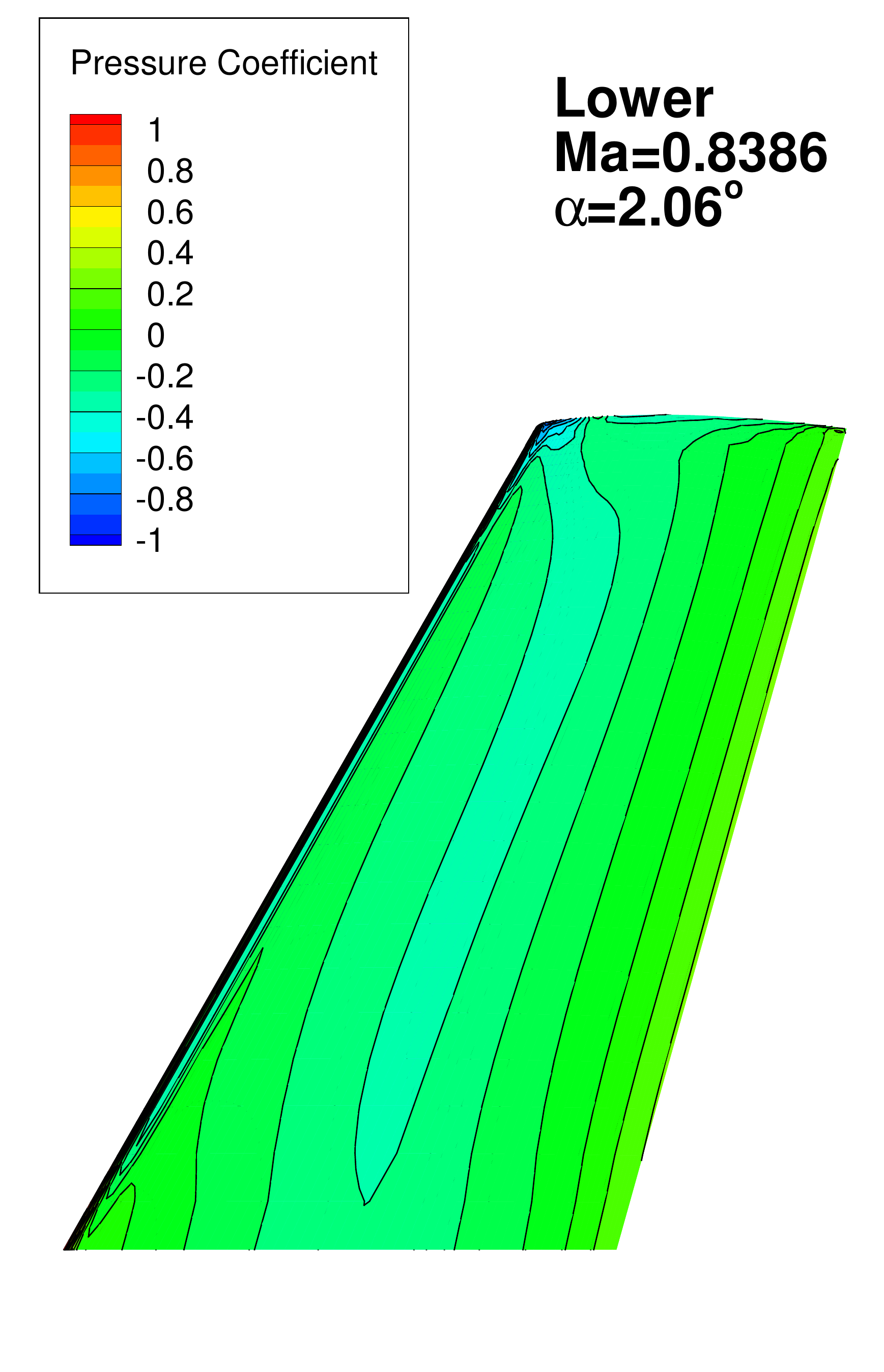} \\
        \includegraphics[width=1\textwidth]{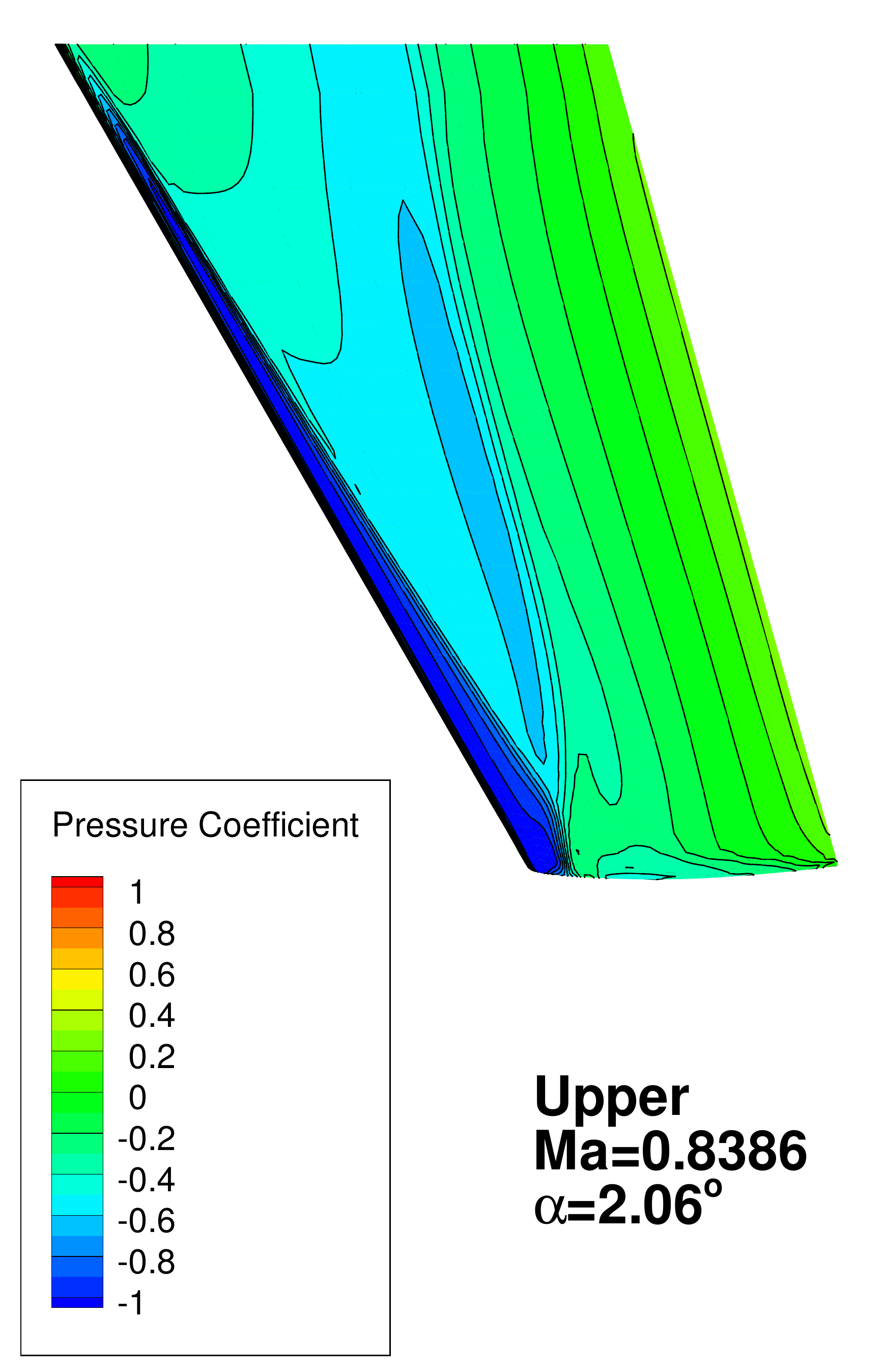}
        \end{minipage}
    }
    \subfigure[$\alpha=4.08^o$]{
        \begin{minipage}[b]{0.3\textwidth}
        \includegraphics[width=1\textwidth]{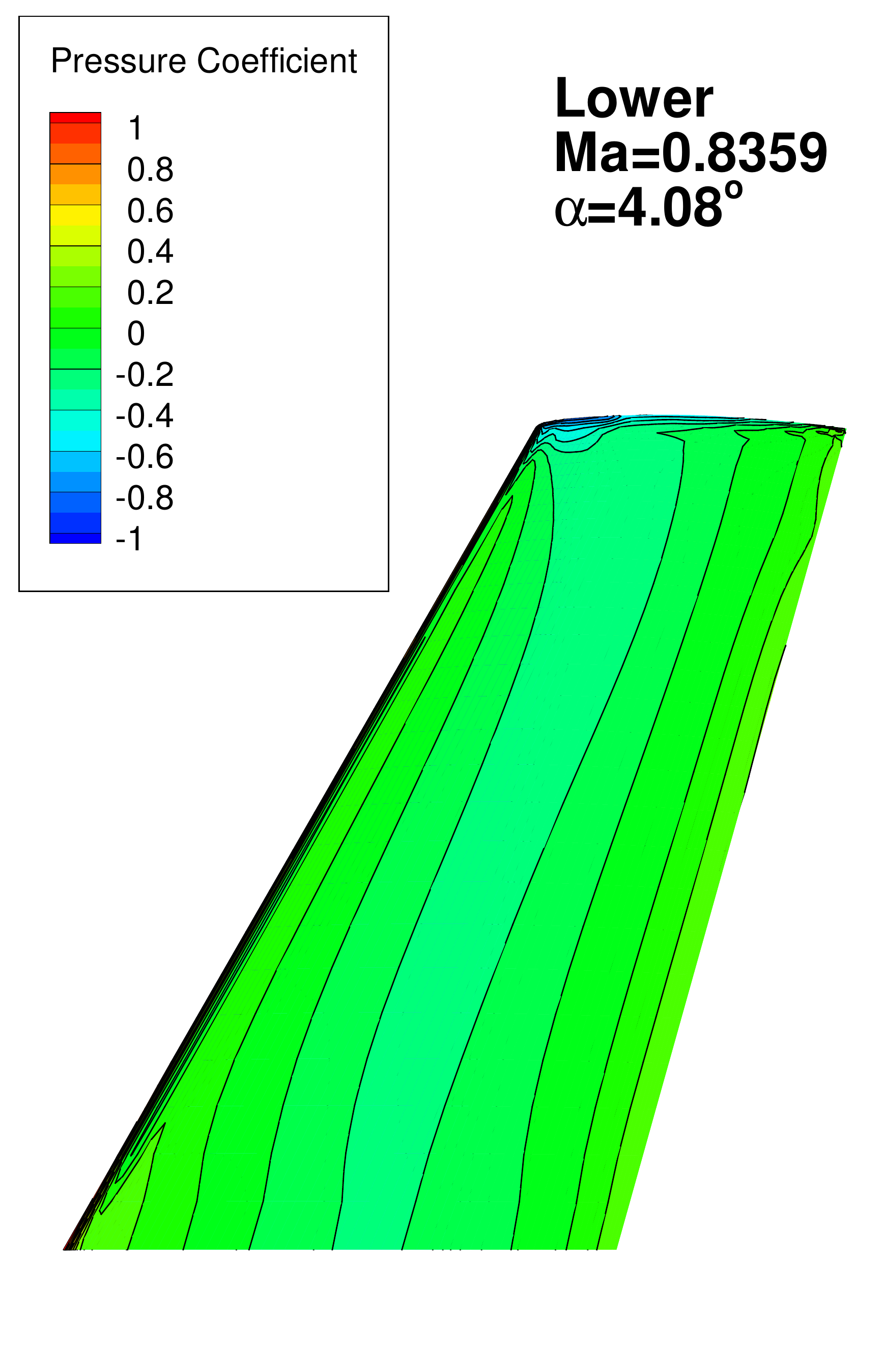} \\
        \includegraphics[width=1\textwidth]{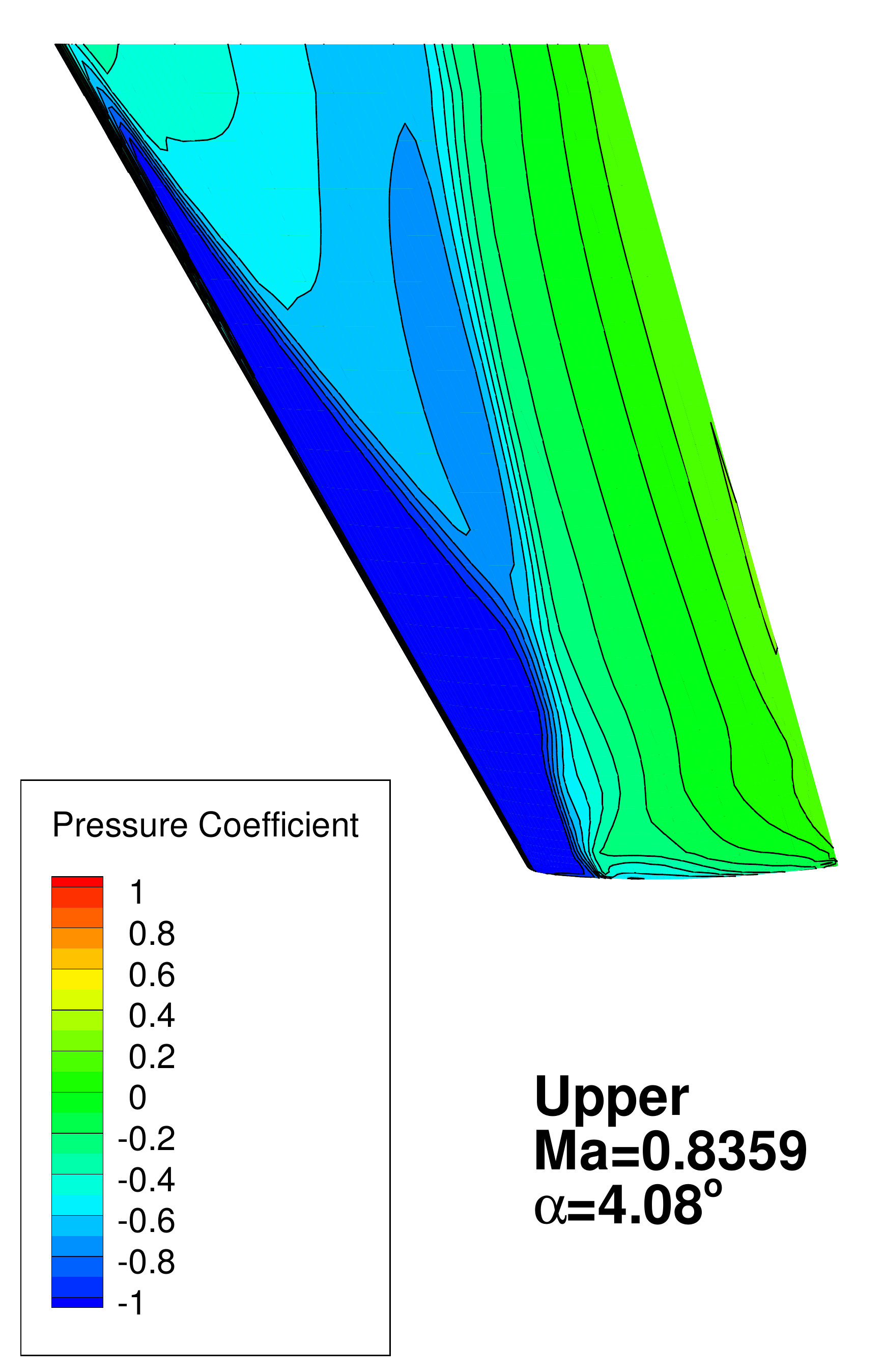}
        \end{minipage}
    }
    \caption{The computed steady pressure distributions at different $\alpha$.}
    \label{fig:Fig5}
\end{figure}

\begin{figure}[htbp]
\centering
\includegraphics[width=0.6\textwidth]{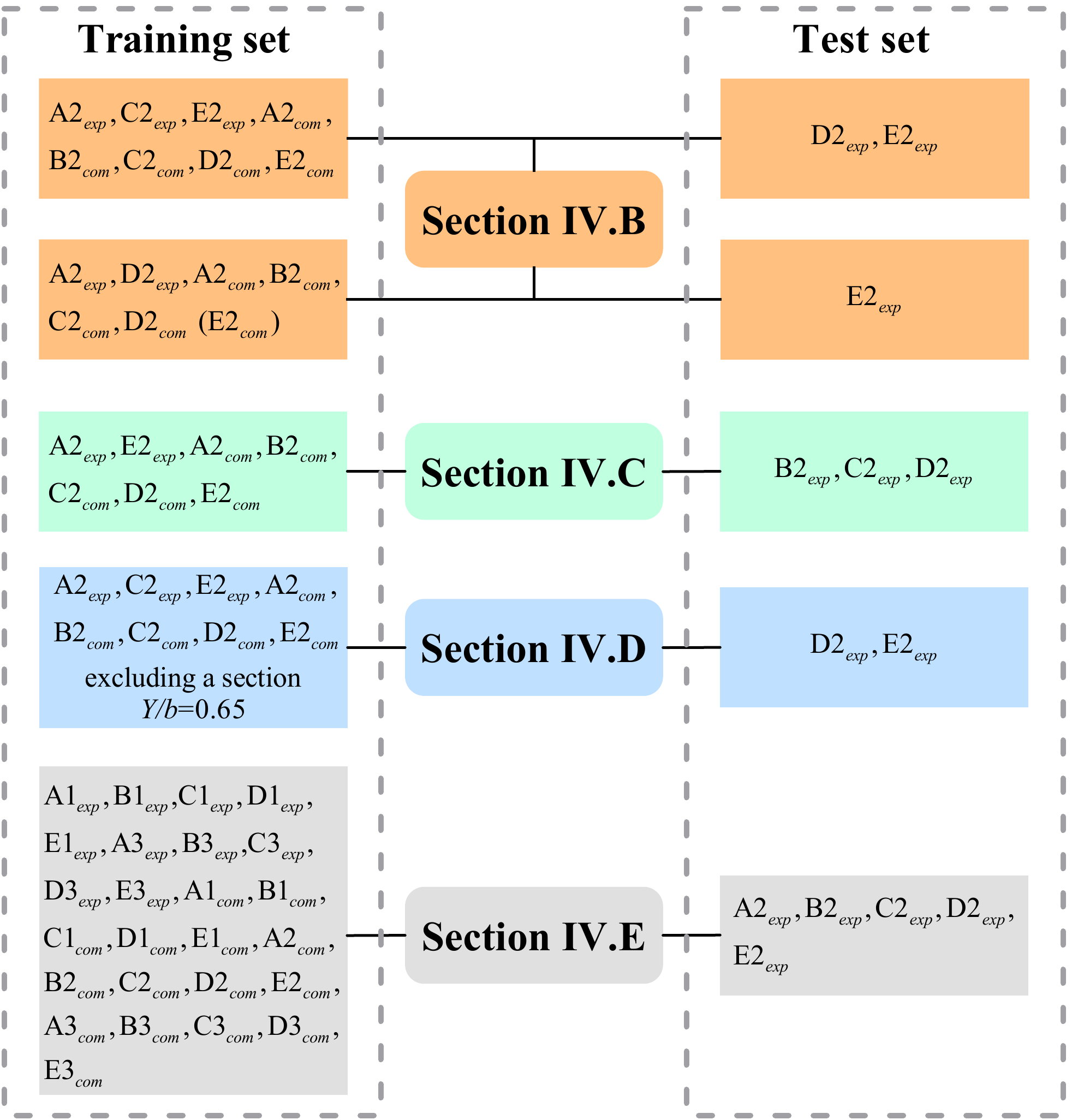}
\caption{Training and test cases for each section.}
\label{fig:Fig6}
\end{figure}

\begin{figure}[htbp]
\centering
\includegraphics[width=0.6\textwidth]{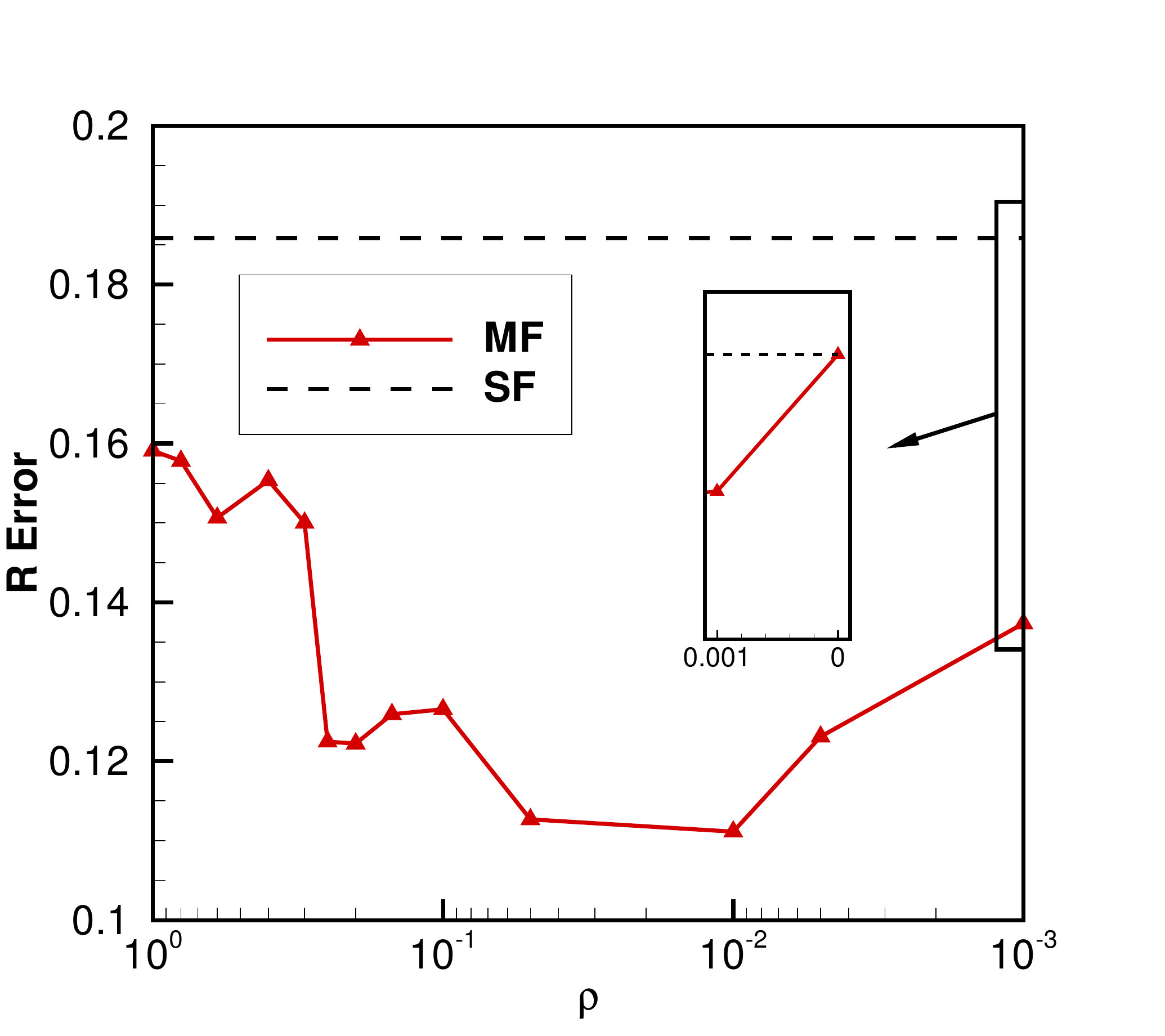}
\caption{$R$ of the reference data set with decreasing $\rho$.}
\label{fig:Fig7}
\end{figure}

In this section, the propose approach is validated by modeling and predicting the surface pressure distribution of ONERA M6 wing at transonic regime. Firstly, the data sets are briefly introduced. Then, the performance of the proposed approach is investigated for several cases with increasing complexity. In order to evaluate the accuracy, three errors are defined as follows:

\begin{equation}
\label{equ:Equ6}
\begin{array}{l}
R = \frac{{\left\| {{{\bf{y}}_{Model}} - {{\bf{y}}_{Exp}}} \right\|_F^2}}{{\left\| {{{\bf{y}}_{Exp}}} \right\|_F^2}} \times 100\% \\
MAE = \frac{1}{m}\sum\limits_{i = 1}^m {\left| {(y_{Model}^i - y_{Exp}^i)} \right|} \\
RMSE = \sqrt {\frac{1}{m}\sum\limits_{i = 1}^m {{{(y_{Model}^i - y_{Exp}^i)}^2}} } 
\end{array}
\end{equation}
where $R$, $MAE$, and $RMSE$ is the relative error, mean absolute error, and root mean squared error, respectively. 

\subsection{Data Sets} \label{section:sec3.1}

\begin{table}[htbp]
\caption{\label{tab:table1} Flow conditions}
\centering
\begin{tabular}[b]{ccccc}
\hline
Experimental case& Computational case& $Ma$& $\alpha(/^o)$& $Re$\\
\hline
A1$_{exp}$          & A1$_{com}$           & 0.6998 & 0.04   & \multirow{15}{*}{$1.172\times10^7$} \\
B1$_{exp}$          & B1$_{com}$           & 0.7003 & 1.08   &                             \\
C1$_{exp}$          & C1$_{com}$           & 0.7001 & 2.06   &                             \\
D1$_{exp}$          & D1$_{com}$           & 0.6990 & 3.06   &                             \\
E1$_{exp}$          & E1$_{com}$           & 0.7009 & 4.08   &                             \\
A2$_{exp}$          & A2$_{com}$           & 0.8399 & 0.04   &                             \\
B2$_{exp}$          & B2$_{com}$           & 0.8398 & 1.07   &                             \\
C2$_{exp}$          & C2$_{com}$           & 0.8386 & 2.06   &                             \\
D2$_{exp}$          & D2$_{com}$           & 0.8395 & 3.06   &                             \\
E2$_{exp}$          & E2$_{com}$           & 0.8359 & 4.08   &                             \\
A3$_{exp}$          & A3$_{com}$           & 0.8840 & 0.03   &                             \\
B3$_{exp}$          & B3$_{com}$           & 0.8833 & 1.08   &                             \\
C3$_{exp}$          & C3$_{com}$           & 0.8803 & 2.05   &                             \\
D3$_{exp}$          & D3$_{com}$           & 0.8809 & 3.06   &                             \\
E3$_{exp}$          & E3$_{com}$           & 0.8831 & 4.07   &                             \\
\hline
\end{tabular}
\end{table}


The ONERA M6 wing experiment was originally illustrated in an AGARD report by Schmitt and Charpin \cite{schmitt1979pressure}, which has been widely used for CFD validation. The Reynolds number was $1.172\times10^7$ based on the mean aerodynamic chord $c=0.64607$m and the semi-span was $b=1.1963$m. The experimental dataset consists of surface pressure coefficients at seven spanwise locations ($Y/b$=0.20, 0.44, 0.65, 0.80, 0.90, 0.95, 0.99). The cases used in this work are denoted in Table.\ref{tab:table1}.

The computed steady pressure distributions at different $\alpha$ are compared in Fig..\ref{fig:Fig5}. Since the "lambda" shock on the surface of the wing differs greatly, modeling the surface pressure distributions with this flow regime is a tough work.

\subsection{Investigation on angle of attack} \label{section:sec3.2}

In this subsection, the data set with three experimental case A2$_{exp}$, C2$_{exp}$, E2$_{exp}$, and five computational cases A2$_{com}$, B2$_{com}$, C2$_{com}$, D2$_{com}$, E2$_{com}$, is employed to validate the proposed approach as shown in Table.\ref{fig:Fig6}.

To begin with, insights on the selection of weighting parameter rho should be obtained from numerical experiment. 20\% of the preceding data set is randomly selected as validation samples for $\rho$. In Fig.\ref{fig:Fig7}, MF and SF denote the results from the multi-fidelity model (see Eq.~\eqref{equ:Equ4}) and  the single-fidelity model (see Eq.~\eqref{equ:Equ3}) only using the high-fidelity data, respectively. As $\rho$ decreases, the error firstly drops and then increase. When $\rho=1$, $R$ error of MF is smaller than that of SF; when $\rho=0$, MF degenerates into SF; $R$ error of MF reaches a minimum around $\rho=0.01$. Therefore, $\rho=0.01$ is used in this subsection.

\begin{figure}[htbp]
\centering
    \subfigure[$Y/b=0.20$]{
    \includegraphics[width=0.31\textwidth]{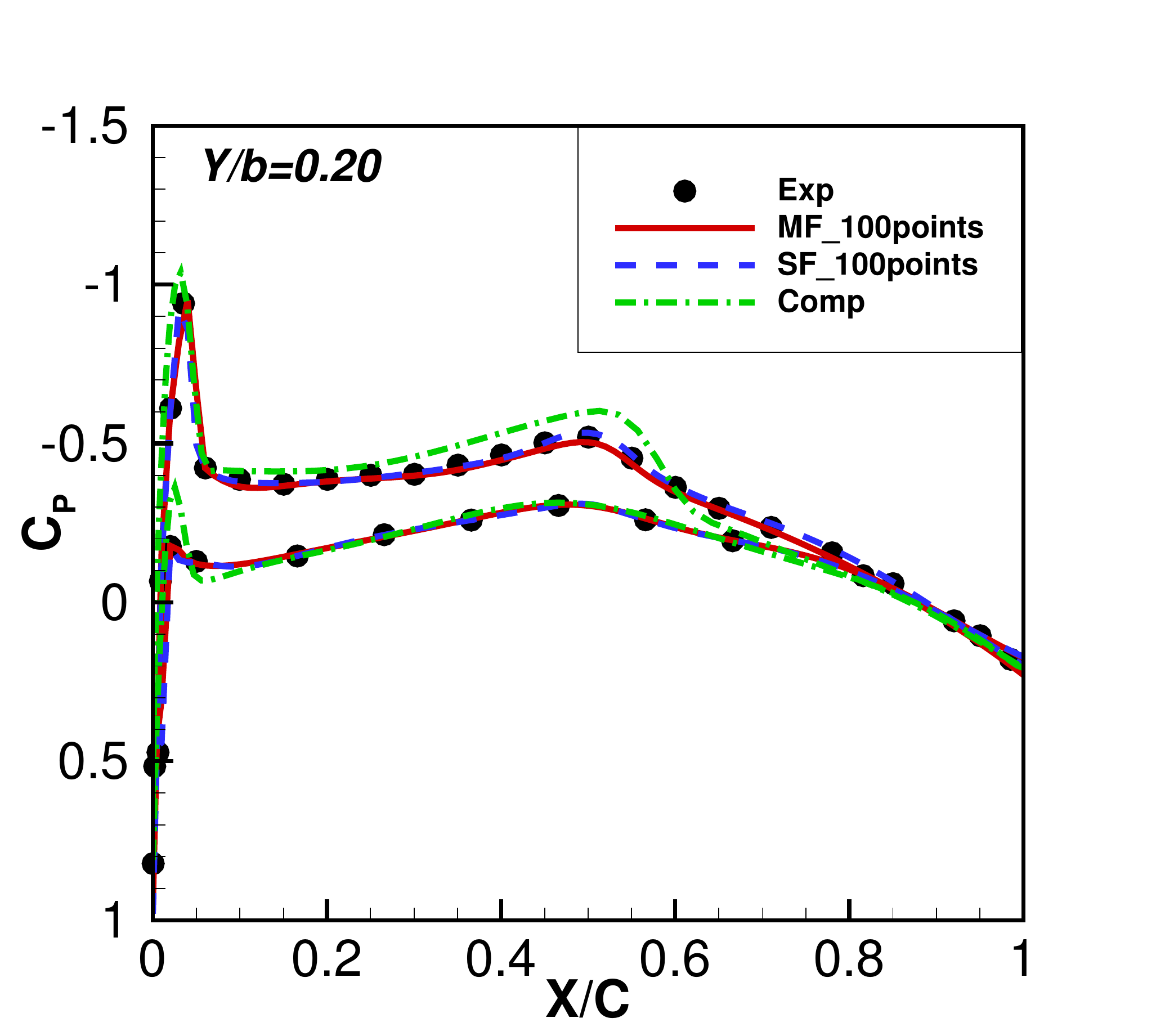}
    }
    \subfigure[$Y/b=0.44$]{
    \includegraphics[width=0.31\textwidth]{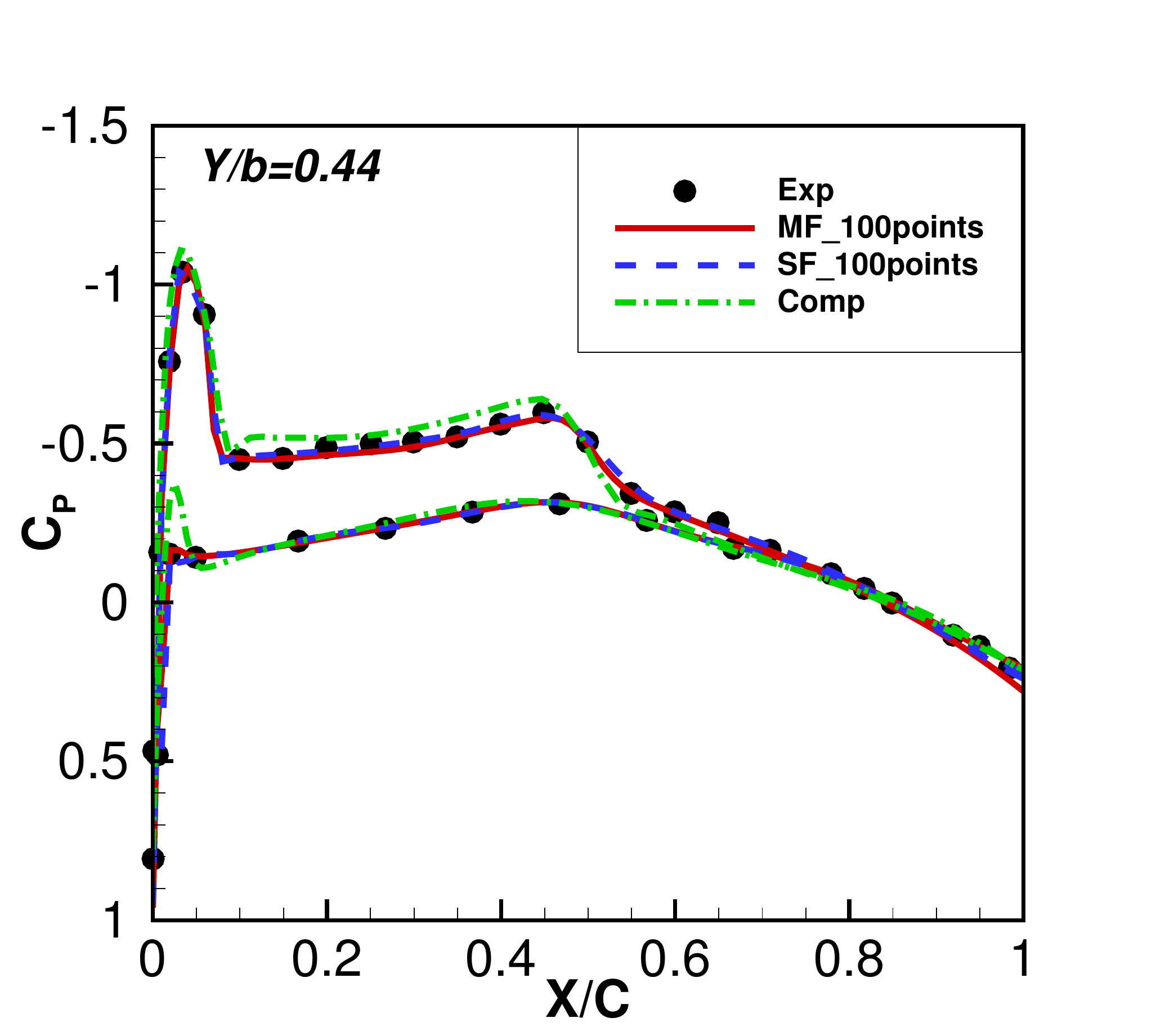}
    }
    \subfigure[$Y/b=0.65$]{
    \includegraphics[width=0.31\textwidth]{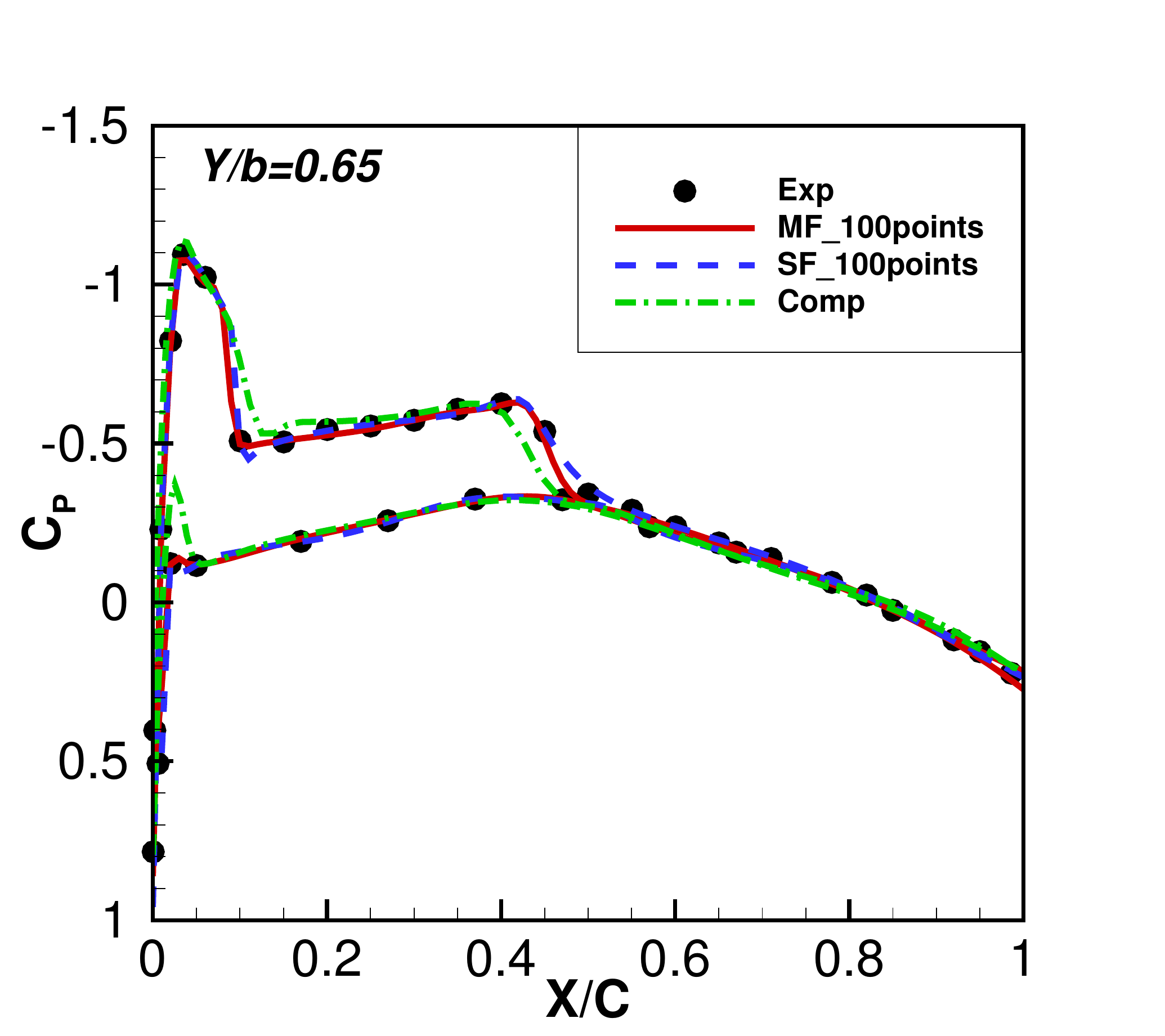}
    }
    \quad
    \subfigure[$Y/b=0.80$]{
    \includegraphics[width=0.31\textwidth]{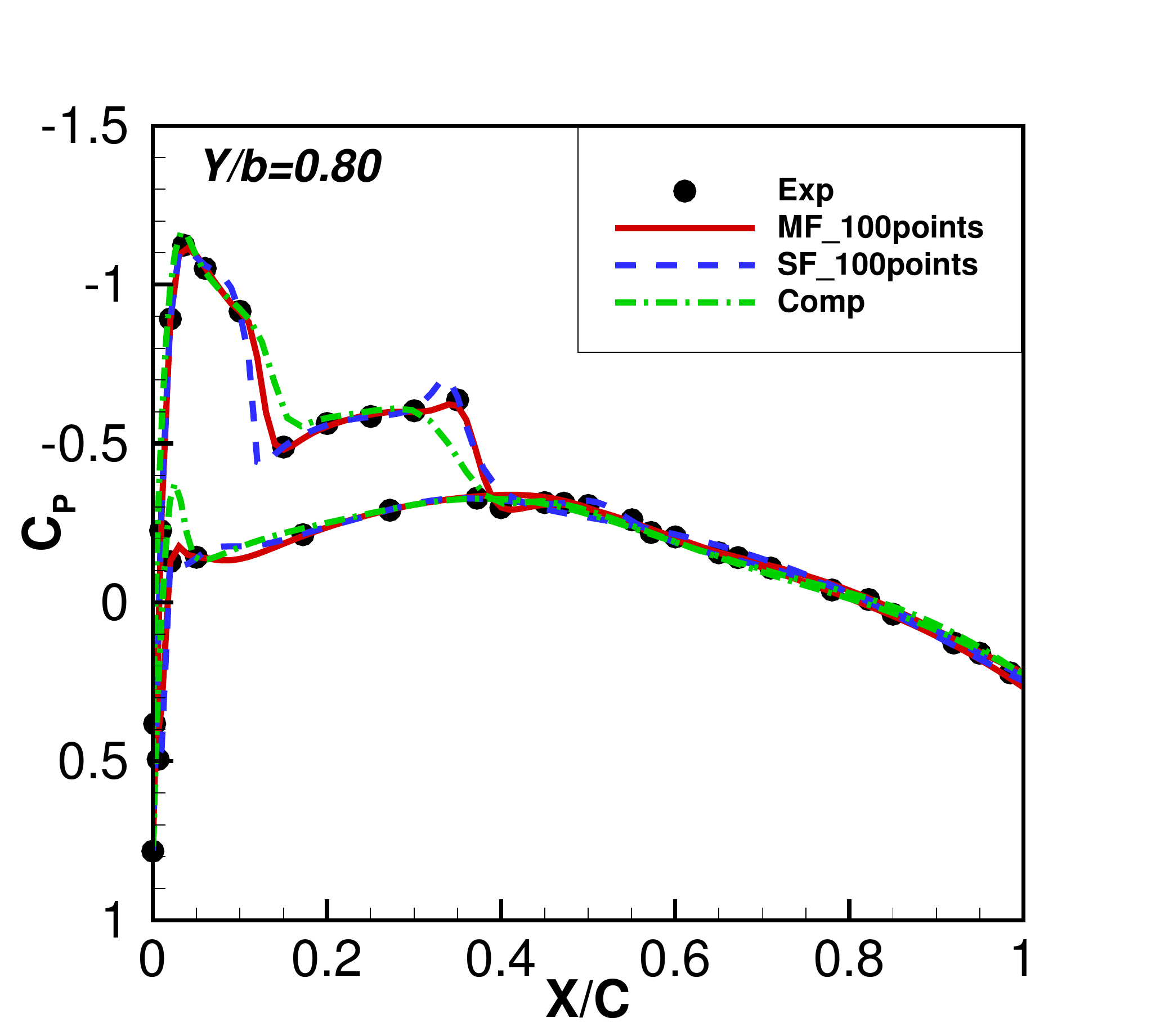}
    }
    \subfigure[$Y/b=0.90$]{
    \includegraphics[width=0.31\textwidth]{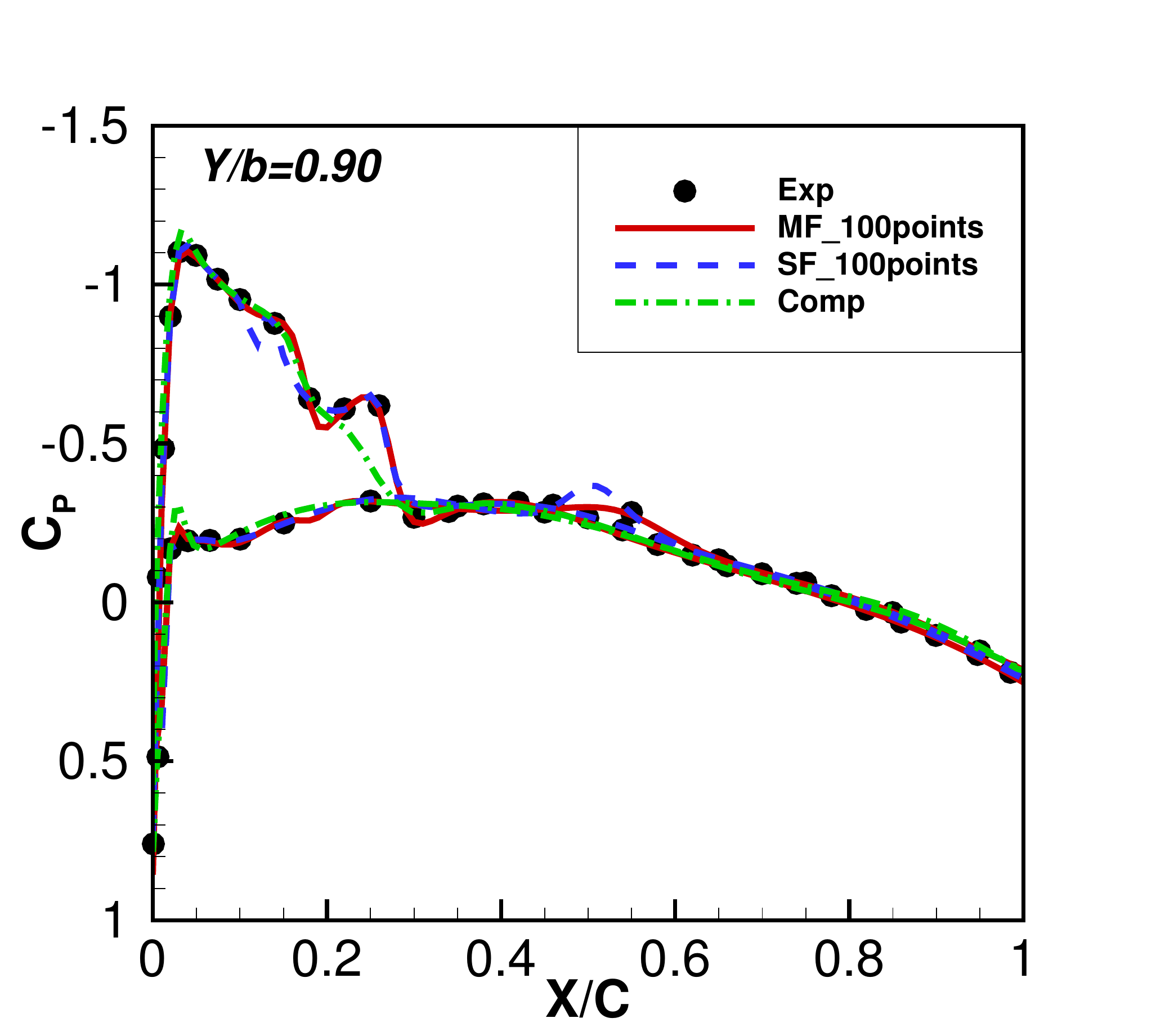}
    }
    \subfigure[$Y/b=0.95$]{
    \includegraphics[width=0.31\textwidth]{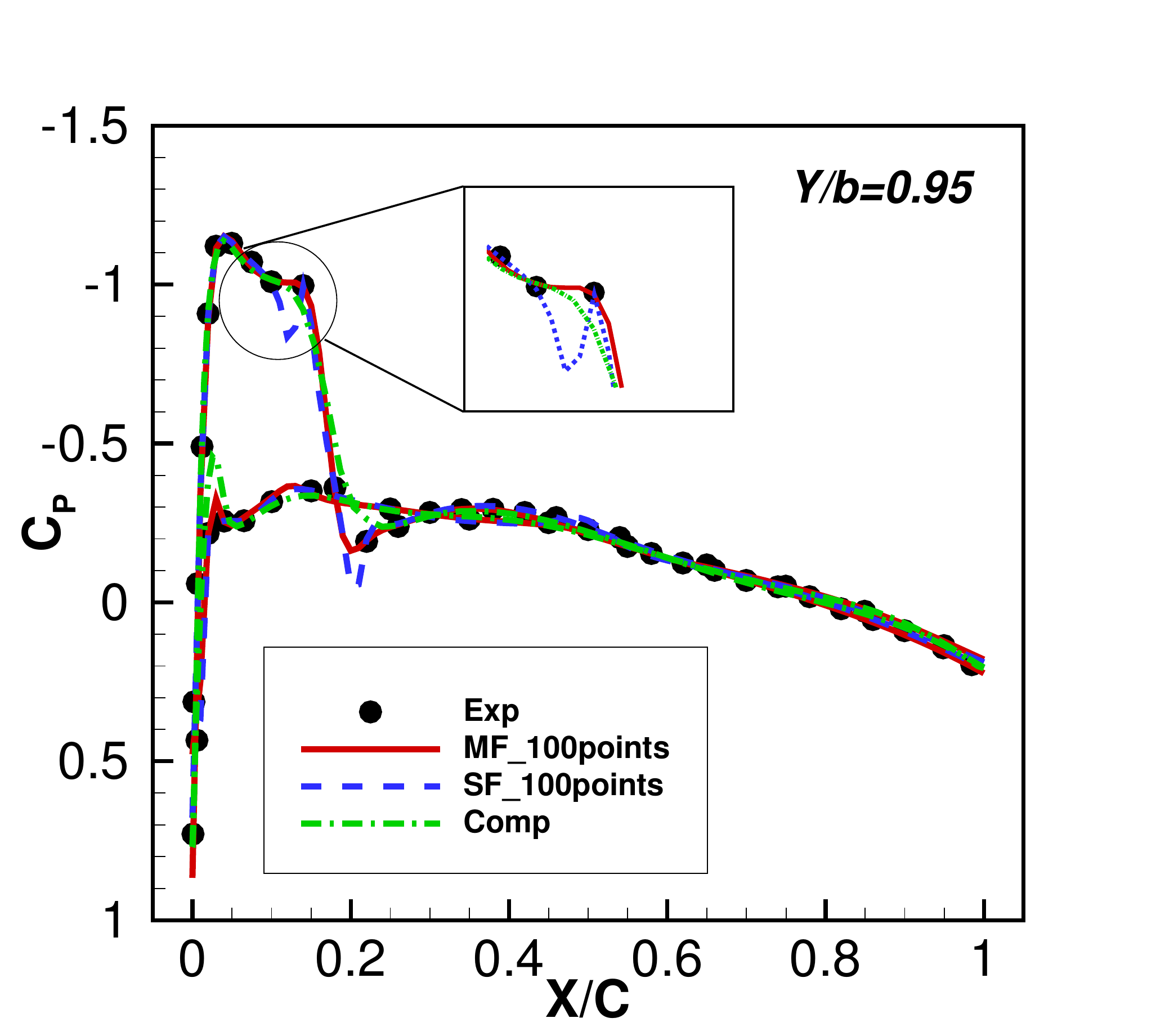}
    }
\caption{$C_p$ at different sections for the training case C2$_{exp}$.}
\label{fig:Fig8}
\end{figure}

The predicted surface pressure distributions for the training case C2$_{exp}$ are shown in Fig.\ref{fig:Fig8}. Due the sparse experimental data, the proposed approach is utilized for each section with 100 points. As a consequece in these figure, MF\_100points and SF\_100points are defined as the multi-fidelity and the single-fidelity DNN model with 100 points for each section, respectively. Exp and Comp denote the experimental and computational results, respectively. It is indicated in Fig.\ref{fig:Fig8} that no significant differences were found between the experimental data and, MF or SF, except for $Y/b=0.90$ and $Y/b=0.95$. By using the low-fidelity data as soft penalty constraint, the proposed model provides more reasonable and accurate predictions.

\begin{figure}[htbp]
\centering
    \subfigure[$Y/b=0.20$]{
    \includegraphics[width=0.3\textwidth]{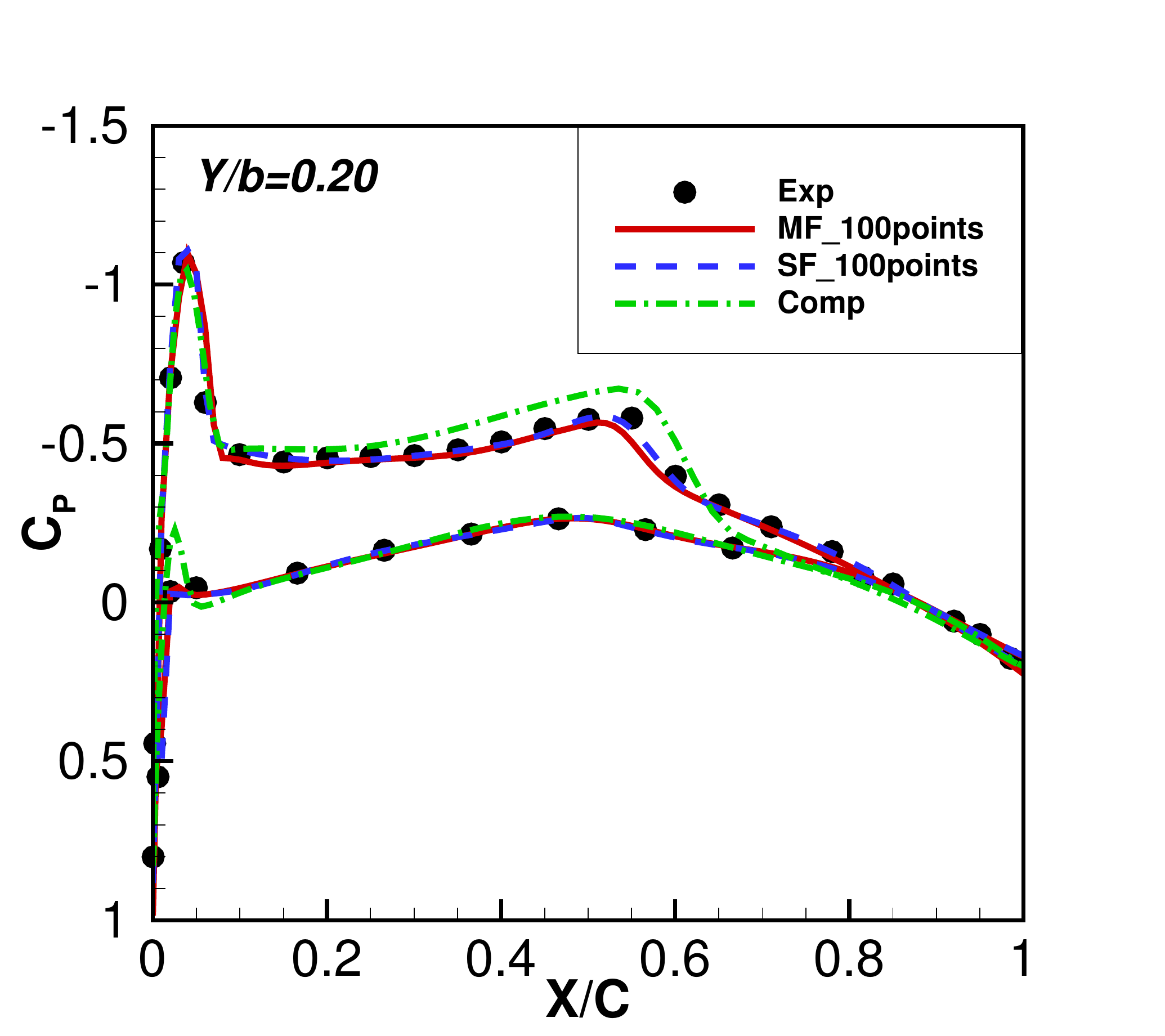}
    }
    \subfigure[$Y/b=0.44$]{
    \includegraphics[width=0.3\textwidth]{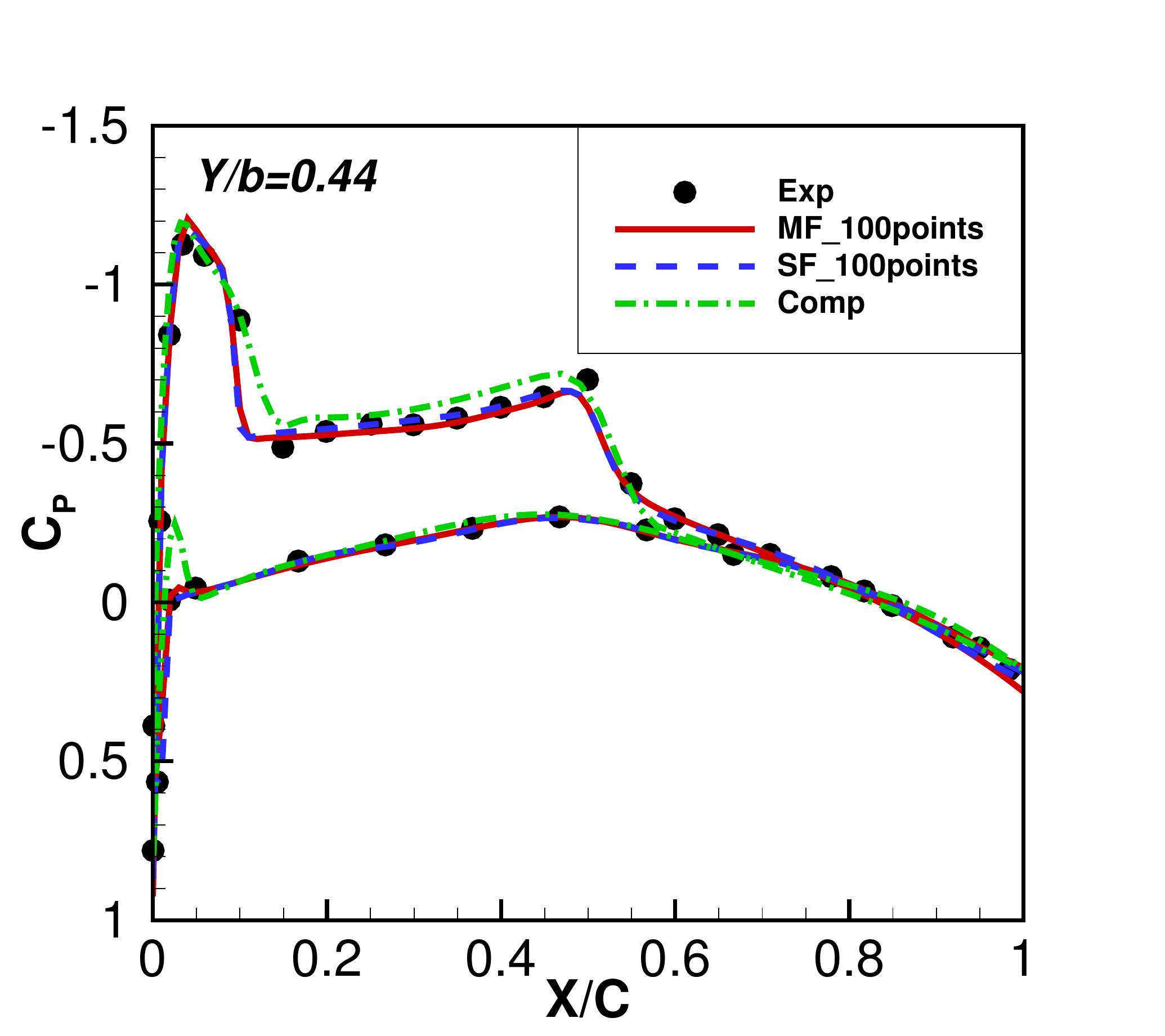}
    }
    \subfigure[$Y/b=0.65$]{
    \includegraphics[width=0.3\textwidth]{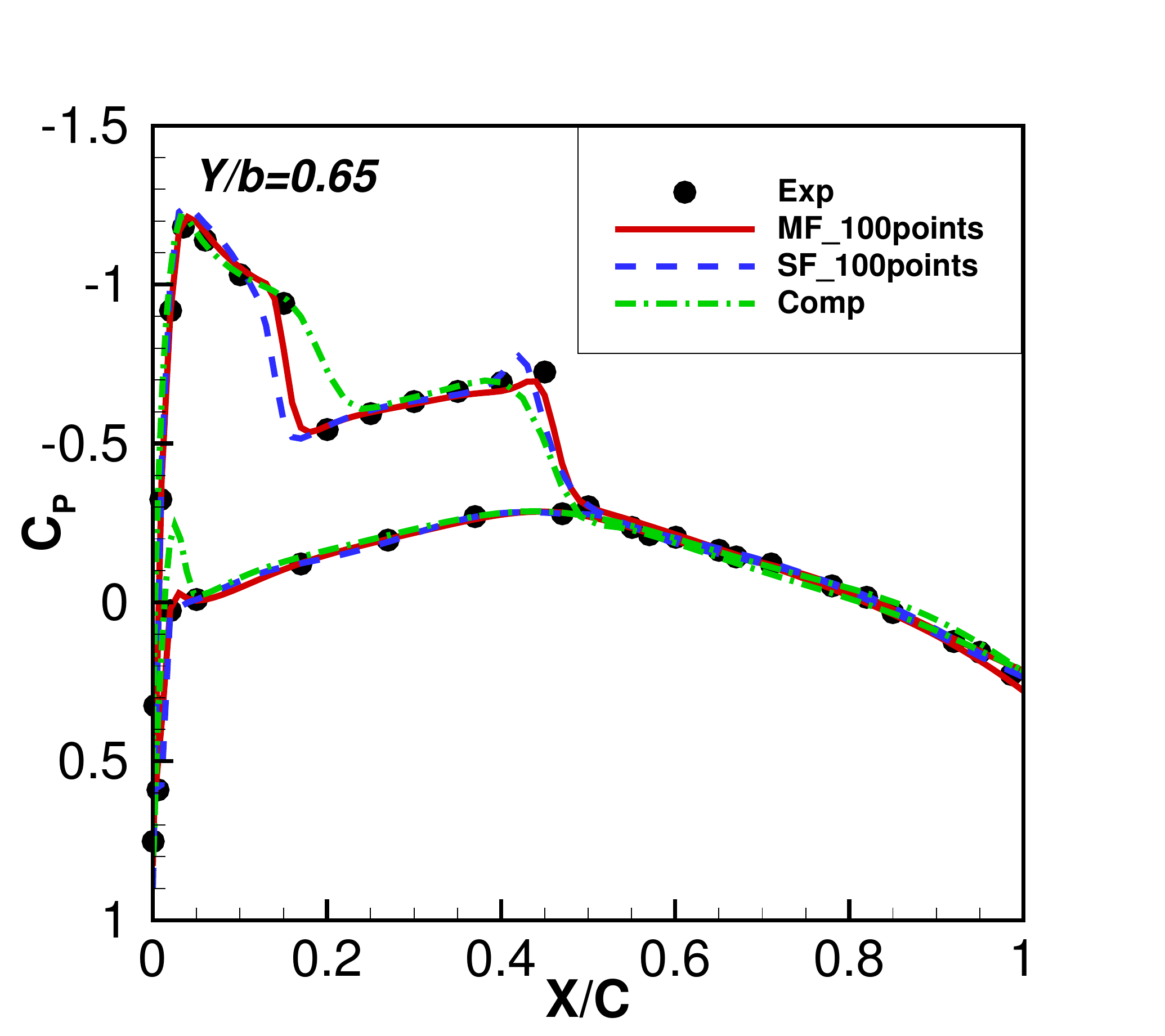}
    }
    \quad
    \subfigure[$Y/b=0.80$]{
    \includegraphics[width=0.3\textwidth]{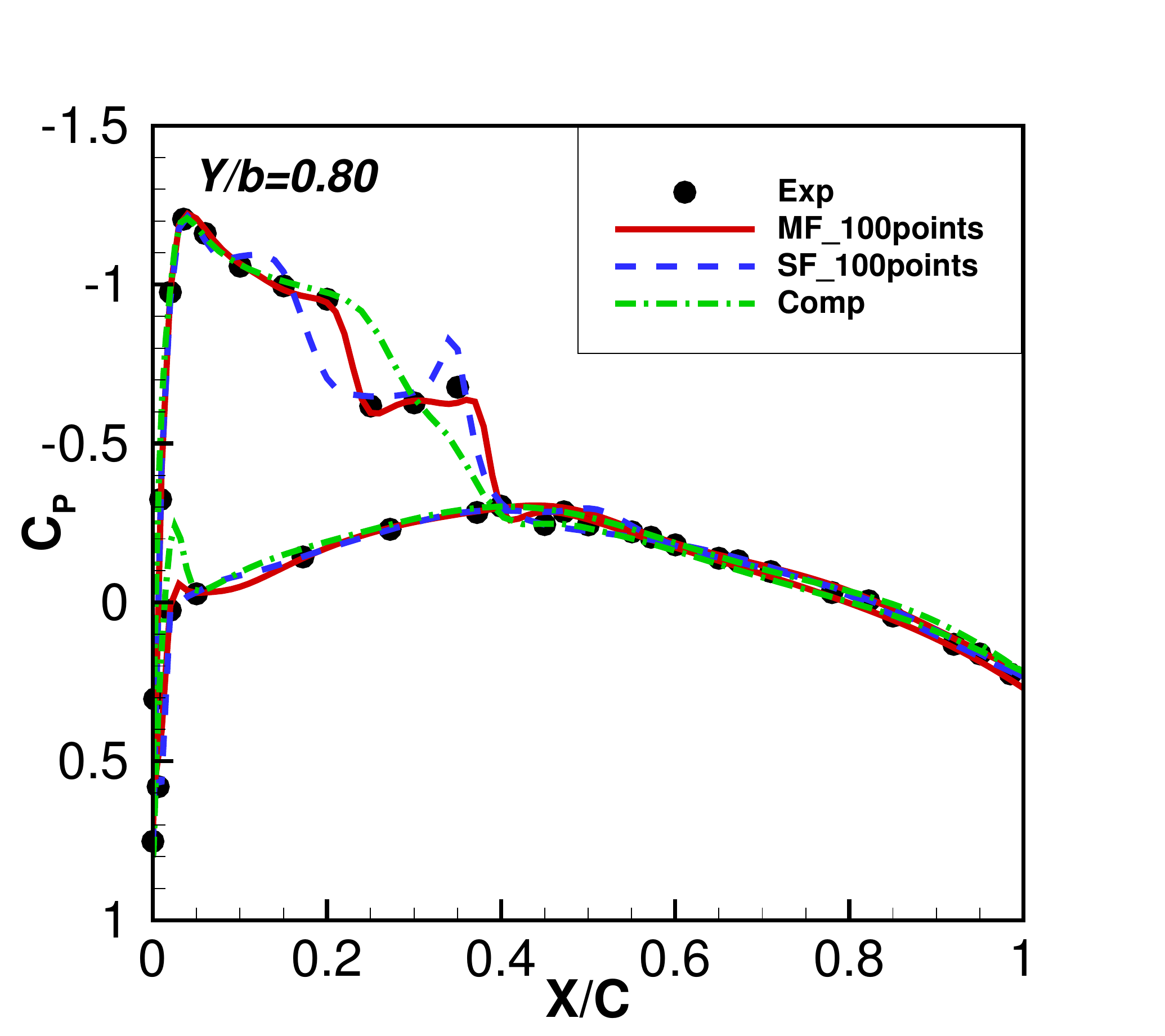}
    }
    \subfigure[$Y/b=0.90$]{
    \includegraphics[width=0.3\textwidth]{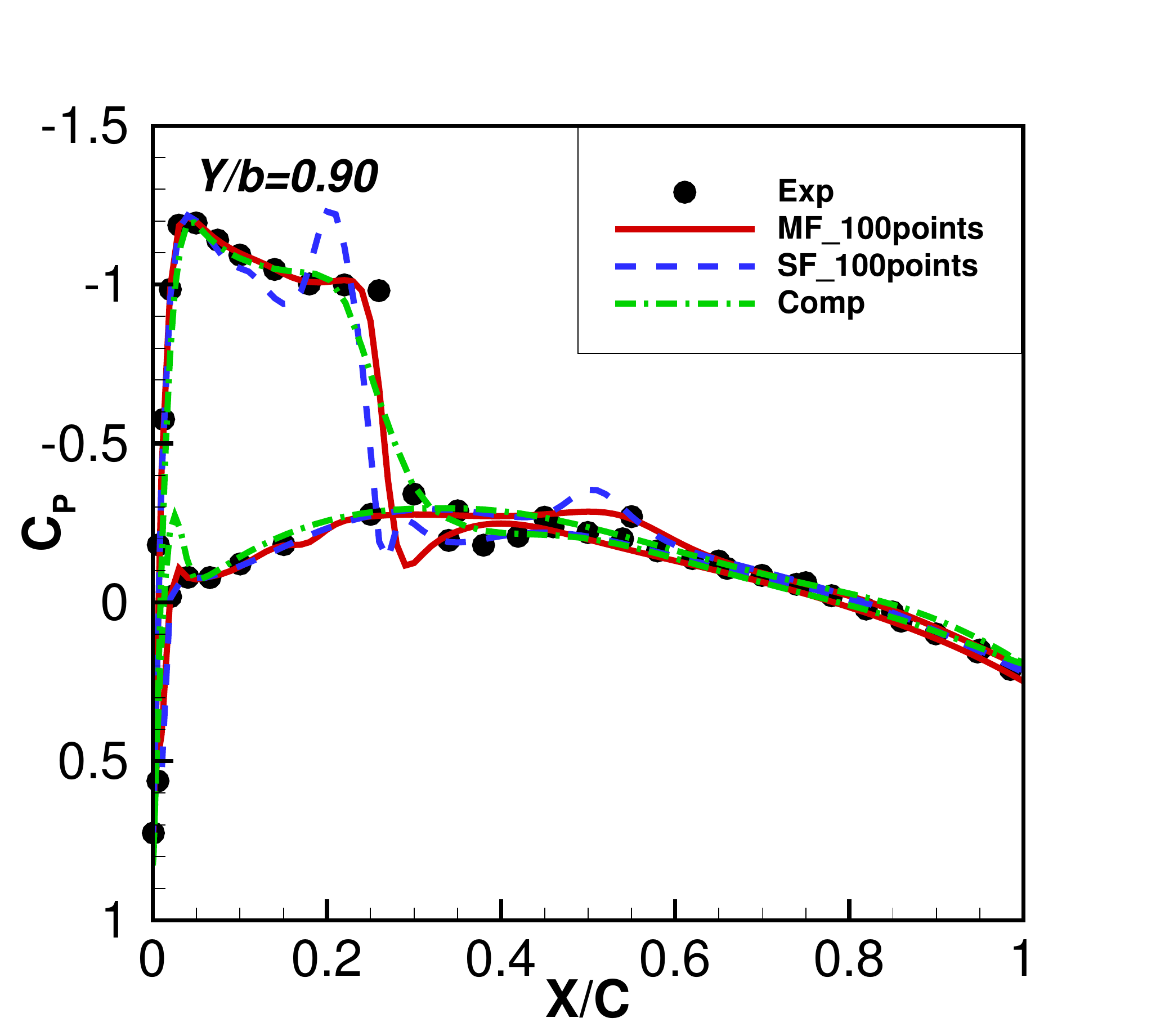}
    }
    \subfigure[$Y/b=0.95$]{
    \includegraphics[width=0.3\textwidth]{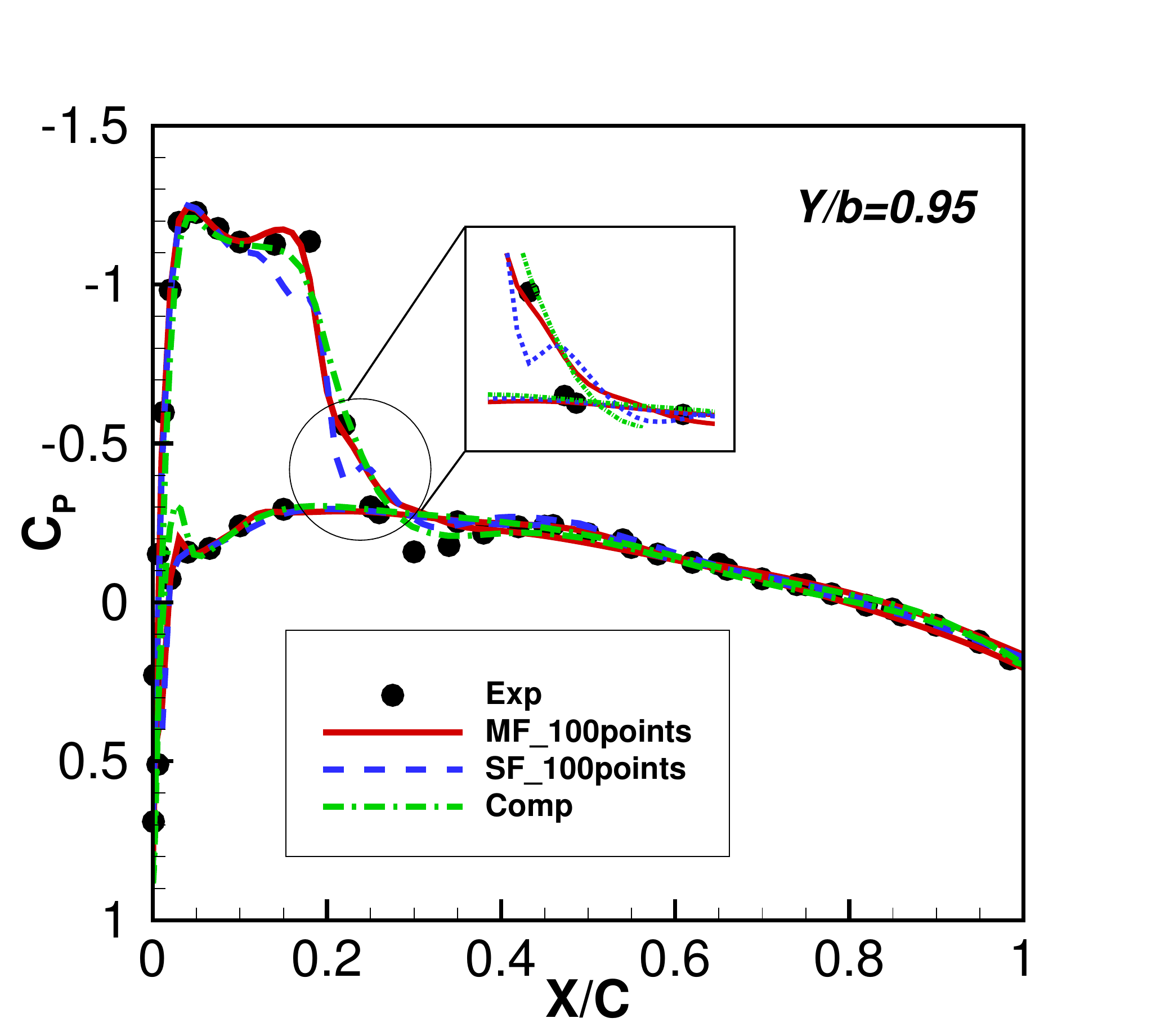}
    }
\caption{$C_p$ at different sections for the test case D2$_{exp}$.}
\label{fig:Fig9}
\end{figure}

Fig.\ref{fig:Fig9} shows the predicted surface pressure distributions for the test case D2$_{exp}$. MF, as well as SF, show a good agreement with the high-fidelity solution at sections $Y/b=0.20, 0.44, 0.65$. However, at sections $Y/b=0.80, 0.90, 0.95$, the accuracy of MF is better than that of SF, indicating that the accuracy and generalization can be enhanced by imposing the low-fidelity data as constraint. In addition, it should be noted that the existing results for M6 wing at section $Y/b=0.80$ are lacking, which can be described accurately by the proposed multi-fidelity model.

\begin{figure}[htbp]
\centering
    \subfigure[MF1:$Y/b=0.20$]{
    \includegraphics[width=0.3\textwidth]{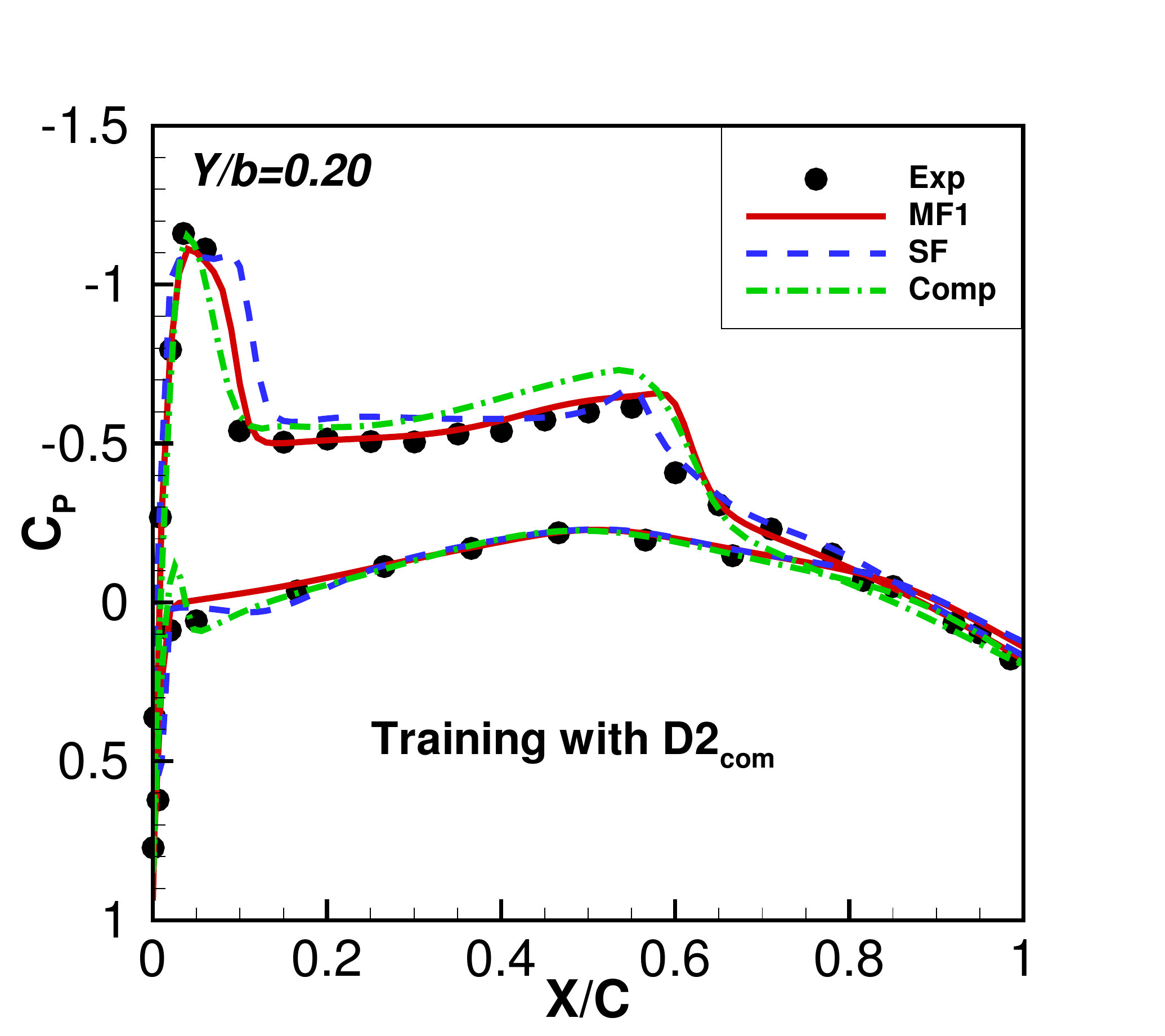}
    }
    \subfigure[MF1:$Y/b=0.44$]{
    \includegraphics[width=0.3\textwidth]{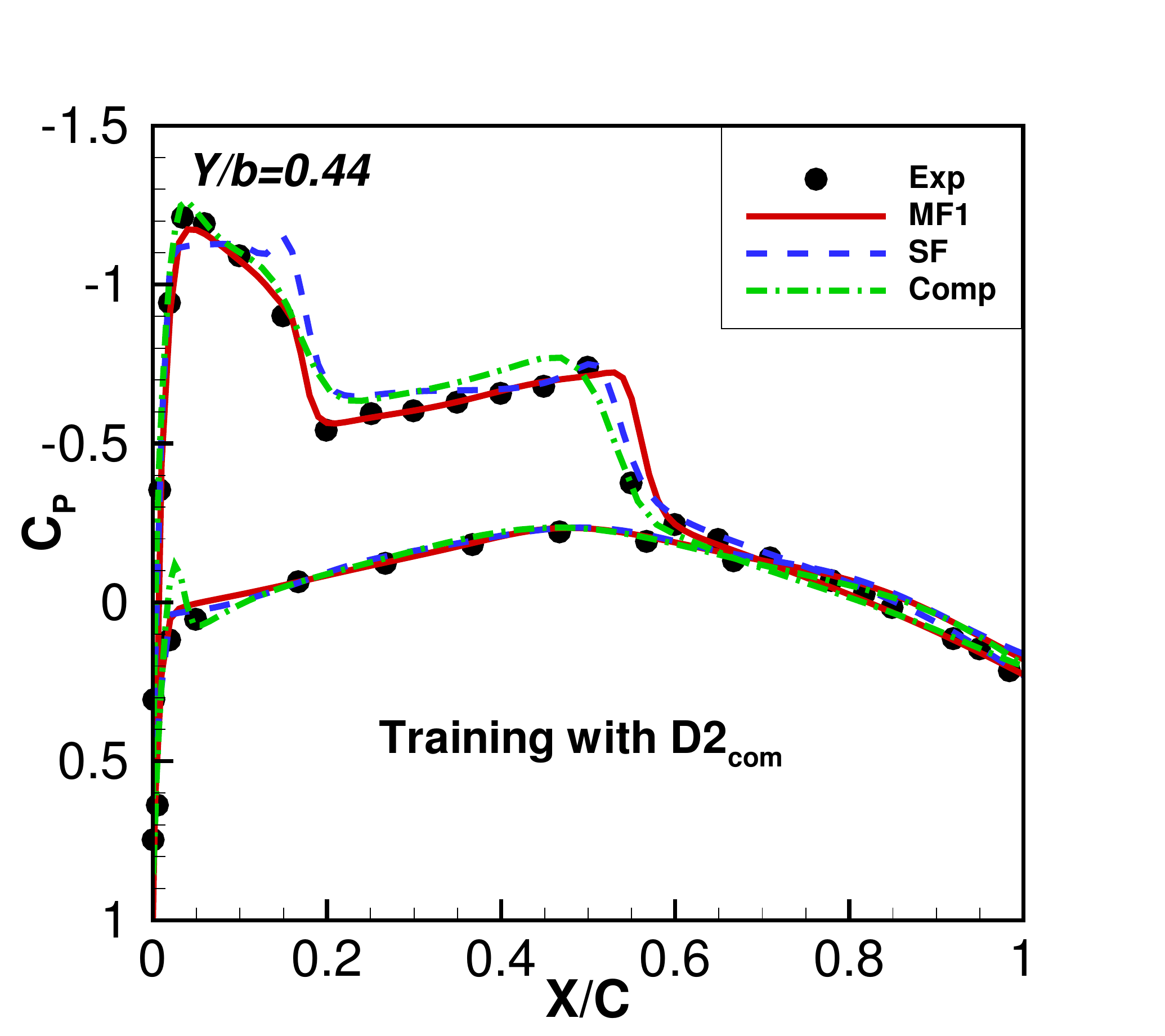}
    }
    \subfigure[MF1:$Y/b=0.80$]{
    \includegraphics[width=0.3\textwidth]{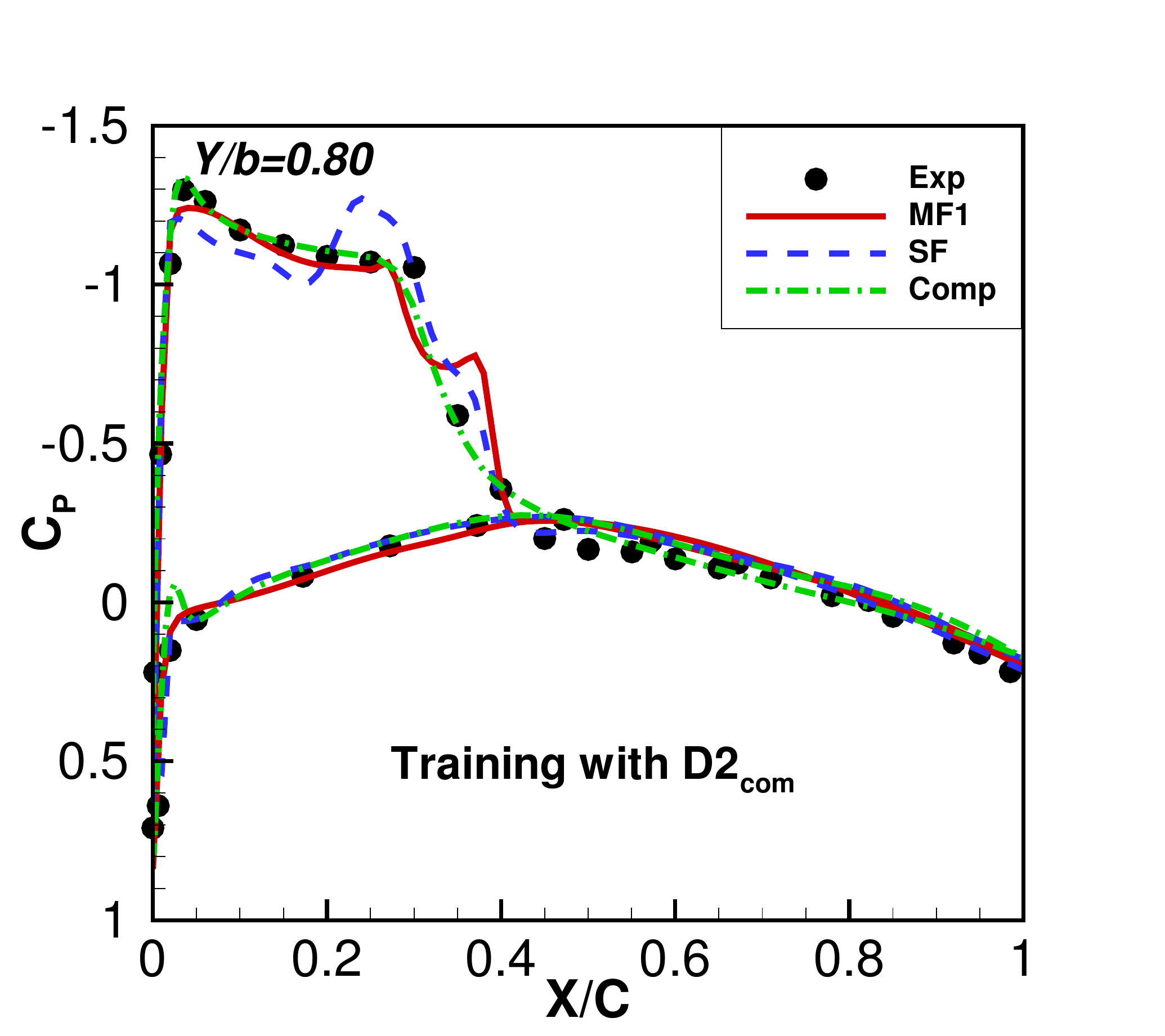}
    }
    \quad
    \subfigure[MF2:$Y/b=0.20$]{
    \includegraphics[width=0.3\textwidth]{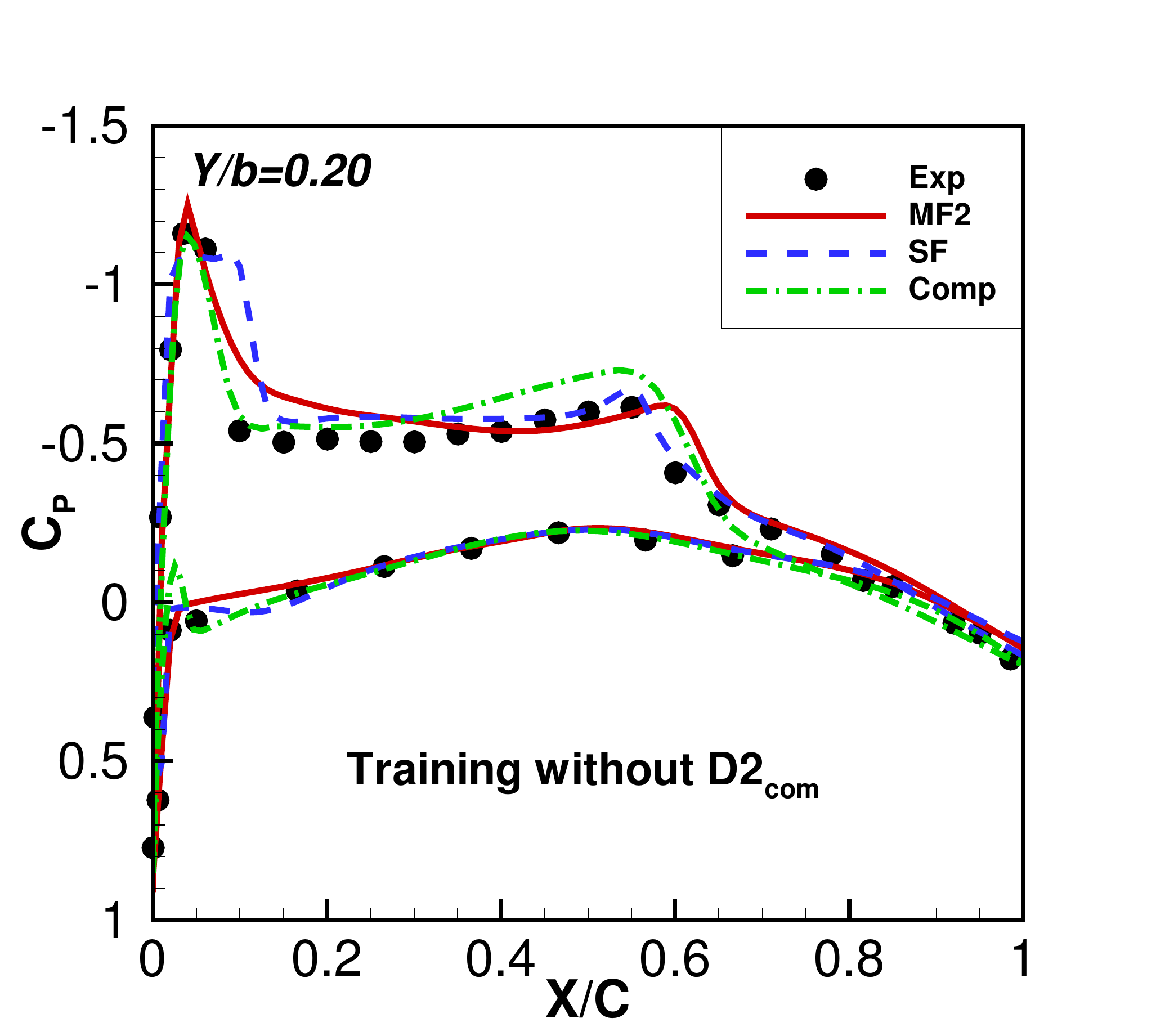}
    }
    \subfigure[MF2:$Y/b=0.44$]{
    \includegraphics[width=0.3\textwidth]{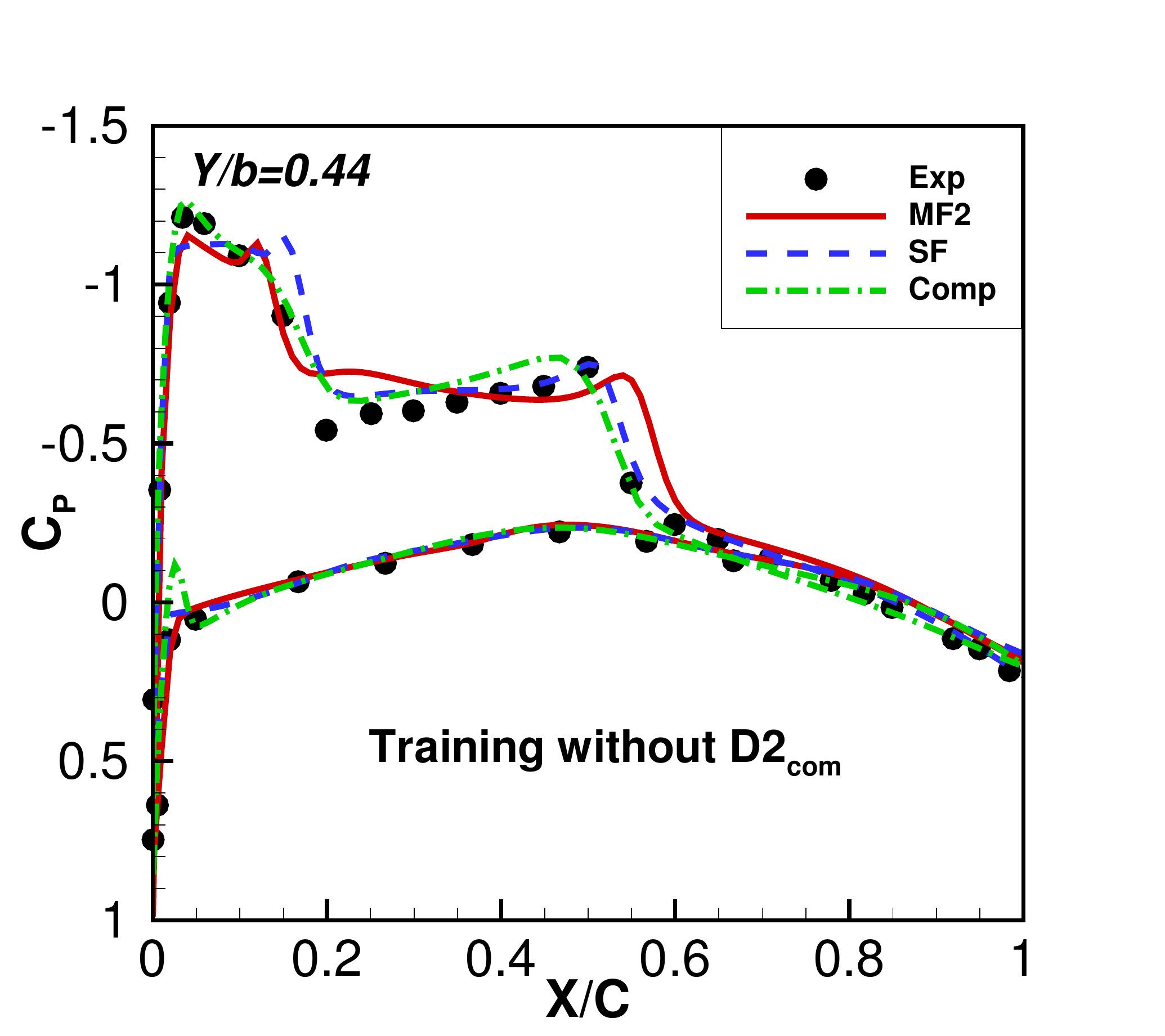}
    }
    \subfigure[MF2:$Y/b=0.80$]{
    \includegraphics[width=0.3\textwidth]{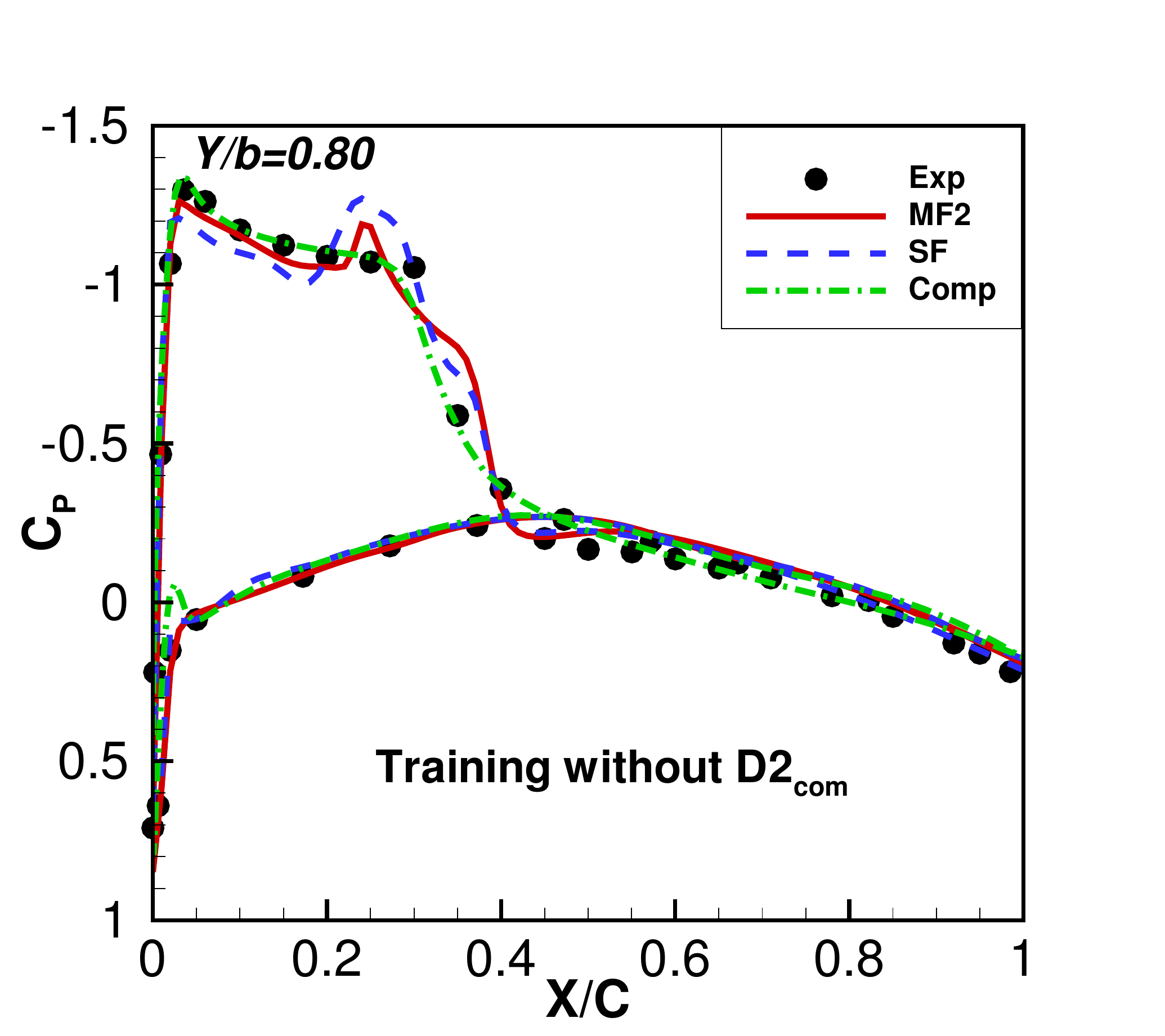}
    }
\caption{$C_p$ computed by MF1 and MF2 at different sections for the test case E2$_{exp}$.}
\label{fig:Fig10}
\end{figure}

\begin{figure}[htbp]
\centering
\includegraphics[width=0.8\textwidth]{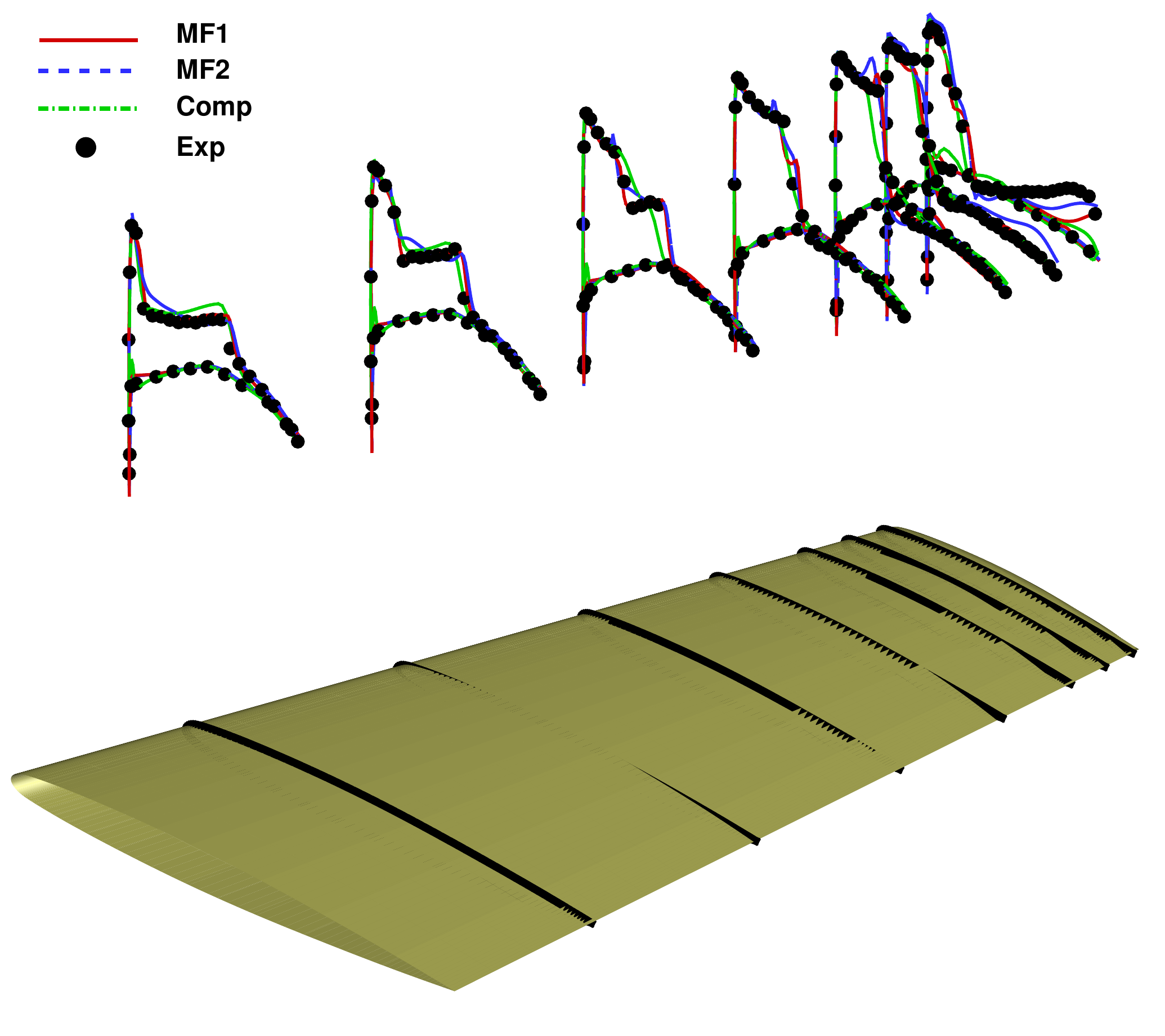}
\caption{Comparison between MF1 and MF2 for the test case E2$_{exp}$.}
\label{fig:Fig11}
\end{figure}

For a further investigation on the proposed model performance, a extrapolated task with respect to $\alpha$ is performed. MF1 is constructed by the data set with two experimental case A2$_{exp}$, D2$_{exp}$, and five computational cases A2$_{com}$, B2$_{com}$, C2$_{com}$, D2$_{com}$, E2$_{com}$. MF2 is built by the data set with two experimental case A2$_{exp}$, D2$_{exp}$, and four computational cases A2$_{com}$, B2$_{com}$, C2$_{com}$, D2$_{com}$. SF is constructed by the data set with two experimental case A2$_{exp}$, D2$_{exp}$. The difference between MF1 and MF2 is whether the training data set includes the computational case E2$_{com}$. As depicted in Fig.\ref{fig:Fig10} and Fig.\ref{fig:Fig11}, this case study confirms the importance of using the computational data for a multi-fidelity model in a extrapolated problem, since the test case E2$_{exp}$ is predicted with larger error by MF2 without computational case E2$_{com}$. In practice, one can include computational data of the state of interest to improve the prediction of high-fidelity solution at the same flow condition.

\begin{figure}[htbp]
\centering
\includegraphics[width=0.9\textwidth]{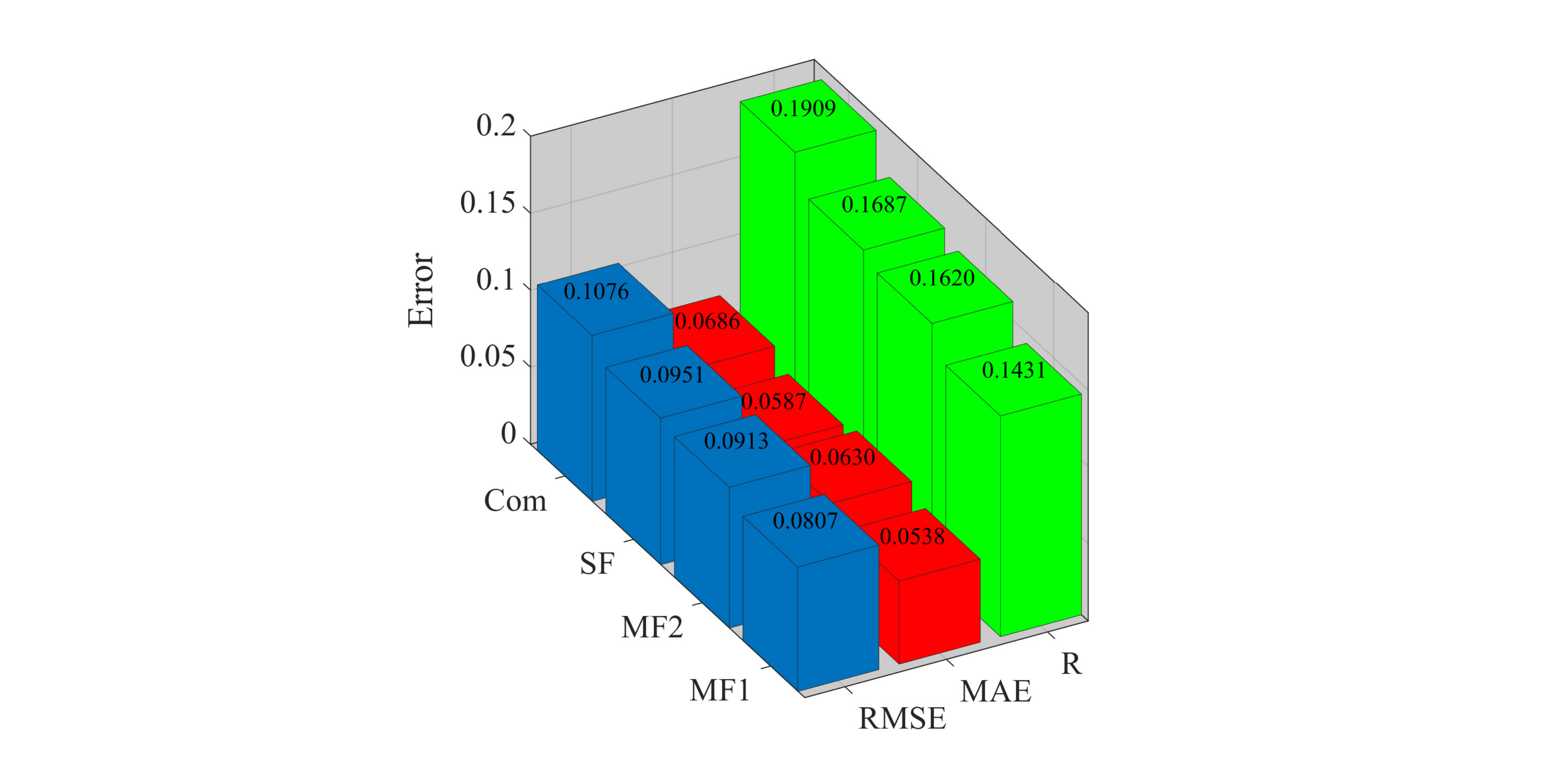}
\caption{Comparison of errors for the test case E2$_{exp}$.}
\label{fig:Fig12}
\end{figure}

From the quantitative comparison in Fig.\ref{fig:Fig12}, the $R$ error of MF1, MF2, SF, and for the test case E2$_{exp}$ are 14.31\%, 16.20\%, and 16.87\%, respectively. The other two errors exhibit similar trend. Hence, MF1 is more useful than MF2 and SF in a extrapolated task. However, It should be also emphasized that NNs are fundamentally interpolative models where accurate prediction is achieved in the range of input parameters used to train the model \cite{brunton2020machine}. Moreover, the generalization of NNs can be facilitated by using the multi-fidelity data in a certain extent, as proposed in this study.

\subsection{Investigation on less experimental data} \label{section:sec3.3}

\begin{table}[htbp]
\caption{\label{tab:table2} $R$ for B2$_{exp}$, C2$_{exp}$, D2$_{exp}$}
\centering
\begin{tabular}[b]{ccc}
\hline
Case  & MF (\%) & SF (\%) \\
\hline
D2$_{exp}$ & 14.20   & 38.86   \\
C2$_{exp}$ & 20.66   & 50.06   \\
B2$_{exp}$ & 18.34   & 38.41   \\
\hline
\end{tabular}
\end{table}

\begin{table}[htbp]
\caption{\label{tab:table3} $R$ at each section for B2$_{exp}$, C2$_{exp}$, D2$_{exp}$}
\centering
\begin{tabular}[b]{ccccccc}
\hline
Case  & \begin{tabular}[c]{@{}l@{}}Y/b=0.20\\    \\ MF/SF (\%)\end{tabular} & \begin{tabular}[c]{@{}l@{}}Y/b=0.44\\    \\ MF/SF (\%)\end{tabular} & \begin{tabular}[c]{@{}l@{}}Y/b=0.65\\    \\ MF/SF (\%)\end{tabular} & \begin{tabular}[c]{@{}l@{}}Y/b=0.80\\    \\ MF/SF (\%)\end{tabular} & \begin{tabular}[c]{@{}l@{}}Y/b=0.90\\    \\ MF/SF (\%)\end{tabular} & \begin{tabular}[c]{@{}l@{}}Y/b=0.95\\    \\ MF/SF (\%)\end{tabular} \\
\hline
D2$_{exp}$ & 18.05/59.87                                                         & 14.27/61.87                                                         & 12.87/47.89                                                         & 10.66/37.13                                                         & 10.76/19.96                                                         & 13.66/27.95                                                         \\
C2$_{exp}$ & 29.68/64.80                                                         & 22.28/54.97                                                         & 22.35/52.48                                                         & 17.29/51.57                                                         & 15.95/26.90                                                         & 18.58/47.99                                                         \\
B2$_{exp}$ & 22.82/41.02                                                         & 20.09/28.92                                                         & 19.71/30.66                                                         & 14.86/38.46                                                         & 17.04/37.62                                                         & 17.09/39.67                                                         \\
\hline
\end{tabular}
\end{table}

In this subsection, the performance of the proposed model using less experimental data is investigated. In particular, the train data consists of two experimental cases A2$_{exp}$, E2$_{exp}$ and five computational cases A2$_{com}$, B2$_{com}$, C2$_{com}$, D2$_{com}$, E2$_{com}$ according to Table.\ref{fig:Fig6}.

\begin{figure}[htbp]
\centering
    \subfigure[$Y/b=0.20$]{
    \includegraphics[width=0.3\textwidth]{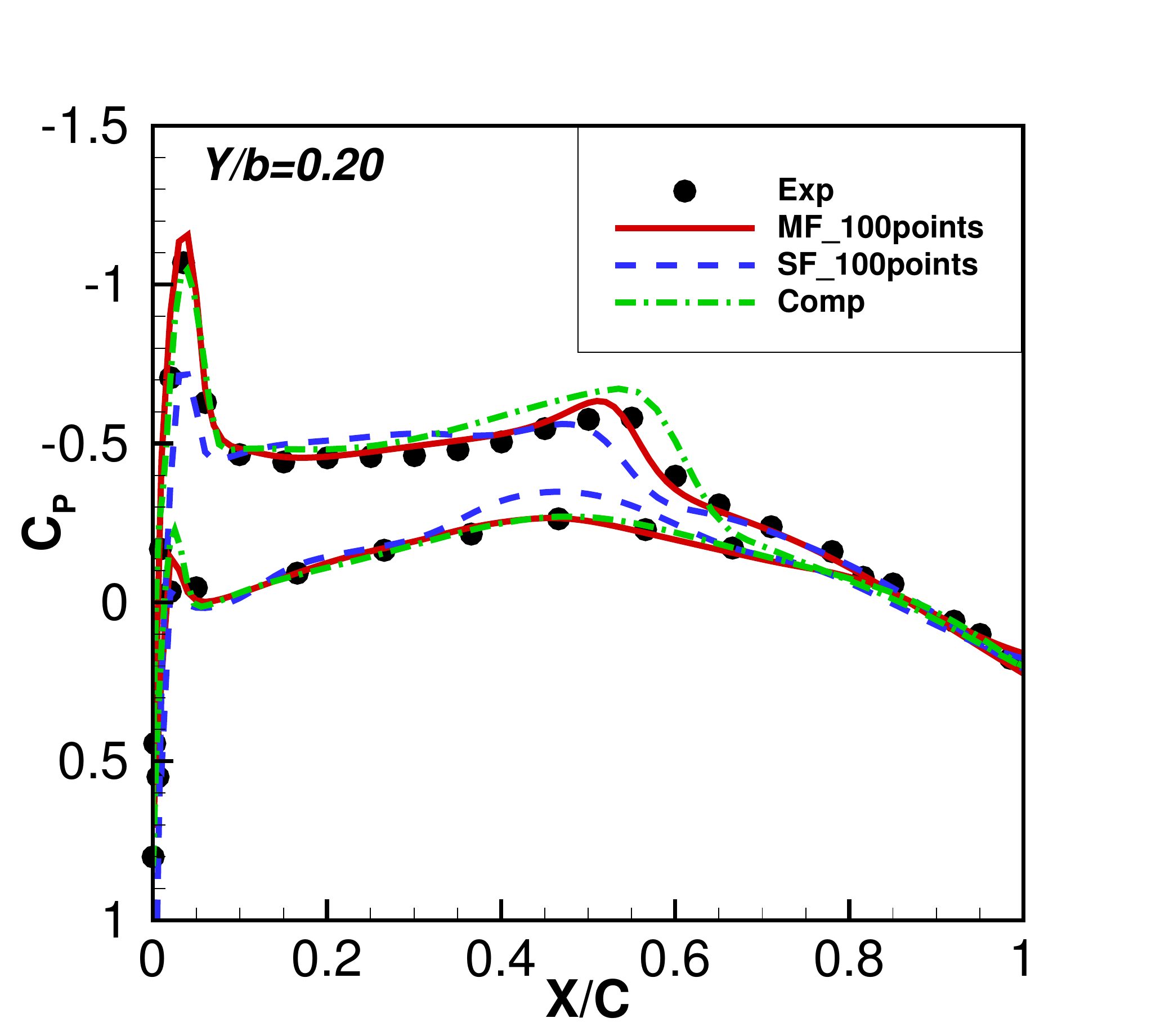}
    }
    \subfigure[$Y/b=0.44$]{
    \includegraphics[width=0.3\textwidth]{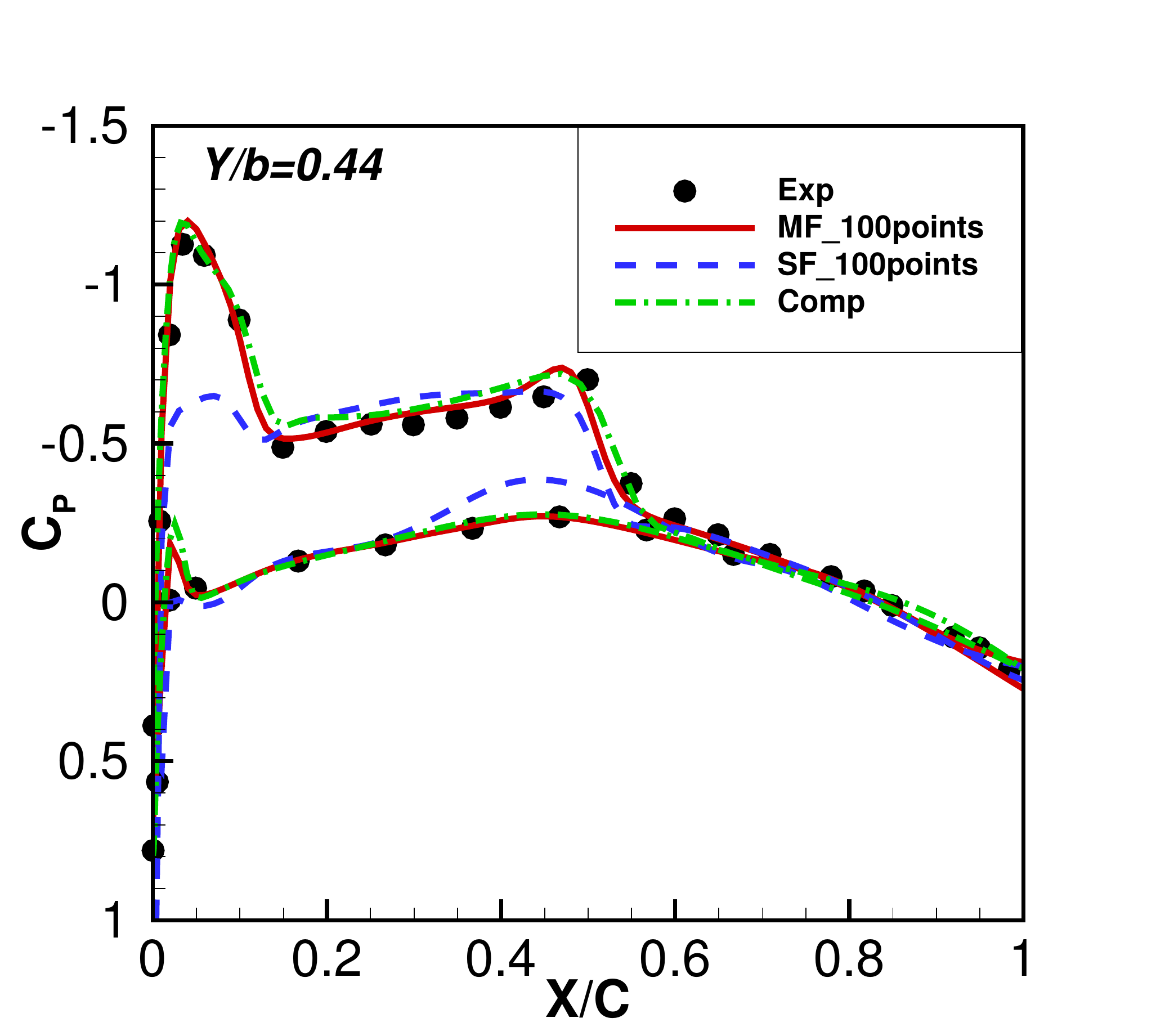}
    }
    \subfigure[$Y/b=0.65$]{
    \includegraphics[width=0.3\textwidth]{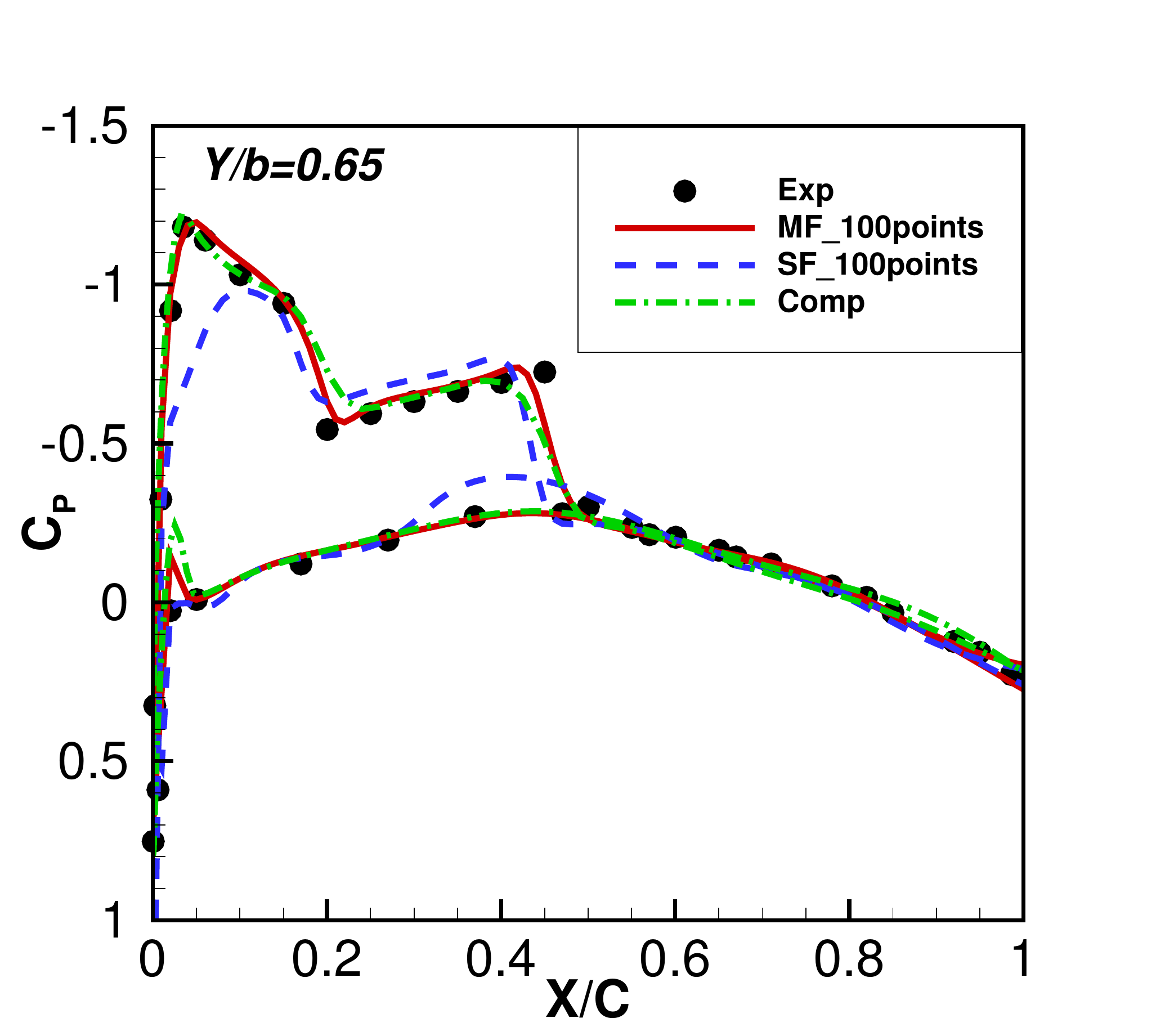}
    }
    \quad
    \subfigure[$Y/b=0.80$]{
    \includegraphics[width=0.3\textwidth]{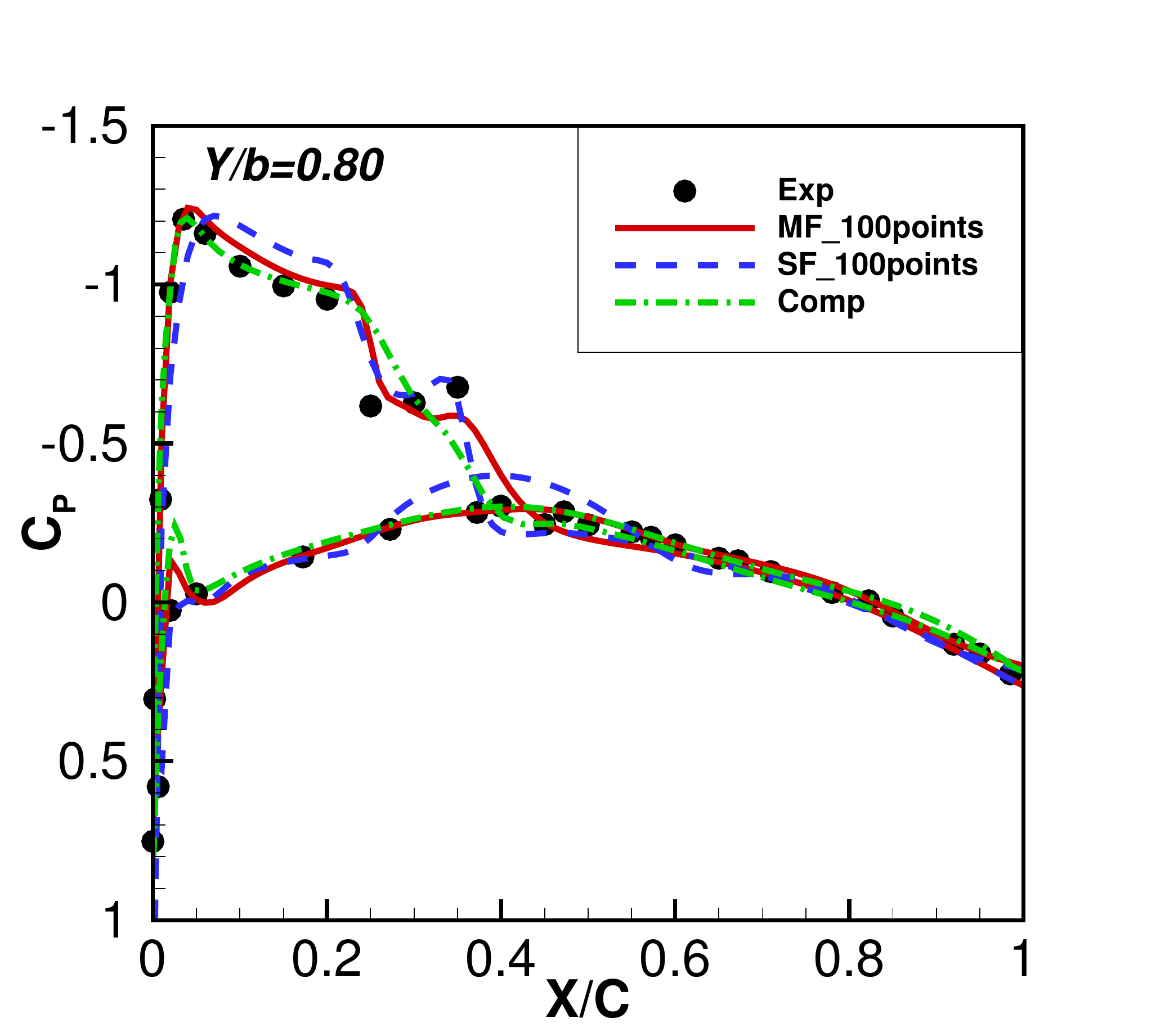}
    }
    \subfigure[$Y/b=0.90$]{
    \includegraphics[width=0.3\textwidth]{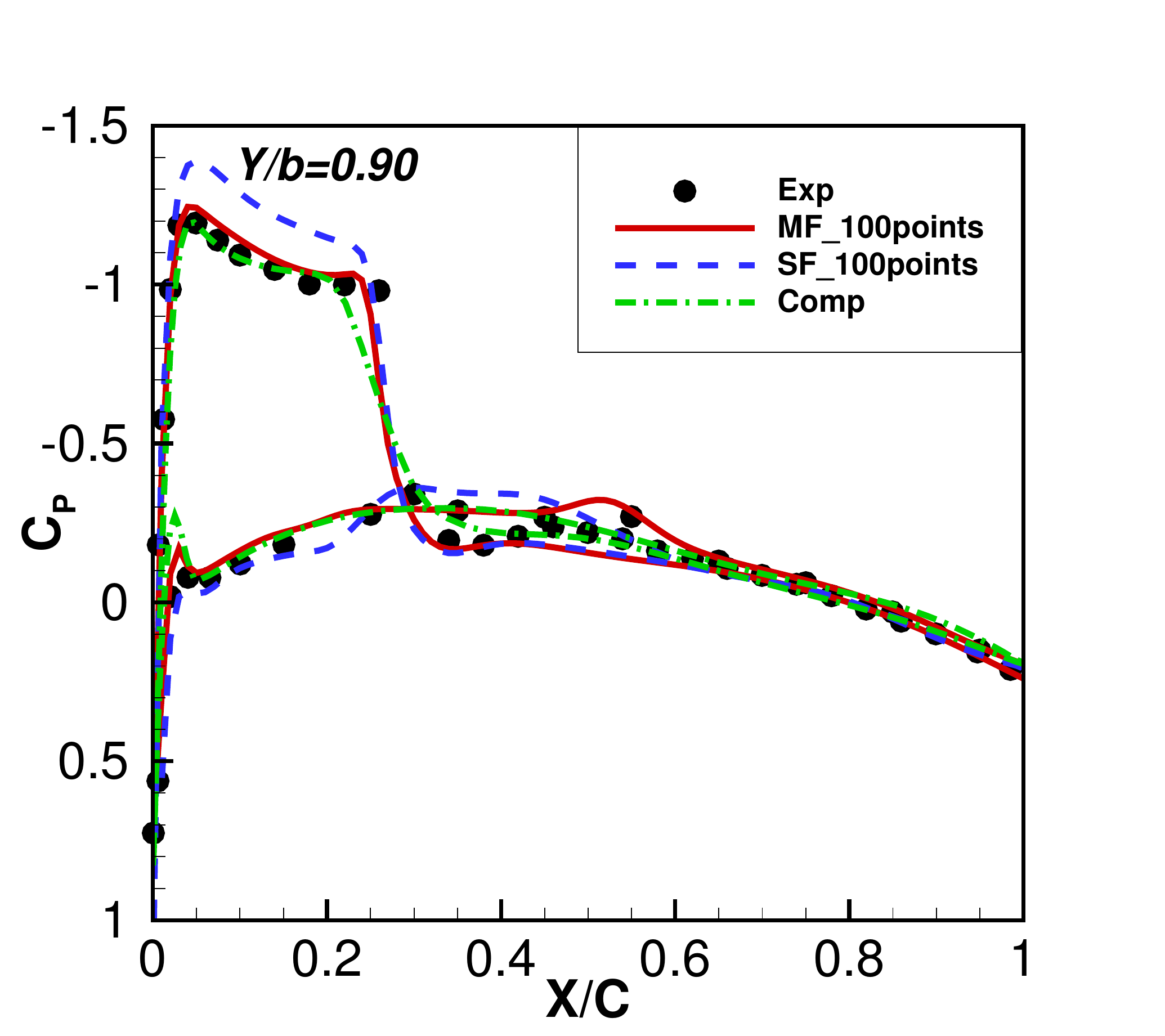}
    }
    \subfigure[$Y/b=0.95$]{
    \includegraphics[width=0.3\textwidth]{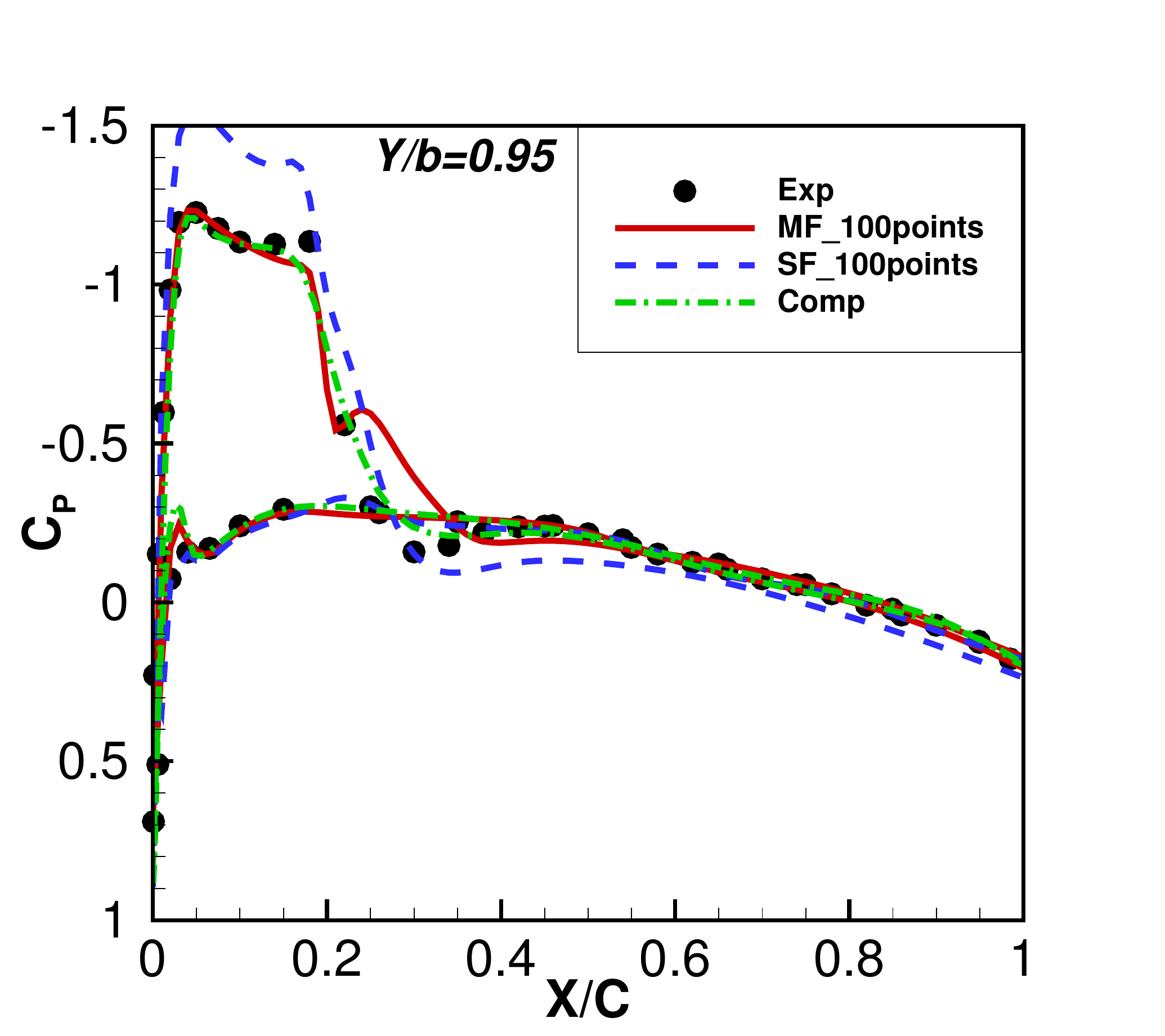}
    }
\caption{For less experimental data, $C_p$ at different sections for the test case D2$_{exp}$.}
\label{fig:Fig13}
\end{figure}

\begin{figure}[htbp]
\centering
    \subfigure[$Y/b=0.20$]{
    \includegraphics[width=0.3\textwidth]{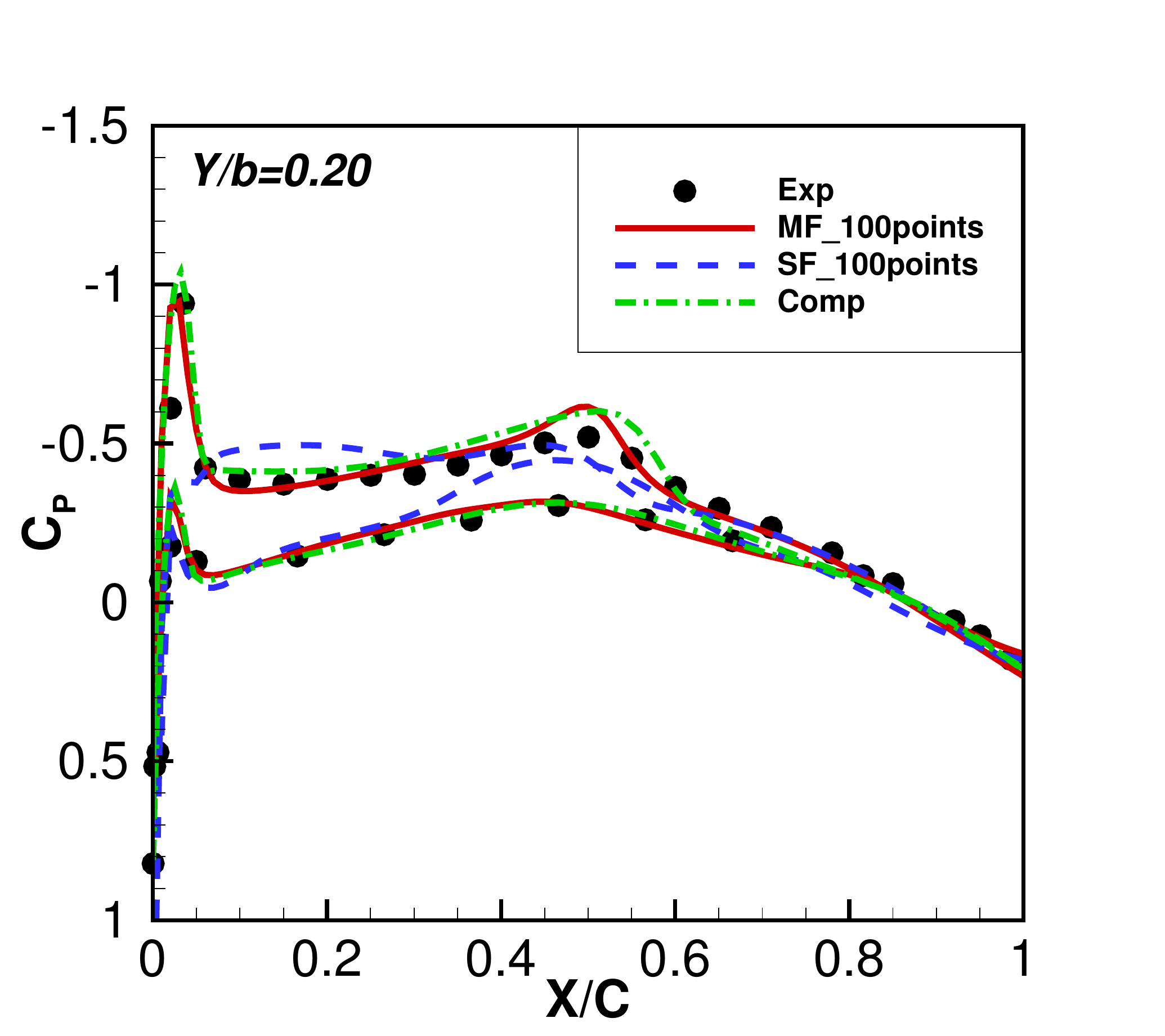}
    }
    \subfigure[$Y/b=0.44$]{
    \includegraphics[width=0.3\textwidth]{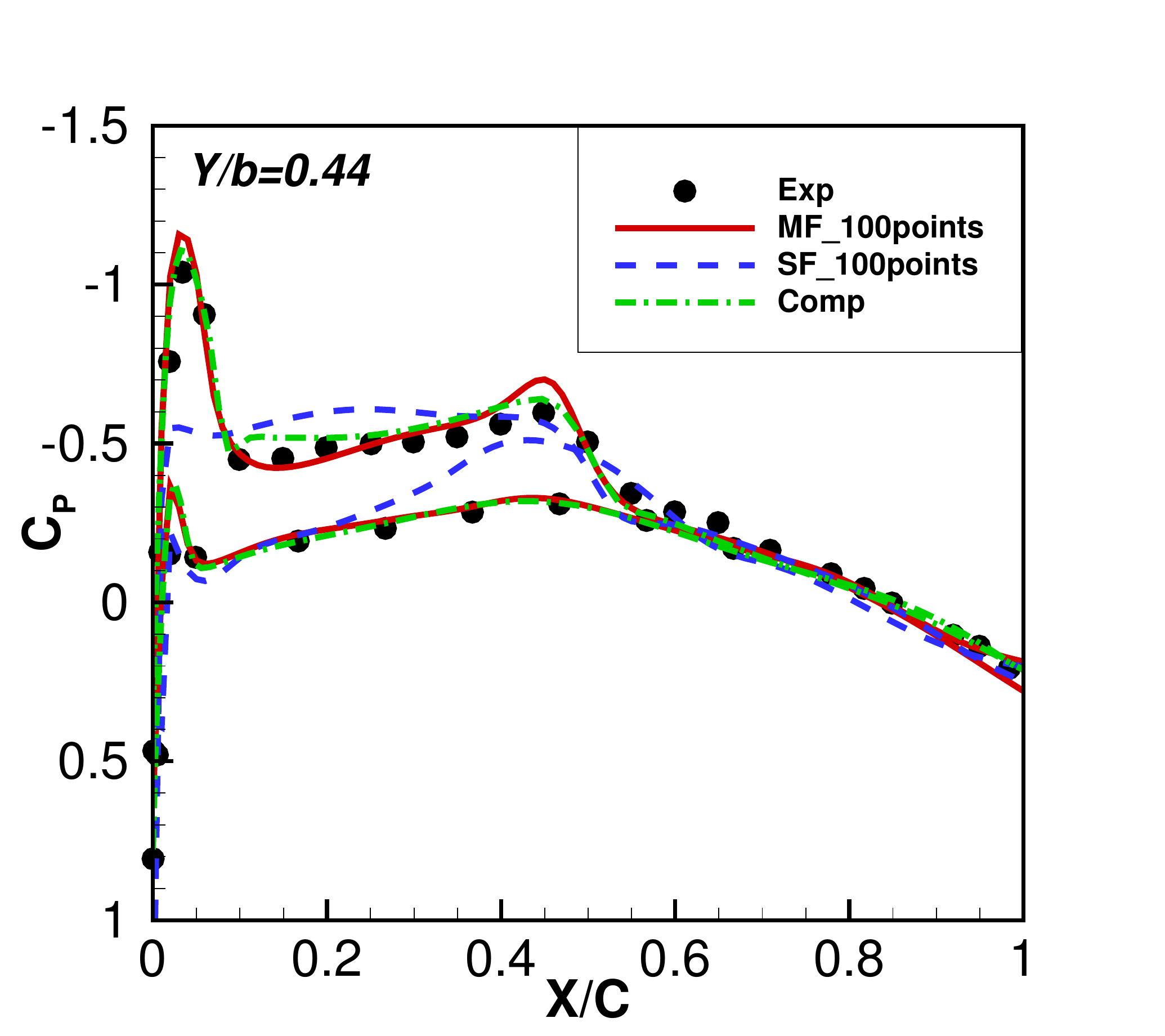}
    }
    \subfigure[$Y/b=0.65$]{
    \includegraphics[width=0.3\textwidth]{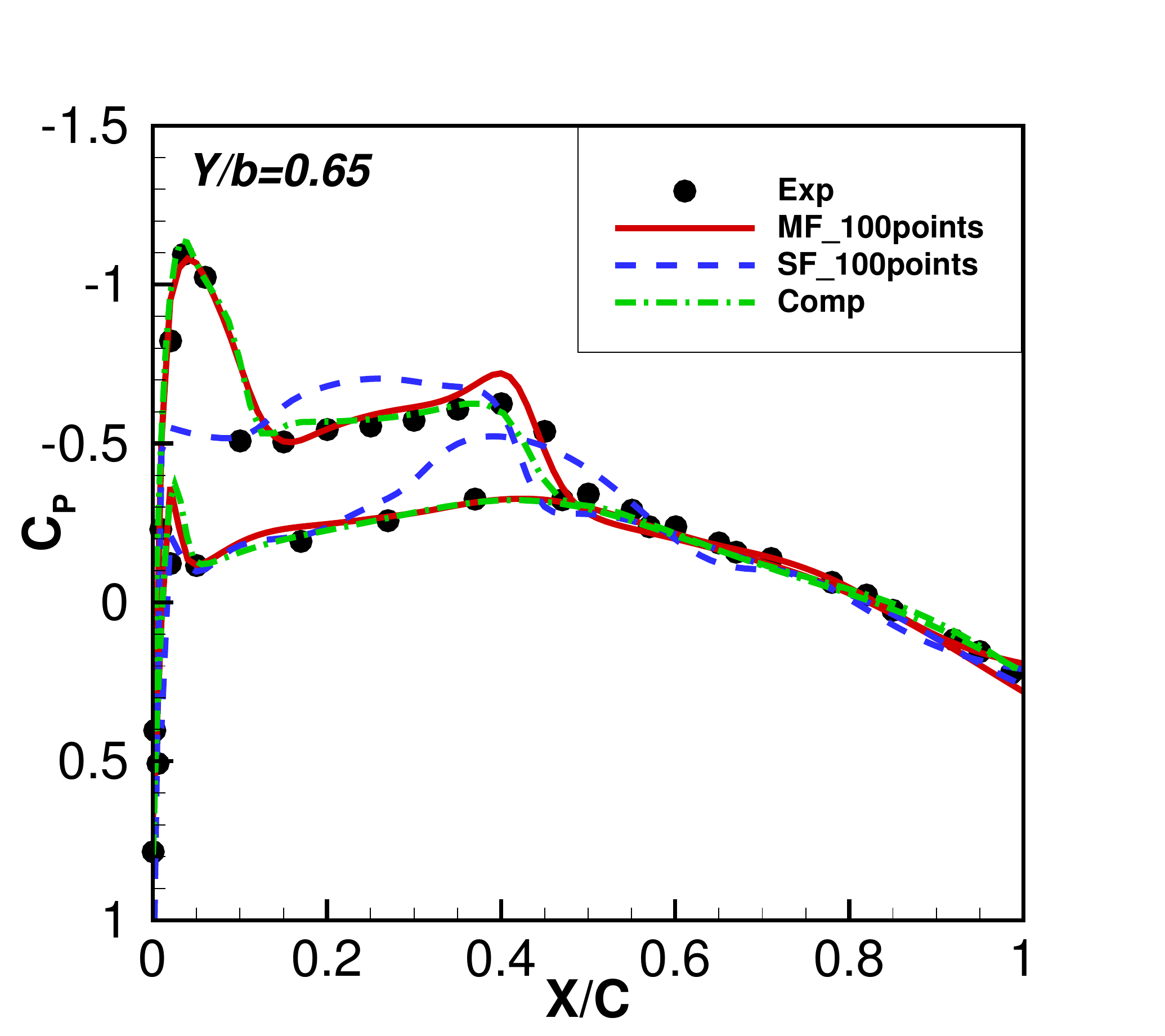}
    }
    \quad
    \subfigure[$Y/b=0.80$]{
    \includegraphics[width=0.3\textwidth]{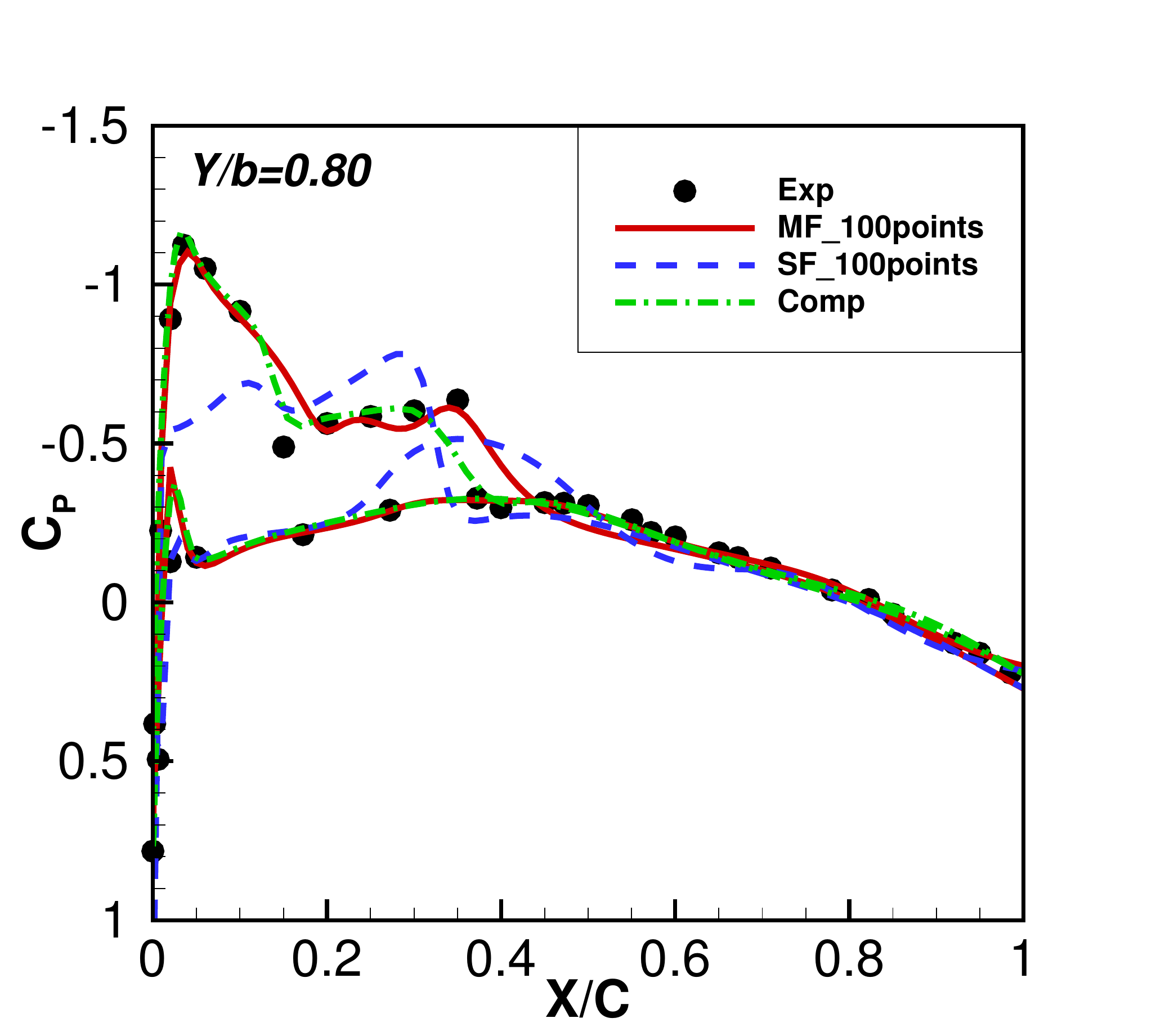}
    }
    \subfigure[$Y/b=0.90$]{
    \includegraphics[width=0.3\textwidth]{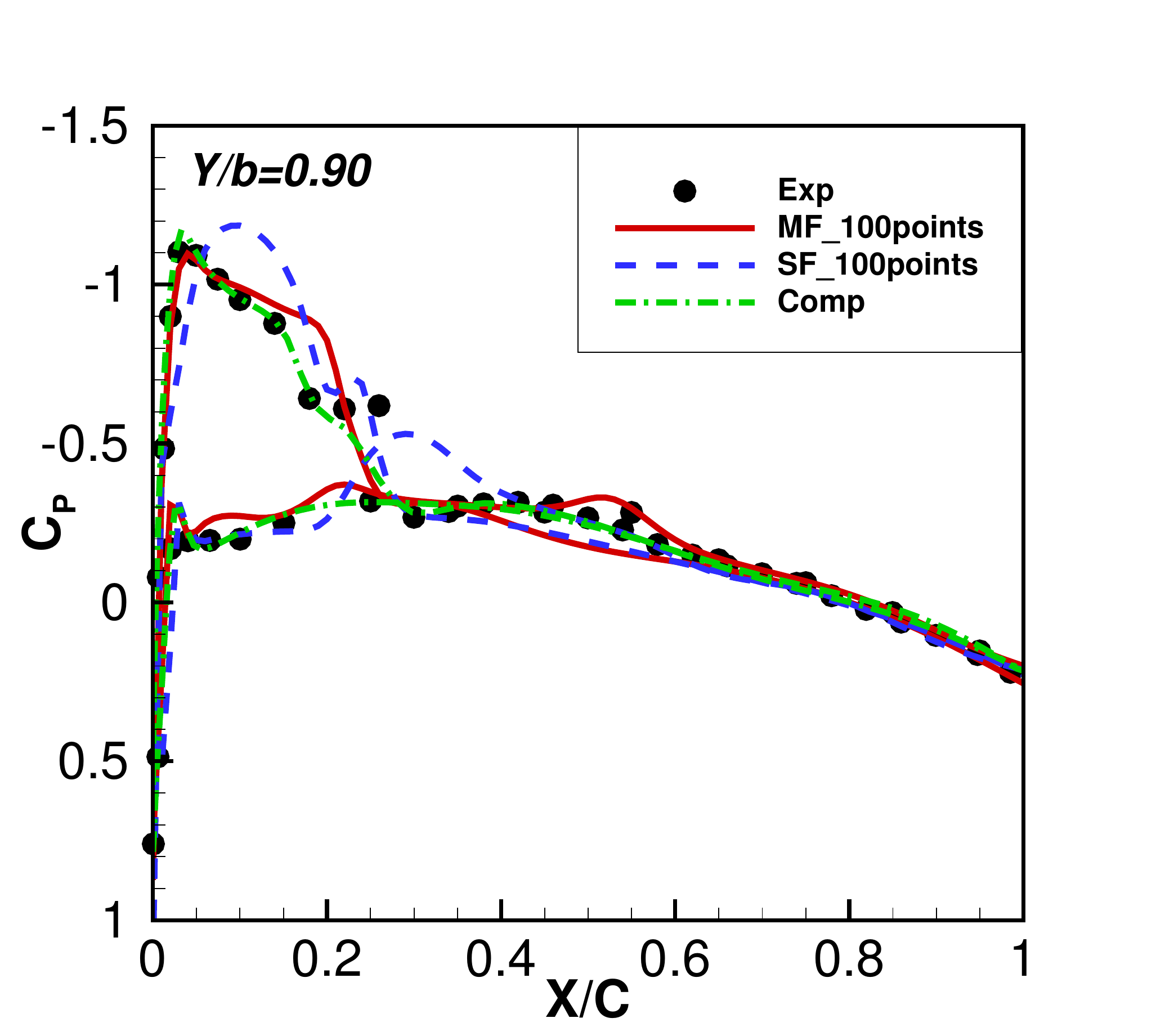}
    }
    \subfigure[$Y/b=0.95$]{
    \includegraphics[width=0.3\textwidth]{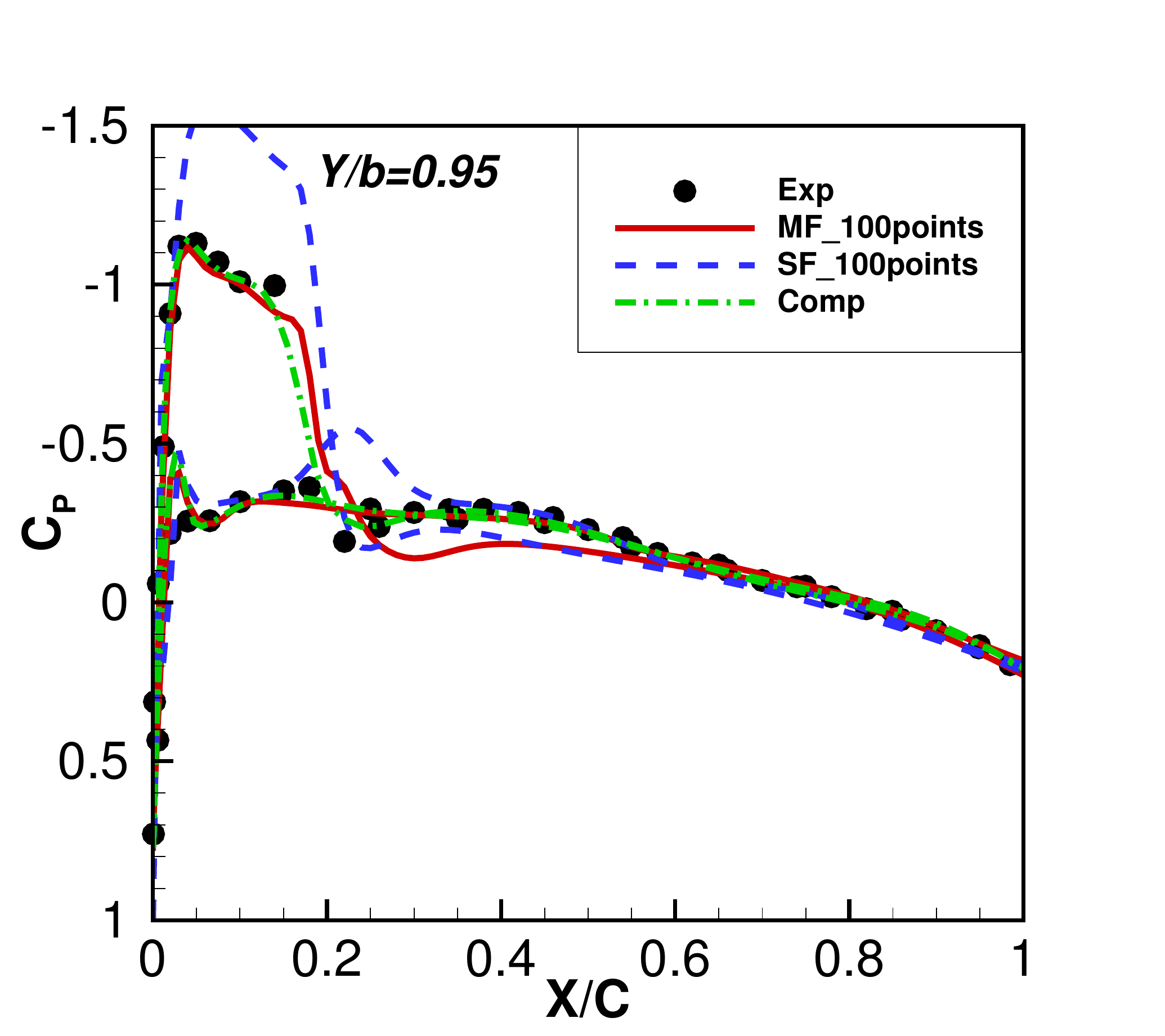}
    }
\caption{For less experimental data, $C_p$ at different sections for the test case C2$_{exp}$.}
\label{fig:Fig14}
\end{figure}

Fig.\ref{fig:Fig13} and Fig.\ref{fig:Fig14} present the predictions of $C_p$ at different sections using less experimental data in the training stage for the test case D2$_{exp}$ and C2$_{exp}$, respectively. As can be seen from these figures, the performance of SF is affected more significantly than MF, and MF can still give acceptable predictions. From the quantitative comparison in Table.\ref{tab:table2} and Table.\ref{tab:table3}, the overall accuracy of the multi-fidelity model is better than the single-fidelity model. In addition, the test case C2$_{exp}$ is predicted with larger error than B2$_{exp}$ and D2$_{exp}$. This is expected because the test case C2$_{exp}$ shows larger difference with the training case A2$_{exp}$ and E2$_{exp}$.

\begin{figure}[htbp]
\centering
    \subfigure[$Y/b=0.20$]{
    \includegraphics[width=0.31\textwidth]{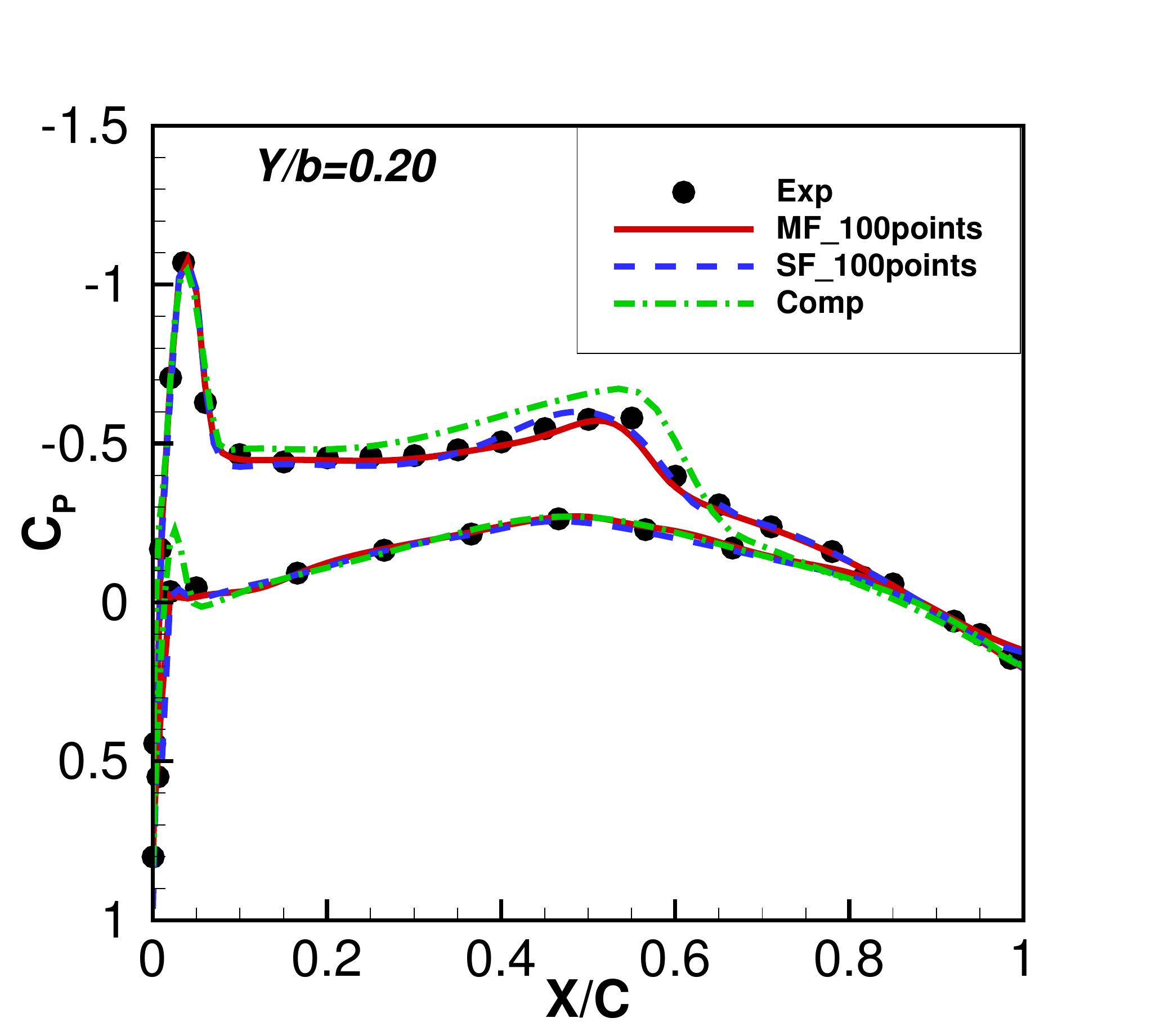}
    }
    \subfigure[$Y/b=0.44$]{
    \includegraphics[width=0.31\textwidth]{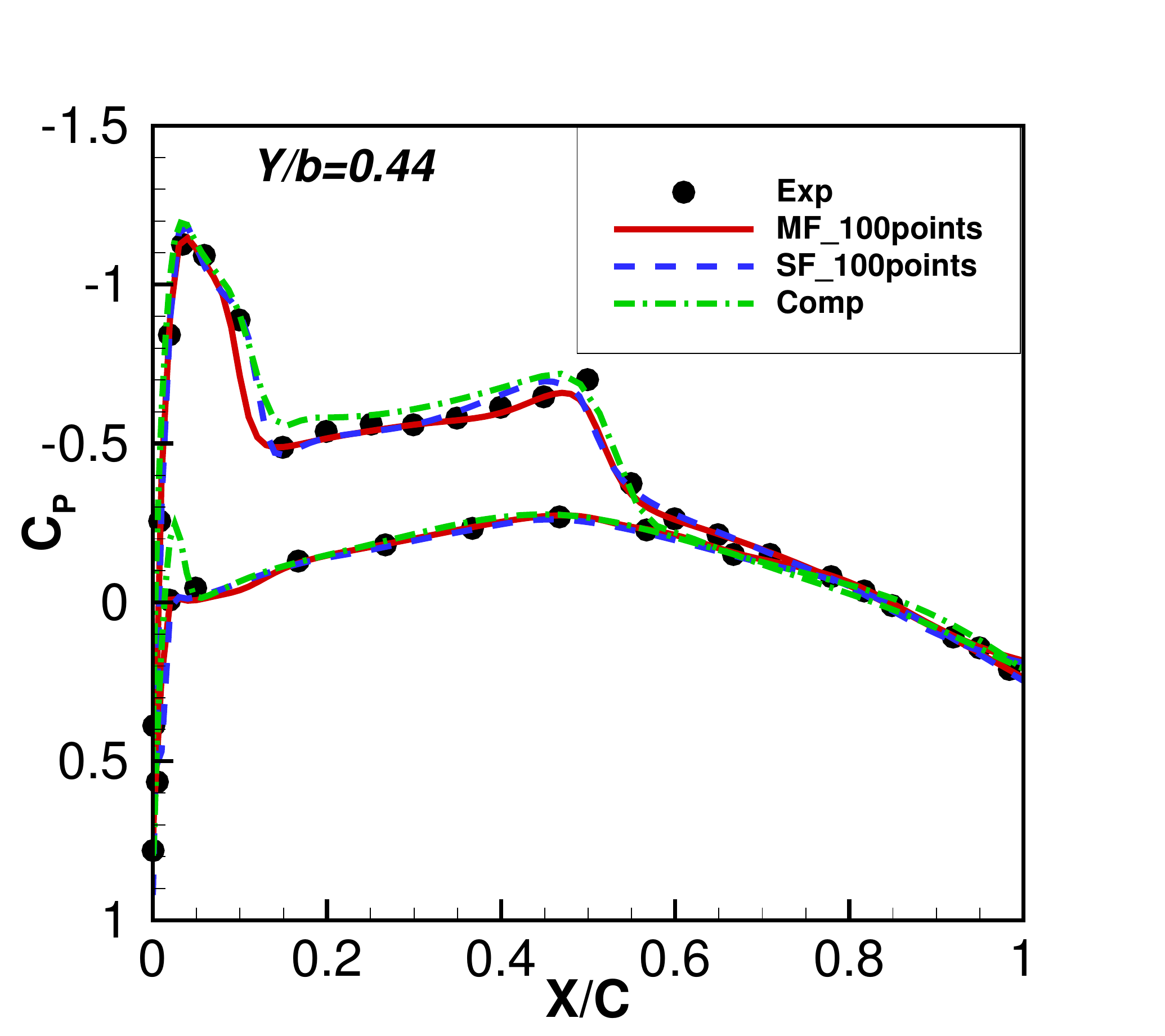}
    }
    \subfigure[$Y/b=0.65$]{
    \includegraphics[width=0.31\textwidth]{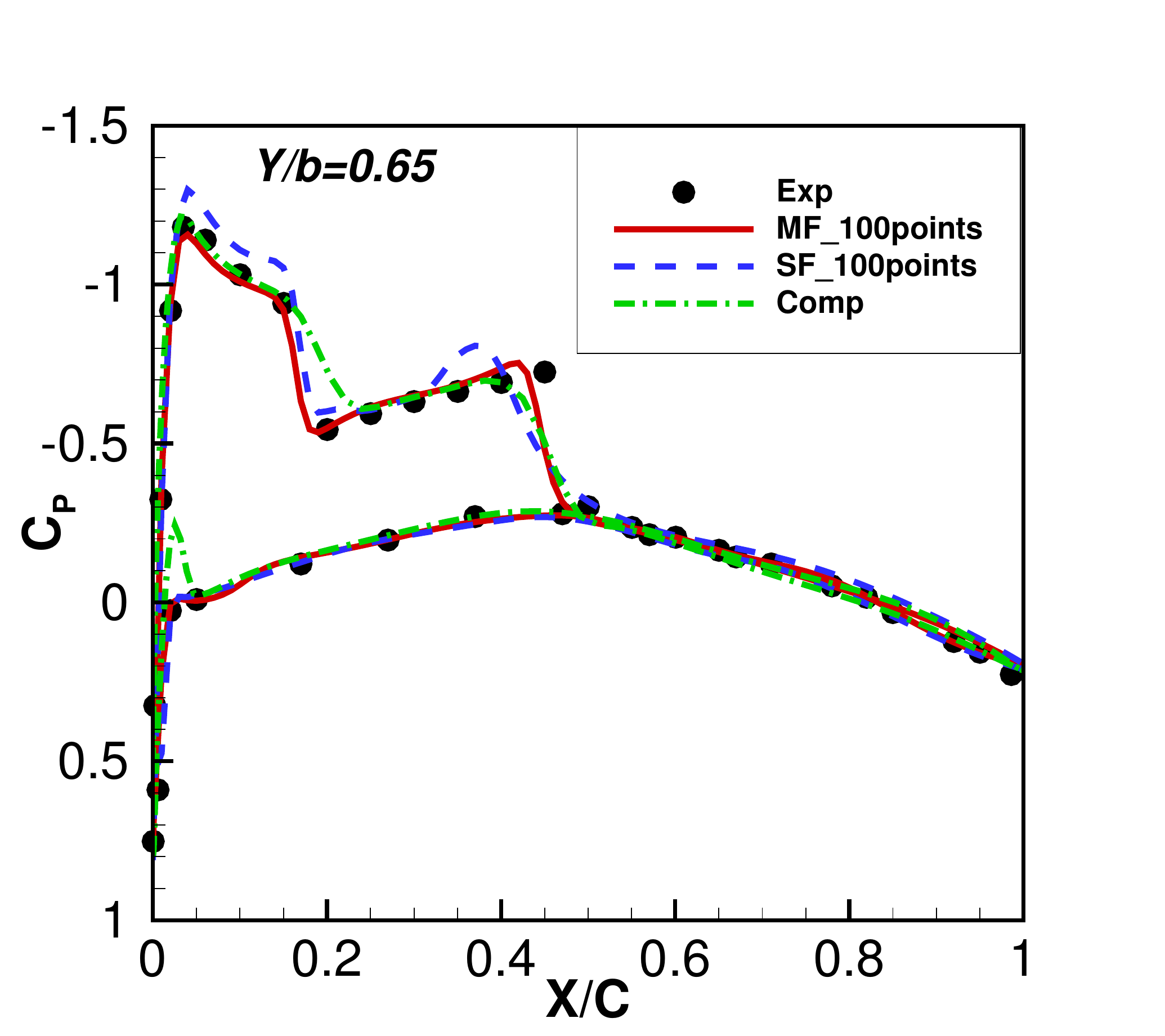}
    }
    \quad
    \subfigure[$Y/b=0.80$]{
    \includegraphics[width=0.31\textwidth]{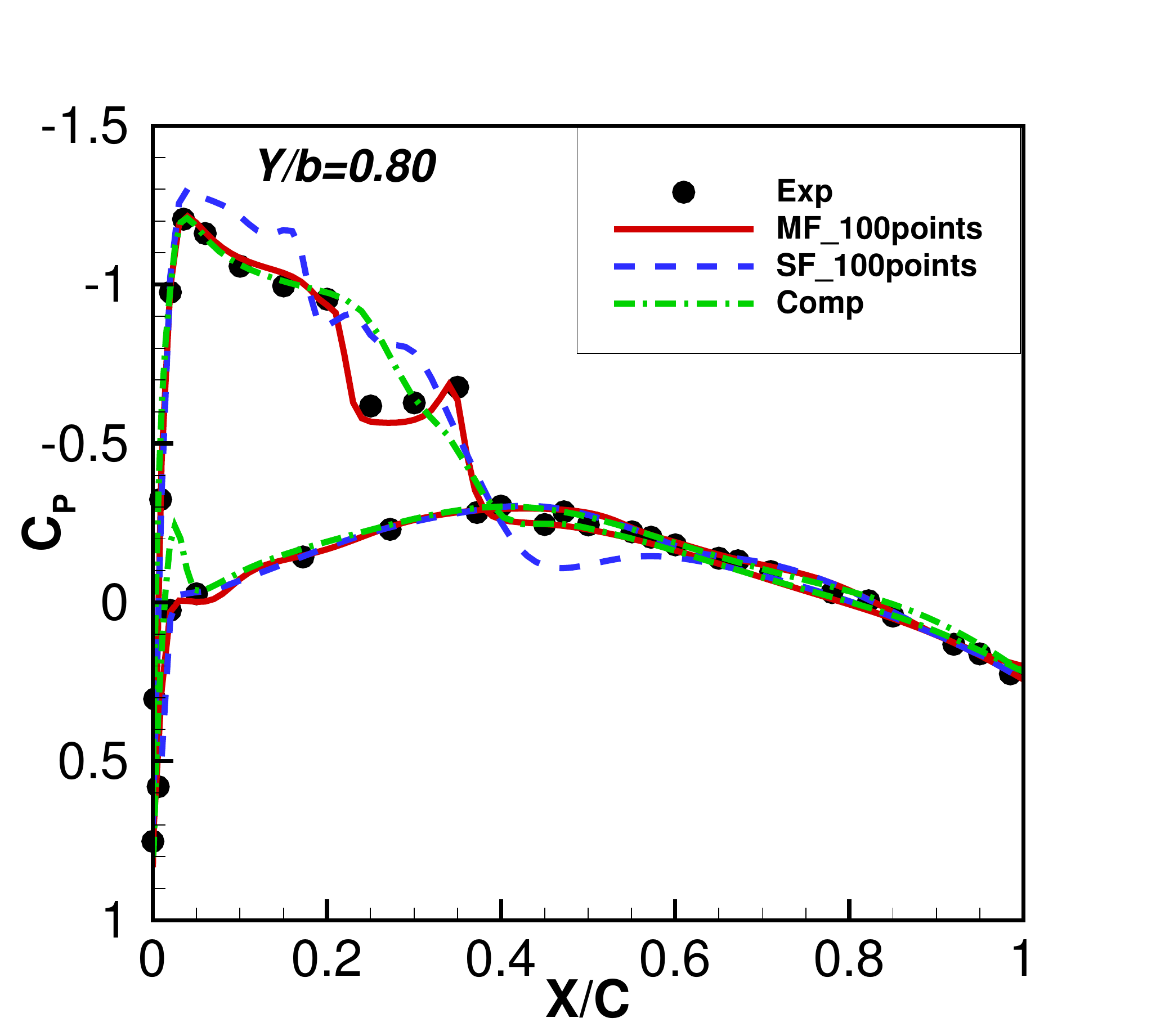}
    }
    \subfigure[$Y/b=0.90$]{
    \includegraphics[width=0.31\textwidth]{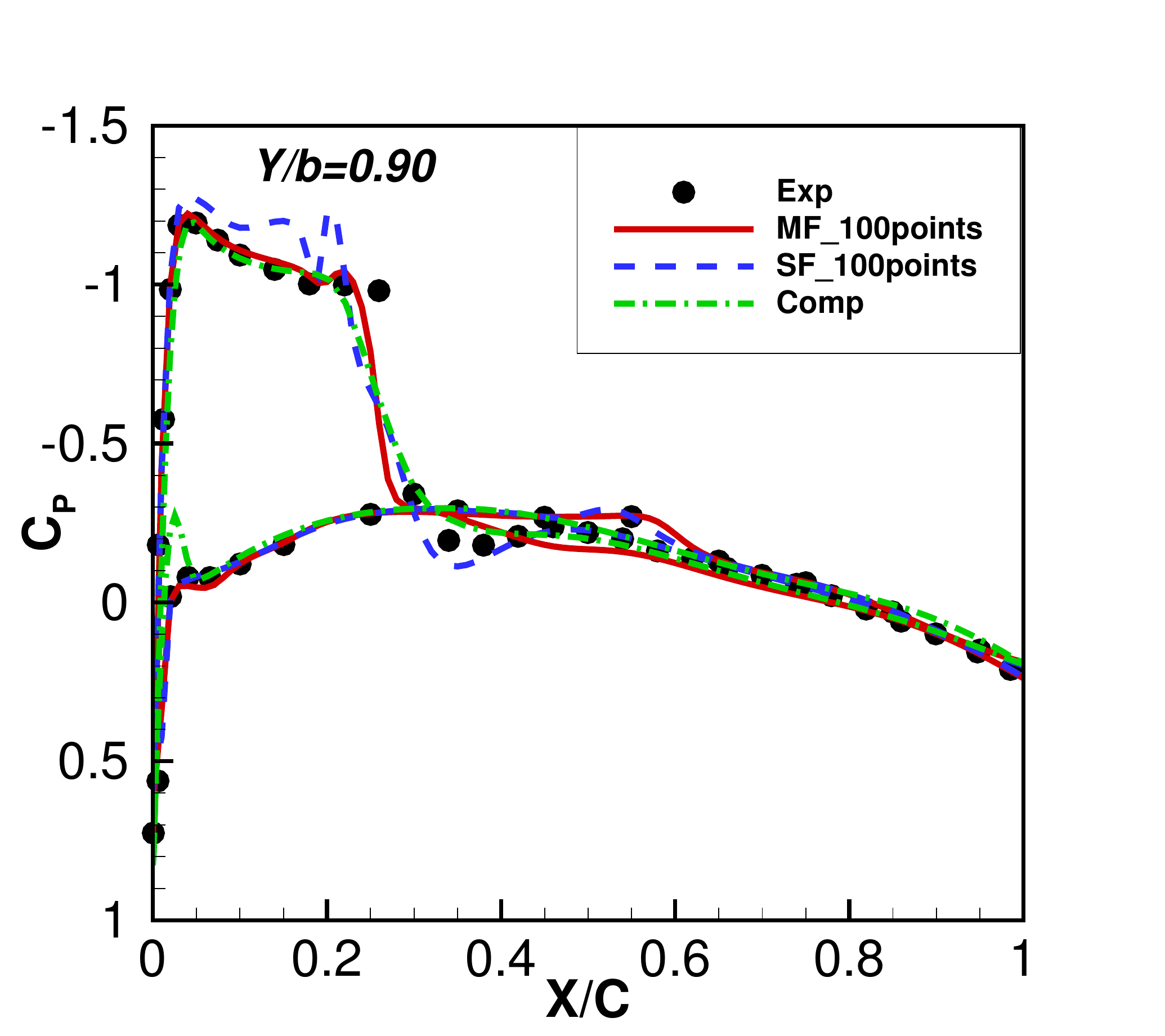}
    }
    \subfigure[$Y/b=0.95$]{
    \includegraphics[width=0.31\textwidth]{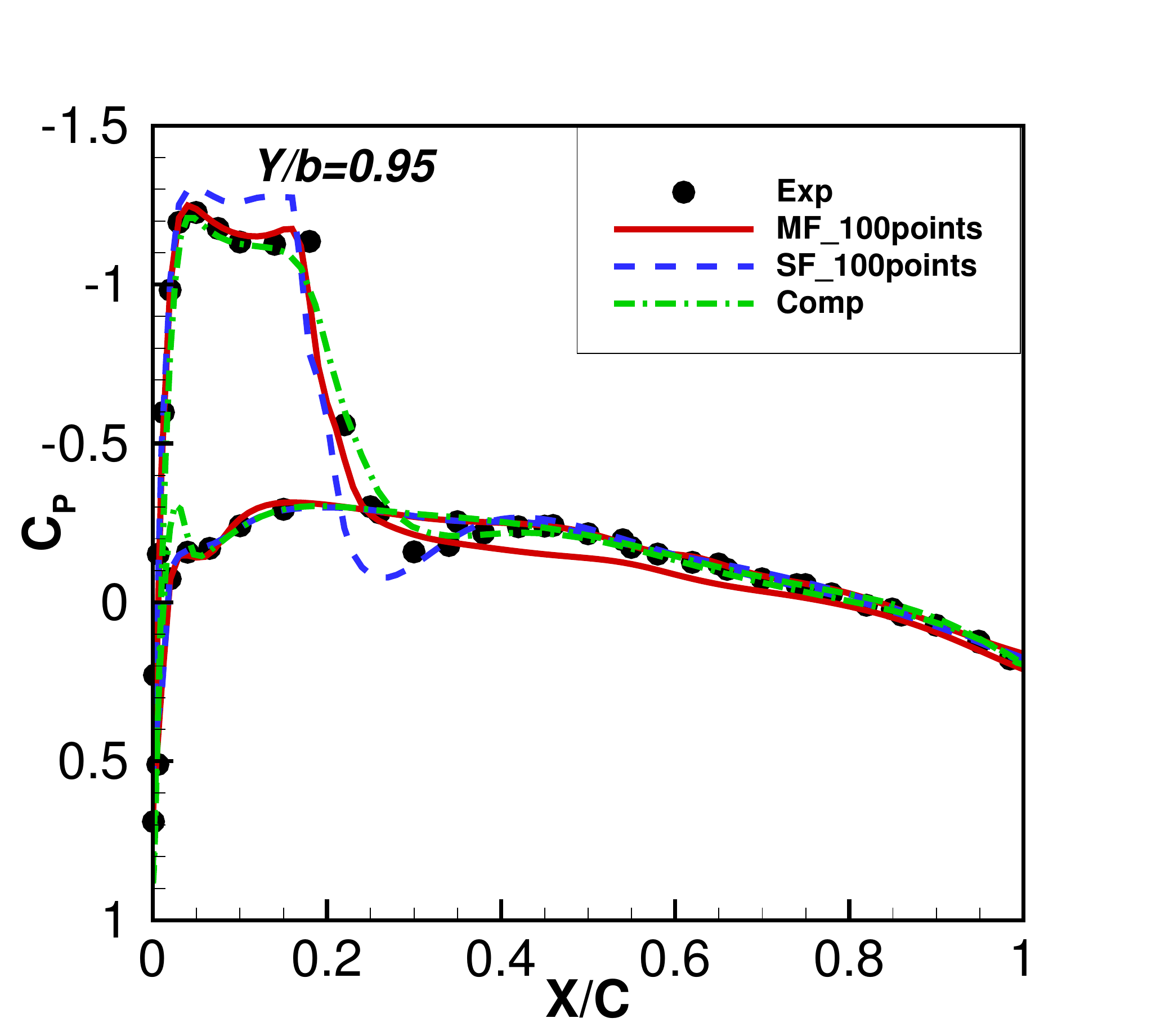}
    }
\caption{For the training data without the section $Y/b=0.65$, $C_p$ at different sections for the test case D2$_{exp}$.}
\label{fig:Fig15}
\end{figure}

\begin{figure}[htbp]
\centering
\includegraphics[width=1\textwidth]{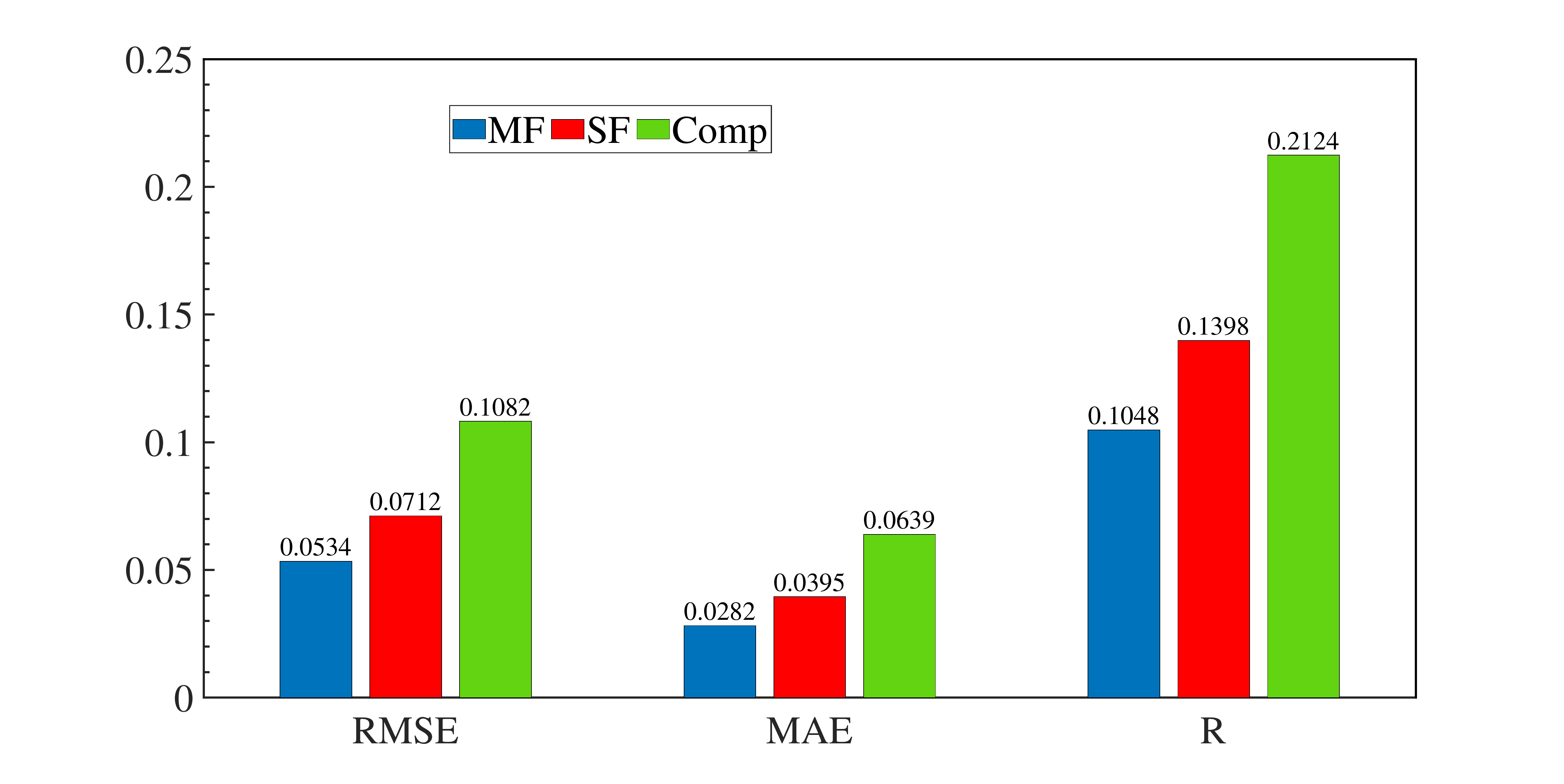}
\caption{Comparison of errors for the test case D2$_{exp}$.}
\label{fig:Fig16}
\end{figure}

\begin{figure}[htbp]
\centering
    \subfigure[$Y/b=0.20$]{
    \includegraphics[width=0.31\textwidth]{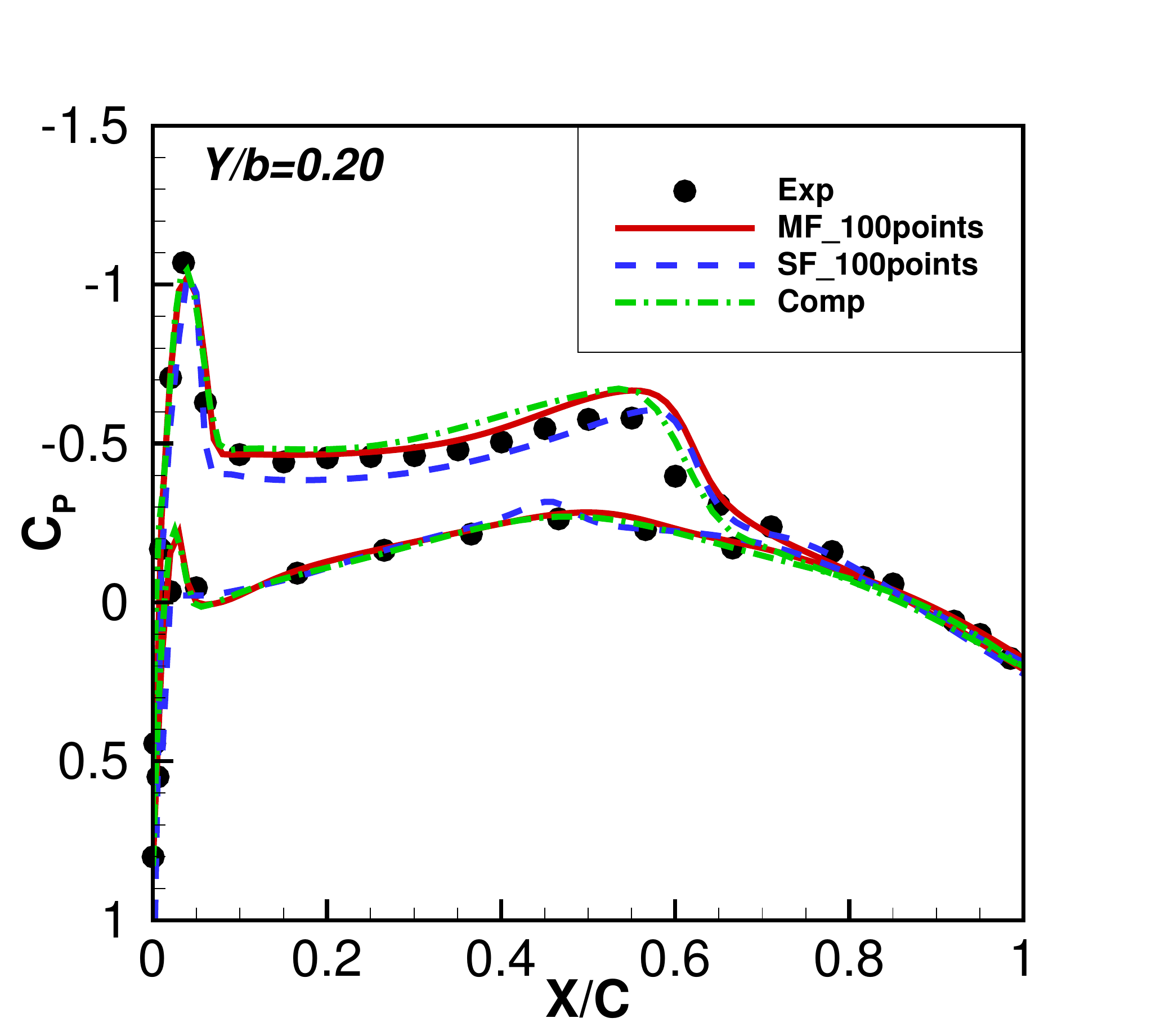}
    }
    \subfigure[$Y/b=0.44$]{
    \includegraphics[width=0.31\textwidth]{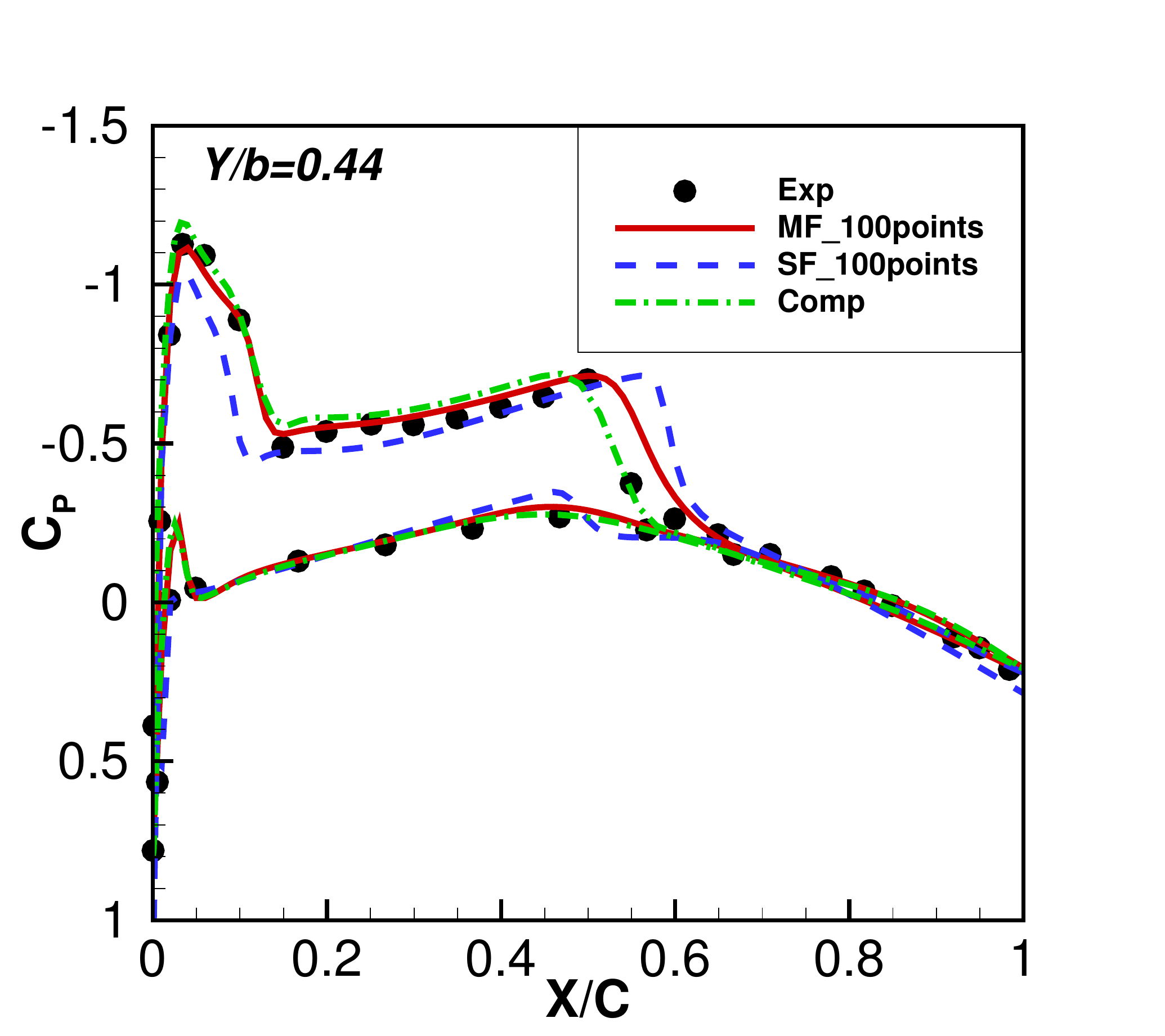}
    }
    \subfigure[$Y/b=0.65$]{
    \includegraphics[width=0.31\textwidth]{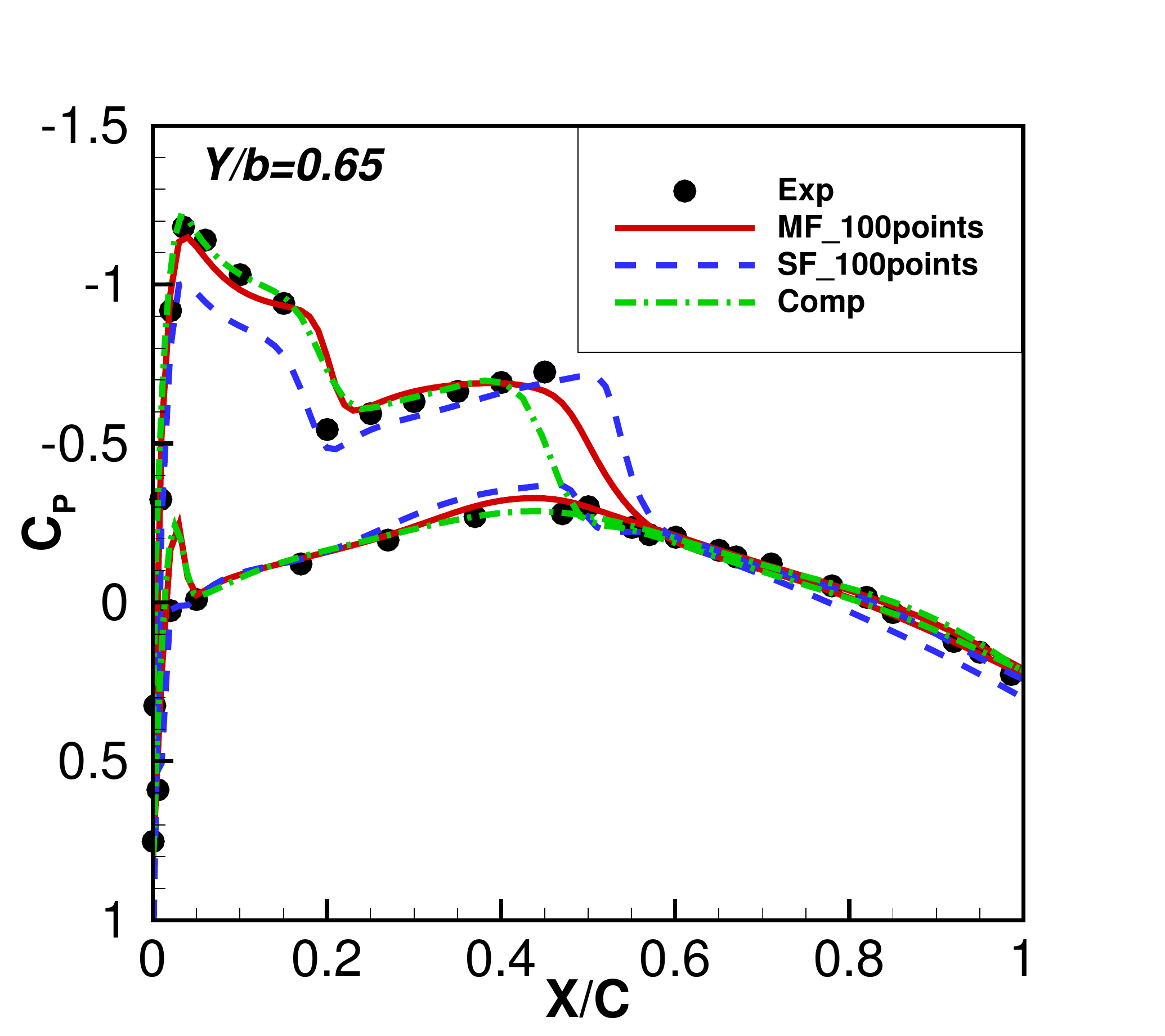}
    }
    \quad
    \subfigure[$Y/b=0.80$]{
    \includegraphics[width=0.31\textwidth]{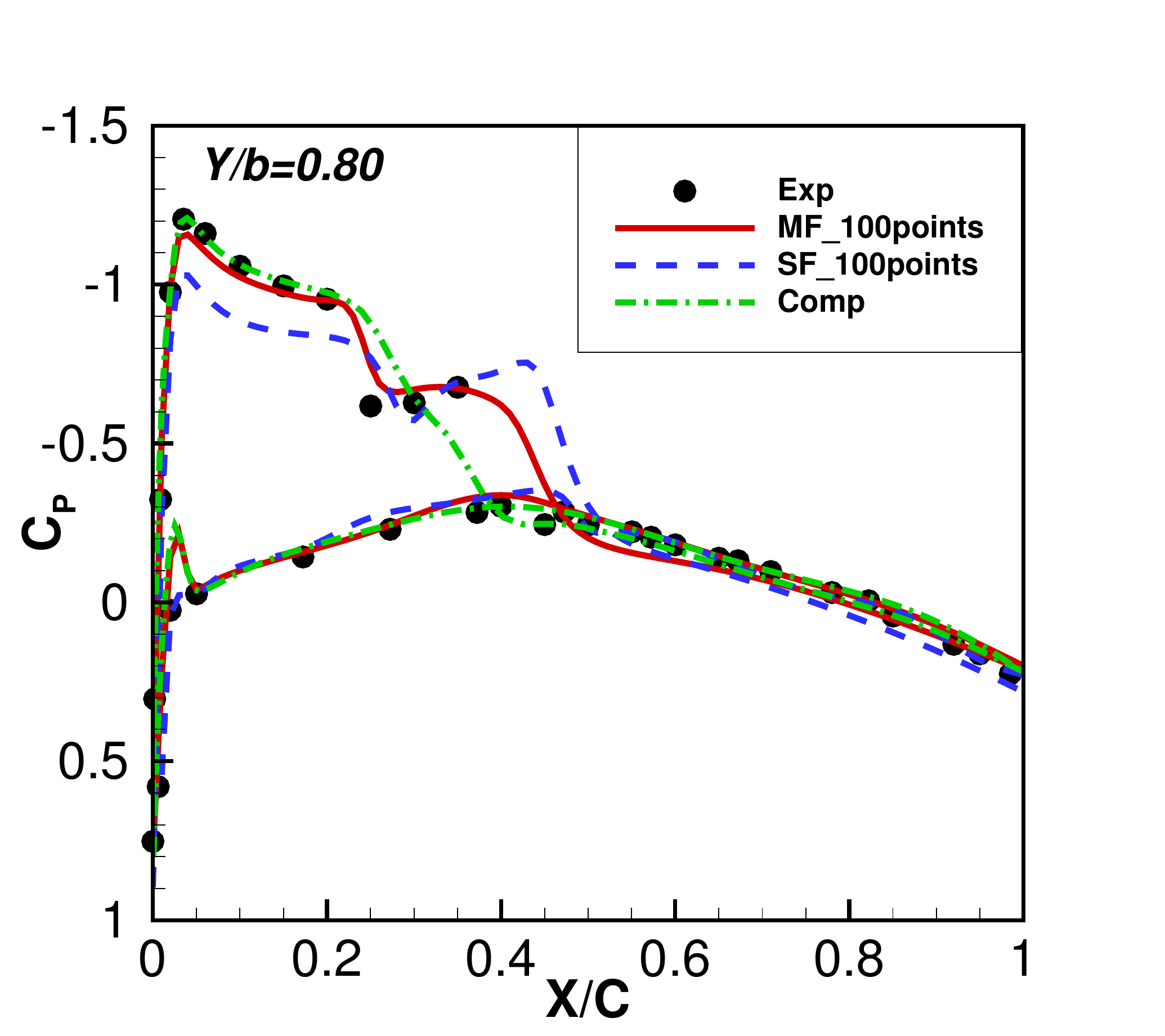}
    }
    \subfigure[$Y/b=0.90$]{
    \includegraphics[width=0.31\textwidth]{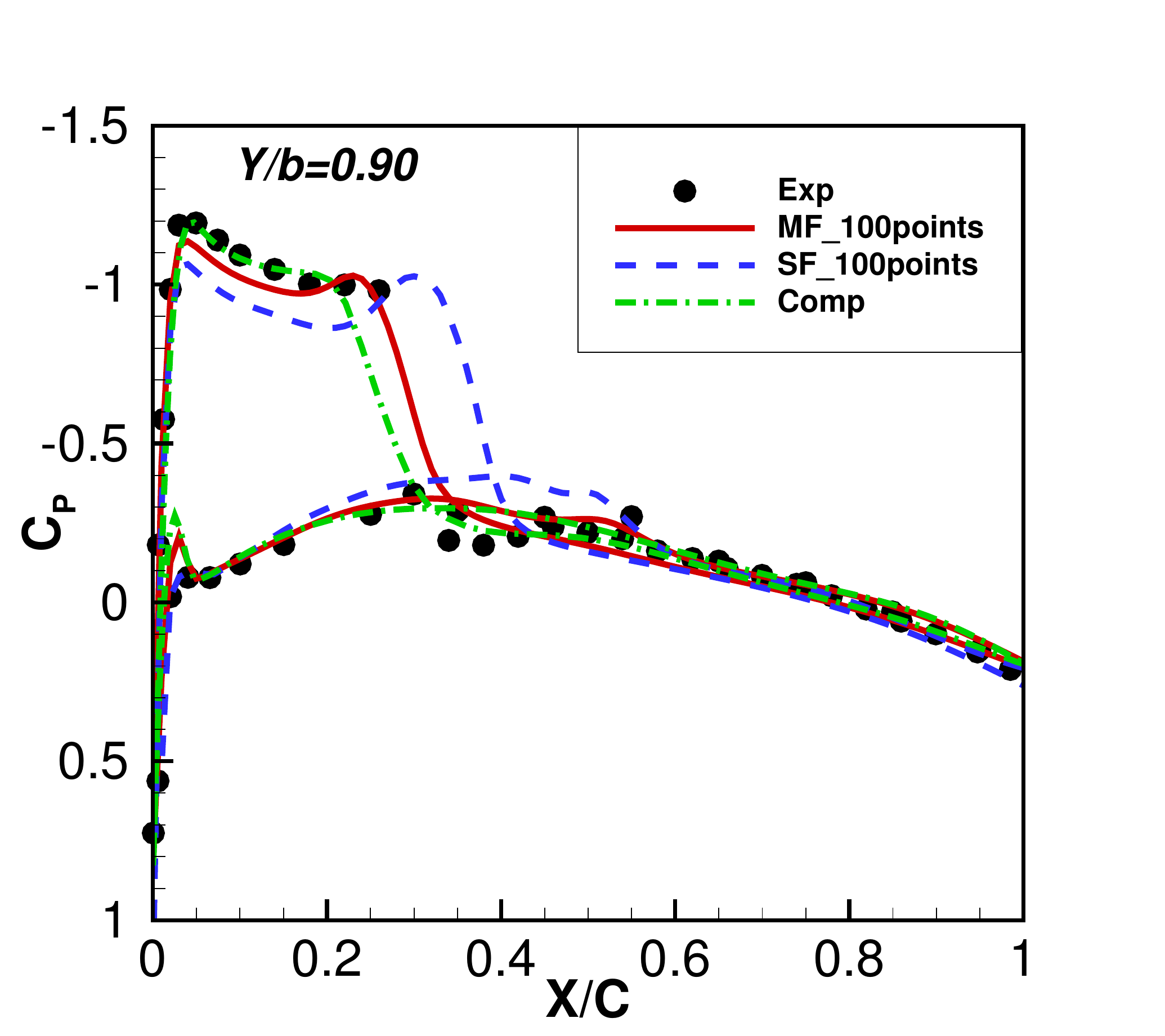}
    }
    \subfigure[$Y/b=0.95$]{
    \includegraphics[width=0.31\textwidth]{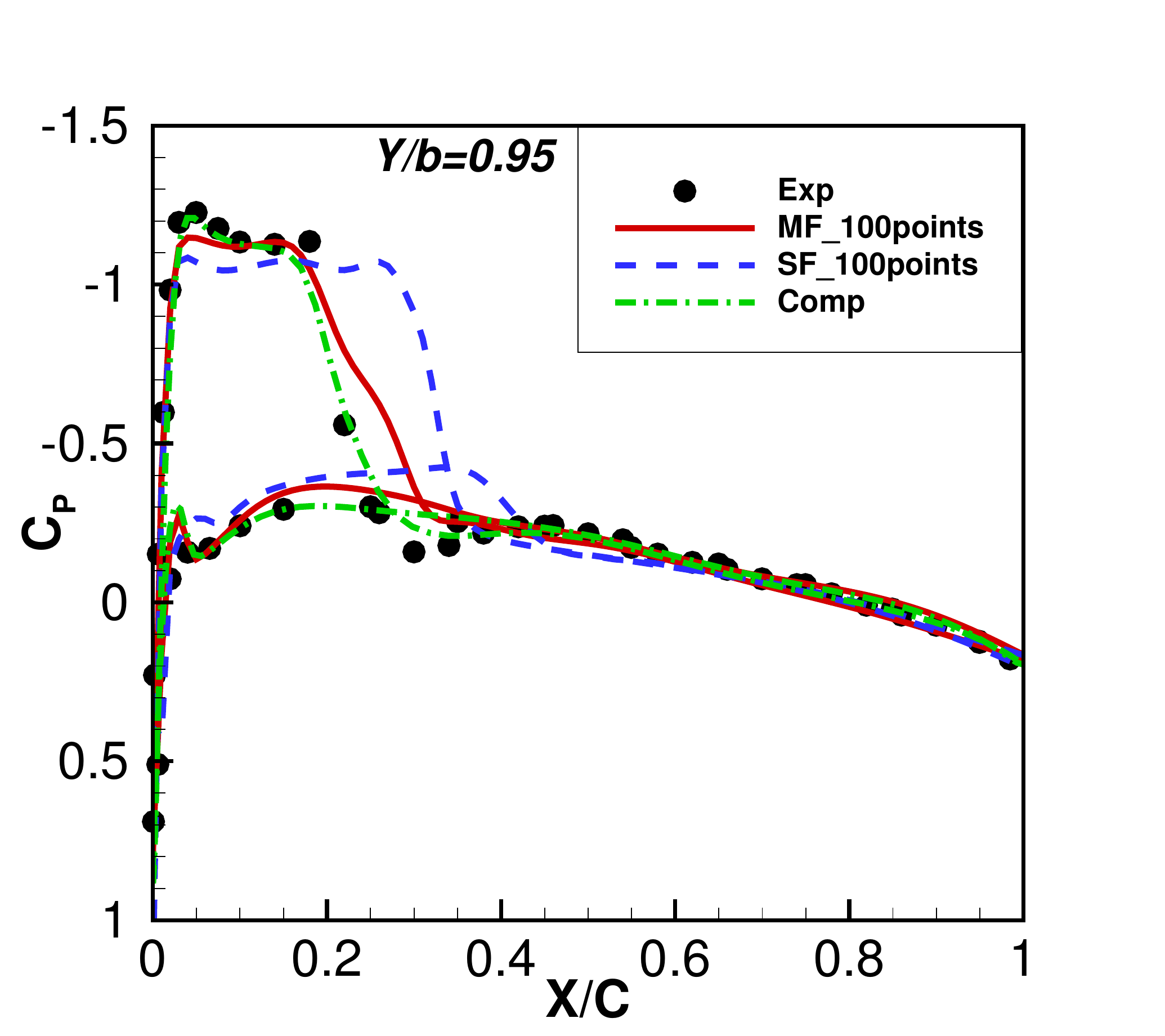}
    }
\caption{For varying $Ma$, $C_p$ at different sections for the test case D2$_{exp}$.}
\label{fig:Fig17}
\end{figure}

\begin{figure}[htbp]
\centering
\includegraphics[width=1\textwidth]{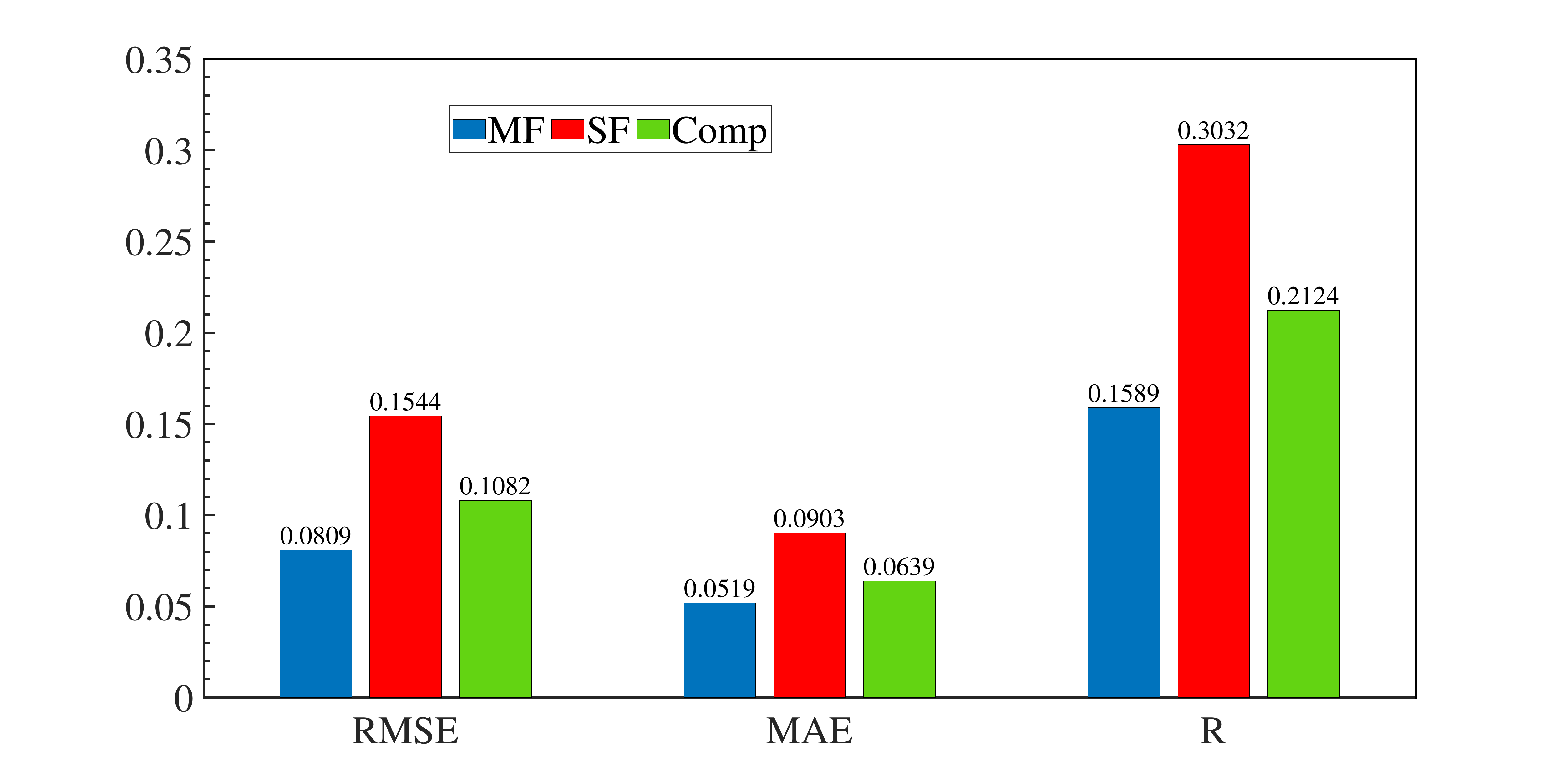}
\caption{Comparison of errors for the test case D2$_{exp}$ with varying $Ma$.}
\label{fig:Fig18}
\end{figure}

\subsection{Investigation on angle of attack and location (excluding a particular section)} \label{section:sec3.4}

To further examine the performance of the proposed model, a study with respect to location and $\alpha$ is undertaken in this subsection. Note that, the train set is same as that in section \ref{section:sec3.2}, except lacking data at section $Y/b=0.65$, as illustrated in Table.\ref{fig:Fig6}.

Using the training data without a section $Y/b=0.65$, $C_p$ at different sections for the test case D2$_{exp}$ is predicted in Fig.\ref{fig:Fig15}. From the graph above we can see that MF predicts $C_p$ at the section $Y/b=0.65$ with smaller error, and there is no dramatic discrepancy between the models trained by the data with and without the section $Y/b=0.65$ in the predictions of the other sections. To directly illustrate the difference between the single-fidelity and multi-fidelity model, comparison between them of errors for the test case D2$_{exp}$ is shown in Fig.\ref{fig:Fig16}. The $R$, $MAE$, and $RMSE$ of MF are smaller than those of SF and computation, thereby showing the ability of the proposed multi-fidelity model. Furthermore, this is reasonable because the low-fidelity data as the prior information at unknown location and $\alpha$ is provided for model training, then avoiding over-fitting and improving the accuracy.

\subsection{Investigation on Mach number} \label{section:sec3.5}

For a more comprehensive evaluation of the multi-fidelity model performance, $Ma$ is adopted as an additional input variable in this subsection. In general, most of the wind-tunnel experiments are to fix $Ma$ with varying $\alpha$. Therefore, the proposed model here is constructed by ten experimental cases A1$_{exp}$, B1$_{exp}$, C1$_{exp}$, D1$_{exp}$, E1$_{exp}$, A3$_{exp}$, B3$_{exp}$, C3$_{exp}$, D3$_{exp}$, E3$_{exp}$ and fifteen computational cases A1$_{com}$, B1$_{com}$, C1$_{com}$, D1$_{com}$, E1$_{com}$, A2$_{com}$, B2$_{com}$, C2$_{com}$, D2$_{com}$, E2$_{com}$, A3$_{com}$, B3$_{com}$, C3$_{com}$, D3$_{com}$, E3$_{com}$, as depicted in Table.\ref{fig:Fig6}.

As shown in Fig.\ref{fig:Fig17}, the prediction of the test case D2$_{exp}$ at $Ma=0.84$ is given by the proposed model with the training data at $Ma=0.7$ and $Ma=0.88$. It is revealed in this figure that the trend of pressure coefficient distributions at varying $Ma$ can be captured by MF very well. In addition, the peaks of pressure coefficient at most of the sections are predicted by SF inaccurately, while MF in the use of auxiliary low-fidelity data is capable of capturing those precisely. For further quantitative analysis, Fig.\ref{fig:Fig18} gives comparison of errors for the test case D2$_{exp}$. The errors of MF are smallest among those of the others, and the errors of SF are even larger than the results from CFD, which has a good agreement with the results in Fig.\ref{fig:Fig17}. The $R$ error for $Ma=0.84$ is depicted in Table.\ref{tab:table4}.

\begin{table}[htbp]
\caption{\label{tab:table4} $R$ at $Ma=0.84$ for B2$_{exp}$, C2$_{exp}$, D2$_{exp}$}
\centering
\begin{tabular}[b]{cccc}
\hline
Case  & MF (\%) & SF (\%) & Comp (\%) \\
\hline
E2$_{exp}$ & 14.69   & 22.87   & 19.09     \\
D2$_{exp}$ & 15.89   & 30.32   & 21.24     \\
C2$_{exp}$ & 16.93   & 30.00   & 20.18     \\
B2$_{exp}$ & 17.60   & 26.33   & 21.26     \\
A2$_{exp}$ & 20.64   & 23.04   & 27.85     \\
\hline
\end{tabular}
\end{table}

\section{Conclusion} \label{section:sec4}

In this work, a multi-fidelity model that allows for aerodynamic distribution prediction based on a modified DNN with a decent of amount of low-fidelity data and a small number of high-fidelity data is proposed. The loss function benefits from re-weighting in the ML community, which is designed to fuse data with different fidelity. This modified DNN model allows modeling based on dataset with multiple fidelities. The proposed approach is validated by modeling aerodynamic distributions of transonic flow past a ONERA M6 wing, where low-fidelity and high-fidelity data are obtained from wind-tunnel experiment and RANS model, respectively. The proposed model can predict high-fidelity results accurately at varying Mach numbers and angles of attack across different wing sections. Furthermore, a extrapolated task is performed to validate the generalization capability of the model, throughing using less high-fidelity data. In summary, three main conclusions are drawn:
\begin{enumerate}
\item The proposed DNN model imposes low-fidelity data as soft constraint in the training stage, thus providing prior information for unknown parameter space and avoiding over-fitting;
\item The proposed model can improve the generalization capability over single-fidelity model, and shows higher accuracy than the results from CFD in predicting aerodynamic distributions at varying flow conditions and unknown location.
\item For a extrapolated task or limited high-fidelity training data, the proposed approach can still provide reasonable predictions. Results also indicate that introducing low-fidelity data to test problem of interest can also improve the corresponding prediction of high-fidelity solution.
\item Future works include the extension of the proposed framework to turbulent flow data, or multiple data sources.
\end{enumerate}

\section*{Acknowledgments}

This work was supported by the National Natural Science Foundation of China (No.12072282 and No.91852115). The authors would like to thank the anonymous reviewers for their invaluable comments.

\bibliography{sample}

\end{document}